\documentclass[numberedappendix]{emulateapj}
\usepackage{subfigure}
\usepackage{color}
\usepackage{amsmath}
\usepackage{mathrsfs}
\usepackage{multirow}
\usepackage{soul}
\usepackage{amssymb} 
\def\be{\begin{equation}}
\def\ee{\end{equation}}
\def\bd{\begin{displaymath}}
\def\ed{\end{displaymath}}
\def\ba{\begin{aligned}}
\def\ea{\end{aligned}}
\def\nms{\mathsurround=0pt}
\def\oversim#1#2{\lower 4pt\vbox{\baselineskip 0pt \lineskip 1pt
    \ialign{$\nms#1\hfil##\hfil$\crcr#2\crcr\sim\crcr}}}
\def\ga{\mathrel{\mathpalette\oversim>}}
\def\la{\mathrel{\mathpalette\oversim<}}
\def\arcdeg{{^{\circ}}}

\def\bh{M_{\bullet}}

\def\kpc{\rm ~kpc}
\def\msun{M_{\odot}}
\def\AU{{\rm AU}}
\def\kms{{\rm km\,s^{-1}}}

\def\rg{r_{\rm g}}

\begin{document}

\title{On Testing the Kerr Metric of the Massive Black Hole in the
Galactic Center via Stellar Orbital Motion: Full General
Relativistic Treatment}

\author{FUPENG ZHANG$^{1,2,\ast}$, YOUJUN LU$^{2,\dagger}$ \& QINGJUAN
YU$^{1,\ddagger}$} 
\affil{ $^1$\,Kavli Institute for Astronomy and Astrophysics, Peking
University, Beijing, 100871, China; $^\ast$\,zhangfupeng@pku.edu.cn; $^\ddagger$\,yuqj@pku.edu.cn\\
$^2$\,National Astronomical Observatories, Chinese Academy of
Sciences, Beijing, 100012, China; ~$^\dagger$luyj@nao.cas.cn }

\begin{abstract}
The S-stars, discovered in the vicinity of the massive black hole
(MBH) in the Galactic center (GC), are anticipated to provide unique
dynamical constraints on the MBH spin and metric, in addition to the
mass. In this paper, we develop a fast full general relativistic
method to simultaneously constrain the MBH mass, spin, and spin
direction by considering both the orbital motion of a star close to
the GC MBH and the propagation of photons from the star to a distant
observer.  Based on the current observations and dynamical model
predictions, we assume six example stars with different semimajor axes
($a_{\rm orb}$) and eccentricities ($e_{\rm orb}$) and numerically
calculate their projected trajectories in the sky plane and redshift
curves. Two of those stars are set to have orbital configurations
similar to that of S0-2/S2 and S0-102. We find that the spin-induced
effects on the projected trajectory and redshift curve of a given
star, including the leading term by the Lense-Thirring precession and
the frame dragging, and the high-order precession due to the quadruple
moment, depend on both the absolute value and the direction of the
spin.  The maximum values of the spin-induced position displacement
and the redshift differences of the star over a full orbit may differ
by a factor of several to more than one order of magnitude for two
cases with significantly different spin directions. The dependence
patterns of the position displacements and redshift differences on the
spin direction are different, and thus the position and the redshift
data are complementary for constraining the MBH spin and its
direction.  Adopting the Markov Chain Monte Carlo fitting technique,
we illustrate that the spin of the GC MBH is likely to be well
constrained by using the motion of S0-2/S2 over a period of $\sim 45$
years if the spin is close to one and if the astrometric and
spectroscopic precisions can be as high as $(\sigma_{\rm
p},\,\sigma_Z)\sim (10\mu{\rm as},\,1\kms)$, which is expected to be
realized by future facilities like the GRAVITY on the Very Large
Telescope Interferometer, the thirty meter telescope, and the European
extremely large telescope.  If $\sigma_{\rm p}$ and $\sigma_Z$ can be
further improved by a factor of several, the MBH spin can be well
constrained by monitoring S0-2/S2 over a period of $\sim 15$ years.
In the mean time, the distance from the Sun to the GC and the MBH mass
can also be constrained to an unprecedented accuracy
($0.01\%$-$0.1\%$). If there exists a star with a semimajor axis that is
a few times smaller than, and eccentricity larger, than those of
S0-2/S2, the MBH spin and its direction can be constrained with high
accuracy over a period of $\la 10$ year by future facilities, even if
the spin is only moderately large.  Our results suggest that long-term
monitoring of the motions of stars in the vicinity of the GC MBH by
the next generation facilities is likely to provide a dynamical test,
for the first time, to the spin and metric of the GC MBH.
\end{abstract}


\section{Introduction}

The Kerr black hole \citep{Kerr63} is one of the most simple and
elegant solutions to the Einstein field equation of general relativity
(GR). It is widely accepted that all astrophysical black holes (BHs),
if existing, can be described by the Kerr metric with only two
parameters, i.e., the mass and the spin (no-hair theorem).
Various lines of evidences for the existence of BHs have been
accumulated in the past several decades, mainly based on dynamical
measurements of the masses of these objects
\citep[e.g.,][]{Magorrian1998, KH13, NM13}. One of the strongest
evidences is provided by the long-term monitoring of the motions of
stars in the Galactic center (GC), which suggests, almost exclusively,
the existence of a massive black hole (MBH; with mass $\sim 4\times
10^6\msun$) in the GC \citep[see][]{Ghezetal08, Gillessenetal09, Meyer12}.
However, it is still not clear whether these objects can be fully
described by the Kerr metric or not.  

A crucial step to check whether an astrophysical BH is described by
the Kerr metric is to accurately constrain/measure the BH spin through
spin-induced GR effects. Recently, some progress has been made in
measuring the spins of both stellar-mass BHs in the Milky Way
\citep[i.e.,][and references therein]{McClintock11} and MBHs in the
centers of active galactic nuclei \citep[e.g.,][and references
therein]{Reynolds13}. These measurements are obtained by modeling the
intrinsic X-ray continuum and its reflection components (e.g., the Fe
K$\alpha$ line and the reflection continuum) from each BH-accretion
disk system, which may suffer from various
uncertainties in accretion disk models.
Therefore, independent determinations
of the spins of BHs are of fundamental importance for testing
the no-hair theorem, GR, and a
deep understanding of the space and time.

The MBH system in the GC, massive and in close proximity to us,
provides a unique laboratory for testing the no-hair theorem and
the GR due to the following reasons. First, the angular size of
the expected shadow/image of the GC MBH is the largest one among those
of MBHs, which makes the GC MBH the best target for the event horizon
telescope \citep[e.g.,][]{Doeleman09}. Second, the GC MBH is the
only MBH which has stars in its vicinity that can be detected individually.
Long-term monitoring of the motions of those stars in the immediate
vicinity of the MBH offers a novel opportunity to probe various GR
effects, including the periastron advancement, the Lense-Thirring
precession, and the frame dragging \citep[see][]{Jaroszynski98, Fragile00, RE01,
Weinberg05, PS09, Angelil10a, Angelil10b, Merritt10, Angelil11,
Iorio11a, Iorio11b}. Currently, the detected S-stars rotating around
the GC MBH that have the smallest semimajor axes are S0-2/S2 and
S0-102, which have semimajor axes of $980$\,AU and $850$\,AU and
pericenter distances of $120$\,AU and $270$\,AU, respectively
\citep[][]{Schodel12, Ghezetal08, Gillessenetal09, Meyer12}. It is
also anticipated that there are some stars (fainter than the S-stars)
and pulsars hiding within the orbit of S0-2/S2 and S0-102
\citep[see][and references therein]{ZLY13, ZLY14}.  The GR effects,
e.g., the Lense-Thirring precession and the frame dragging, which are
inherited from the MBH spin, should hide in the apparent trajectories
of these stars \citep[e.g.,][]{Jaroszynski98} and in the redshift
curve of the star \citep{Angelil10a}, however,
they are almost unreachable by current facilities. 

A number of next-generation facilities are currently
in progress, such as the GRAVITY on the Very Large Telescope
Interferometer (VLTI), the thirty meter telescope (TMT), and the
European extremely large telescope (E-ELT). With these facilities,
not only will there be great advancement in the precisions
of the astrometric and spectroscopic measurements, but also some
stars, fainter and closer to the central MBH than the known S-stars,
may be revealed \citep[e.g.,][]{ZLY13}. It is promising to dynamically
constrain the MBH spin and metric by using the relativistic orbital
motions of those stars.

Various spin-induced GR effects have been investigated by many authors
\citep[e.g.,][]{Jaroszynski98, RE01, Angelil10a, Angelil10b,
Merritt10, Angelil11, Iorio11a, Iorio11b}, however, most of those
studies are based on perturbative approximations
rather than full general relativistic calculations and do not consider
how tight the constraint on the MBH spin could be obtained. In this
paper, we develop a full general relativistic method to investigate
the spin-induced GR effect by considering both the relativistic
orbital motion of a star in the vicinity of the GC MBH and the
propagation of photons from the star to a distant observer.  Using
this method and the Markov Chain Monte Carlo (MCMC) fitting scheme, we
investigate how a constraint on the MBH spin and how accurate the
constraint can be obtained by using the orbital motion of an example
star, e.g., S0-2/S2, for a given set of astrometric and spectroscopic
accuracies of an instrument or a telescope.

This paper is organized as follows. By adopting the Kerr metric to
describe the curved spacetime around the MBH, we introduce the motion
equations for a star rotating around the MBH and for photons
propagating in the spacetime in Section~\ref{sec:metric}. We introduce
the numerical method to solve the GR motion equations for the star and
the ray-tracing technique to trace back those photons propagating from
the star to a distant observer in Sections~\ref{sec:cal_star} and
\ref{sec:cal_photon}. In Section 5, we summarize the main results of
our calculations on the orbital motions of some example stars and the
propagation of photons from those stars to the distant observer. Two
of those example stars have similar orbital configurations as those of
S0-2/S2 and S0-102; and the others are set according to the
predictions on the probability distributions of the semimajor axis and
eccentricity of a star, which is expected to be the closest one to the
GC MBH as predicted by some dynamical models.  In
Section~\ref{sec:fit}, we describe the procedures of an MCMC fitting
to the mock observations of the example stars, assuming
astrometric and spectroscopic accuracies of the future observations.
By doing this, we further investigate how accurate the constraint on
the MBH spin and metric can be obtained by using the relativistic
orbital motions of a star, e.g., S0-2/S2, with the next generation
facilities.  Discussions on some related issues are given in
Section~\ref{sec:discussions} and conclusions are summarized in
Section~\ref{sec:conclusions}.

\section{Kerr Metric and Motion Equations}\label{sec:metric}

The curved spacetime around an astrophysical BH may be described by
the Kerr metric \citep{Kerr63}, which can be expressed in the
Boyer-Lindquist coordinates $(t, r, \theta, \phi)$ as \be
ds^2=-e^{2\nu} dt^2 + e^{2\psi}(d\phi-\omega dt)^2+e^{2\mu_1}dr^2
+e^{2\mu_2}d\theta^2,
\label{eq:metric}
\ee where  \be
\left\{
\begin{array}{lcl}
e^{2\nu}  & = & \Sigma \Delta /A, \\ e^{2\psi} & = & A \sin^2\theta
/\Sigma,\\ e^{2\mu_1}& = & \Sigma/\Delta, \\ e^{2\mu_2}& = & \Sigma,
\\ \omega    & = & 2ar/A,\\ \Sigma    & = & r^2+a^2\cos^2\theta, \\
\Delta    & = & r^2-2 r+a^2,\\ A         & = &
(r^2+a^2)^2-a^2\Delta\rm{sin}^2\theta, 
\end{array}
\right.
\ee and $a$ is the dimensionless spin parameter of the MBH
\citep{Boyer67}. For simplicity, we set $G = c =\bh = r_g = G\bh/c^2
=1$ above, $G$, $c$, $\bh$ and $r_g$ are the gravitational constant,
the speed of light, the MBH mass, and the gravitational radius,
respectively.

In the Kerr metric, the motion of a particle, a star, or a photon, is
controlled by the following equations \citep{Bardeen72, Misner73,
Chandra83}
\begin{eqnarray}
\Sigma \dot{r}      & = & \pm \sqrt{R}, \label{eq:motionr} \\
\Sigma \dot{\theta} & = & \pm \sqrt{\Theta}, \label{eq:motiontheta} \\
\Sigma \dot{\phi}   & = & -a+\lambda /\sin^2\theta+ aT / \Delta,
\label{eq:motionphi} \\
\Sigma \dot{t}      & = & -a^2\sin^2\theta+a\lambda+(r^2+a^2) T /
\Delta,
\label{eq:motiont}
\end{eqnarray}
where overdot represents an ordinary derivative with respect to an
affine parameter $\tau$, and $T=r^2+a^2-\lambda a$. The quantities $R$
and $\Theta$ in Equations (\ref{eq:motionr}) and
(\ref{eq:motiontheta}) are given by
\begin{eqnarray}
R  & = & (1-\xi^2) r^4+2 \xi^2 r^3+[a^2(1-\xi^2)-q^2-\lambda^2]r^2
\nonumber \\
   &  &  + 2[(a-\lambda)^2+q^2]r-a^2q^2,
\label{eq:Rr} 
\end{eqnarray}
and 
\be
\Theta  =  q^2-[a^2(\xi^2-1)+\lambda^2/\sin^2\theta] \cos^2\theta ,
\label{eq:RT}
\ee
respectively, where $\xi=m/E$ ($\xi=0$ for photons), $\lambda=L_z/E$,
and $q^2=Q/E^2$. The constants of motion are the rest mass $m$, the
energy at infinity $E$, the azimuthal angular momentum $L_z$, and the
Carter's constant $Q$.

For a star in the immediate vicinity of an MBH that is monitored by a
distant observer, the GR effects encoded in the measurements can be
divided into two parts. First, the orbital motion of the star rotating
around the MBH is affected by the GR effects, such as the periastron
advancement, the Lense-Thirring precession, and the frame dragging. Second,
the propagation of photons from the star to the distant observer is
also affected by the GR effects, including the gravitational redshift
for those photons to climb out of the MBH potential, and the
gravitational bending of the photon trajectories if the distance of
those photons to the central MBH is small. Therefore, it is necessary
to consider both the relativistic motion of the star and the
propagation of photons from the star to the distance observer,
simultaneously.  Below, we introduce the full general relativistic
numerical methods to calculate the relativistic orbits of stars and
the trajectories of photons in the Kerr metric, respectively.


\section{Relativistic Orbits of Stars}\label{sec:cal_star}

The orbital motion of a star in the vicinity of the central MBH, with
a fixed mass ($M_{\bullet}$), spin ($a$) and its direction,\footnote{
The MBH mass, spin, and its direction are assumed to be fixed values
because the time duration of the orbital motion considered in this study is
short and no more than several decades, and thus there should be no
significant evolution in the MBH mass and spin vector.} is determined
by equations~(\ref{eq:motionr})-(\ref{eq:motiont}) once the initial
conditions are known. The initial conditions of the star can be given
by its position and tetrad velocity at a time $t_{\star,0}$ in the
Boyer-Lindquist coordinates, i.e., $\mathbf{r}_{\star,0}=
(t_{\star,0},\, r_{\star,0},\, \theta_{\star,0},\, \phi_{\star,0})$
and $\mathbf{u}_{\star,0} = (u^t_{\star,0},\, u^r_{\star,0},\,
u^{\theta}_{\star,0},\, u^{\phi}_{\star,0})$. Hereafter, $t_{\star,0}$
is set to be zero for simplicity, if not otherwise stated. At any given
time $t_{\star}$, the three-velocity $\mathbf{v}_{\star} =
(v^r_{\star}, v^\theta_{\star}, v^\phi_{\star})$ in the space of the
local non-rotating rest frame (LNRF) associated with $(r_{\star},
\theta_{\star}, \phi_{\star})$ can be transformed from the tetrad
velocity $\mathbf{u}_{\star} = (u^t_\star, u^r_\star, u^\theta_\star,
u^\phi_\star) = (\dot{t}, \dot{r}, \dot{\theta}, \dot{\phi})$ as
\citep[see][]{Bardeen72, Bardeen73, Misner73}
\be
v^r_{\star}=u^r_{\star} e^{\mu_1-\nu}/ u^t_{\star},
\label{eq:vrlnrf}
\ee
\be
v^\theta_{\star}=u^\theta_{\star} e^{\mu_2-\nu}/ u^t_{\star},
\label{eq:v2}
\ee
\be
v^\phi_{\star}=\left(u^\phi_{\star}/u^t_{\star} -\omega\right)
e^{\psi-\nu}.
\label{eq:v3}
\ee
The tetrad velocity $\mathbf{u}_{\star}$ can be reversely transformed
from the three-velocity $\mathbf{v}_{\star}$ in the LNRF frame. 

If the position and the three-velocity of a star in the LNRF frame are
provided, the values of $\xi$ and $\lambda$ can be derived as
\be
\xi = 1\left/[ \gamma (e^{\nu}+\omega e^\psi v^\phi_{\star})]\right. 
\label{eq:xi}
\ee
and
\be
\lambda =A v^\phi_{\star} \left/ (2arv^\phi_{\star}+\Gamma)\right.,
\label{eq:lambda} 
\ee
respectively, where
\be
\gamma  =(
1- \vec{\mathbf v}_\star \cdot \vec{\mathbf v}_\star
)^{-1/2}, 
\ee
and
\be
\Gamma    =[A (\Delta-a^2 \sin^2\theta )+4a^2r^2\sin^2\theta]
\left/\Sigma\sqrt{\Delta}\sin\theta\right.. 
\ee
Substituting $\xi$ and $\lambda$ into Equations (\ref{eq:RT}) and
(\ref{eq:v2}), we have 
\be
q^2=\Sigma \left(\gamma v^\theta_{\star} \right)^2+\cos^2\theta
\left[a^2(\xi^2-1) +\lambda^2 / \sin^2\theta \right].
\label{eq:Q}
\ee

\begin{figure*}
\center
\includegraphics[scale=0.80]{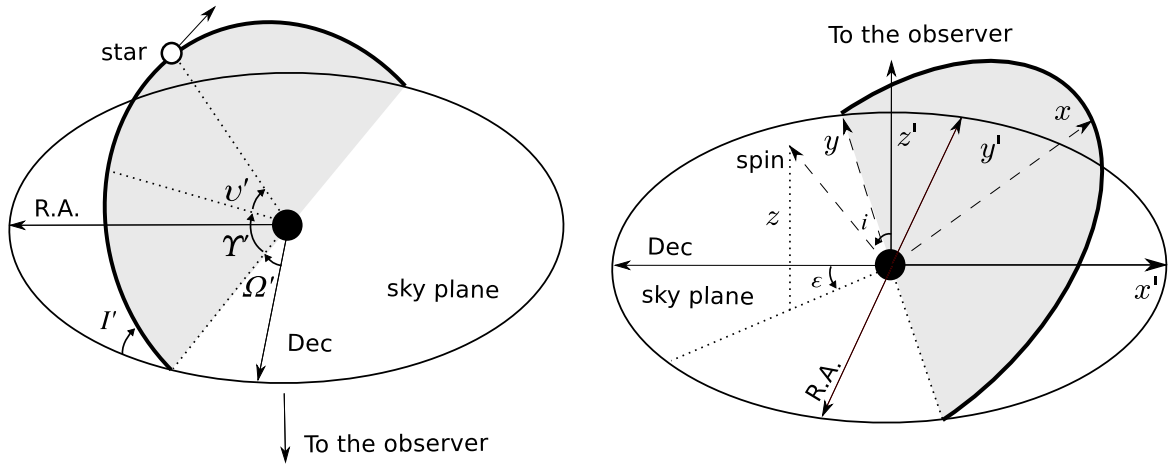}
\caption{ Schematic diagram for the star-MBH system and the coordinate
systems.  Right panel: a pseudo-Cartesian coordinate $(x', y', z')$ is
defined in the rest frame of a distant observer (i.e., the observer's
frame), where the $x'y'$ plane represents the sky plane of the
observer and is taken as the reference plane, the $z'$ axis represents
the line of sight, and the $x'$ axis is taken as the reference
direction on the sky plane.  Another pseudo-Cartesian coordinate $xyz$
is defined to relate the Boyer-Lindquist coordinates $(r, \theta,
\phi)$ to orthogonal coordinates $(x, y, z)$, where the $z$ axis
represents the spin direction of the MBH and the $y$ axis represents
the intersection line of the MBH equatorial plane with the observer's
sky plane.  The direction of the MBH spin in the observer's frame is
therefore described by two angles, i.e., the angle the $z$ axis and
the $z'$ axis ($i$) and the angle between the projection vector of the
$x$ axis on the $x'y'$ plane and the $x'$ axis ($\epsilon$). The
directions of the R.A. ($-\vec{y}'$) and Dec.
($-\vec{x}'$) are marked in this figure.  Left panel: the orbit of
the star may be approximated as a Newtonian ellipse in the
observer's frame, especially when the semimajor axis of the star is
large, which is described by six orbital elements, i.e., the semimajor
axis $a_{\rm orb}$, eccentricity $e$, the longitude of ascending node
$\Omega'$, argument of periapsis $\Upsilon'$, true anomaly $\upsilon'$
and inclination $I'$, respectively.  }
\label{fig:ills}
\end{figure*}

We define two pseudo-Cartesian coordinate systems for the convenience
of relating the motion of a star in the vicinity of the MBH with the
Boyer-Lindquist coordinates $(r, \theta, \phi)$ to that seen by a
distant observer.  The first pseudo-Cartesian coordinate system
$x'y'z'$ is defined in the rest frame of the distant observer (denoted
as the observer's frame hereafter). The direction of the $z'$-axis is
pointing from the MBH to the distant observer, and the $x'y'$ plane
represents the sky plane of the observer. The second pseudo-Cartesian
coordinate system $xyz$ is defined to relate the Boyer-Lindquist
coordinate $(r, \theta, \phi)$ to an orthogonal coordinate system $(x,
y, z)$ with $\vec{z}$ representing the direction of the spin vector
(see a similar definition in \citealt{Angelil10a}). Because the star is
usually far away from the event horizon of the central MBH, the
coordinates of the star in the $xyz$ frame may be approximated by
\be
\left\{
\begin{array}{lcl}
x &= & r\sin\theta\cos\phi, \\
y &= & r\sin\theta\sin\phi, \\
z &= &r\cos\theta.
\end{array}
\label{eq:xyz}
\right.
\ee
In this $xyz$ frame, the direction of the $z$-axis represents the spin
direction of the MBH, the $y$-axis represents the intersection line of
the equatorial plane of the MBH with the sky plane of the distant
observer.  The LNRF frame is rotating with an angular velocity of
$\omega$ relative to the distant observer's frame. However, $\omega$
is relatively small since the star is usually far away from the event
horizon of the MBH.  Therefore, the rotation of the LNRF frame with
respect to the distant observer's frame can be neglected when relating
the initial conditions to the Newtonian approximations.

Figure~\ref{fig:ills} shows the configuration of the stellar-MBH
system and the two pseudo-Cartesian coordinate systems defined above.
As seen from Figure~\ref{fig:ills}, the spin direction $\vec{z}$ is
determined by two angles in the observer's frame $x'y'z'$, i.e., the
angle ($i$) between $\vec{z}$ and $\vec{z}'$, and the angle
($\epsilon$) between the projection of $\vec{z}$ on the $x'y'$ plane
and $-\vec{x}'$.  The angle between the $y$ axis and the $y'$ axis is
the same as $\epsilon$. Therefore, a vector defined in the observer's
frame can be approximately transformed to the LNRF frame (or the MBH's
frame) by first rotating the $x'y'z'$ frame around the $z'$ axis
counter-clockwise by an angle of $\epsilon$ and then rotating it
around the new $y'$ axis clockwise by an angle of $i$.  For example,
a vector ${\mathbf{n}'}$ in the observer's frame can be transformed to
that in the LNRF frame ${\mathbf{n}}$ by $\mathbf{n}=\mathbf{M}
\mathbf{n}'$, where $\mathbf{M}$ is a rotation matrix given by
\be
\mathbf{M}
=
\begin{pmatrix}
\cos i\cos \epsilon & \cos i \sin \epsilon & \sin i \\
-\sin \epsilon      & \cos \epsilon        & 0      \\
-\sin i\cos\epsilon & -\sin i\sin\epsilon  & \cos i 
\end{pmatrix}.
\label{eq:coordrot}
\ee

In the Newtonian case, the orbital motion of a star rotating around a
massive object is determined by six orbital elements, i.e., the
semimajor axis $a_{\rm orb}$, the eccentricity $e_{\rm orb}$, the
longitude of the ascending node $\Omega'$, the argument of the
pericenter $\Upsilon'$, the true anomaly $\upsilon'$, and the orbital
inclination angle $I'$ (see Figure~\ref{fig:ills}). For convenience in
comparison with the Newtonian orbits currently determined for a few
GC S-stars with the smallest semimajor axis, in our following
simulations of the orbital motions of some (example) stars, we set
their initial conditions by fixing the six orbital elements at a given
moment in the distant observer's frame.\footnote{The orbital elements
set here are used in a simple way to generate the initial conditions
of a star moving in the Boyer-Lindquist coordinates. The elliptic
orbit defined by these orbital elements is only taken as a Newtonian
approximation to the real orbit of a star in the Boyer-Lindquist
coordinates. These orbital elements may lose their original meaning in
the curved spacetime around a Kerr MBH.} Then, the values for the
four-position $\mathbf{r}_{\star,0}$ and the tetrad-velocity
$\mathbf{u}_{\star,0}$ of a star can be approximately obtained by the
following procedures.

\begin{enumerate}
\item The initial position $(x'_{\star,0}, y'_{\star,0},
z'_{\star,0})$ and the three velocity $(v'^x_{\star,0},
v'^y_{\star,0}, v'^z_{\star,0})$ of the star in the distant observer's
frame are first obtained from the six orbital elements initially set.
\item Transforming the position and the velocity of the star in the
distant observer's frame into those in the LNRF frame, i.e.,
$\vec{r}_{\star,0}=(x_{\star,0}, y_{\star,0}, z_{\star,0})$ and
$\vec{v}_{\star,0}= (v^x_{\star0}, v^y_{\star,0}, v^z_{\star,0})$, by
the rotation given in Equation~(\ref{eq:coordrot}), and then
transforming these two vectors from that using the $xyz$ coordinates
into that using the $r\theta\phi$ coordinates.
\item Transforming $\vec{r}_{\star,0}$ and $\vec{v}_{\star,0}$ to the
four-position $\mathbf{r}_{\star,0}$ and the tetrad-velocity
$\mathbf{u}_{\star,0}$ in the Boyer-Lindquist coordinates according to
Equations~(\ref{eq:vrlnrf})-(\ref{eq:v3}).
\end{enumerate} 

Once the initial conditions are given for a star, the motion
parameters $\lambda$, $\xi$, and $q$ can be obtained by Equations
(\ref{eq:xi}), (\ref{eq:lambda}), and (\ref{eq:Q}), respectively.
With those constants of motion, the orbital motion of a star, in
principle,  can be obtained by integrating the motion Equations
(\ref{eq:motionr})-(\ref{eq:motiont}). However, it is required to
frequently judge the sign ``$\pm$'' at the left side of the motion
Equations (\ref{eq:motionr}) and (\ref{eq:motiontheta}) in the
numerical integrations. To avoid this complication, one may use the
following equations to replace the motion Equations (\ref{eq:motionr})
and (\ref{eq:motiontheta}).
\be 
\Sigma \dot{r} = \Delta p_r,
\label{eq:rdot}
\ee
\be 
\Sigma \dot{\theta} = p_{\theta},
\ee
\be
\Sigma \dot{p}_r =\frac{1}{2\Delta} \frac{\partial R}{\partial r} -
R\frac{\partial \Theta}{\partial \theta},
\ee
\be 
\Sigma \dot{p}_{\theta} = \frac{1}{2} \frac{\partial \Theta}{\partial
\theta}.
\label{eq:ptheta}
\ee
We use the code DORPI5 based on the explicit fifth (fourth)-order
Runge Kutta method \citep{DP80,Hairer93} to integrate the motion
Equations (\ref{eq:rdot})-(\ref{eq:ptheta}) and
(\ref{eq:motionphi}) and (\ref{eq:motiont}) to obtain the orbit motion of
a star. We set the relative integration errors to be $\leq 10^{-12}$
for all the position and momentum quantities $(r,\, \theta,\, \phi,\,
p_r,\, p_\theta,\, p_\phi)$ in those equations, which are sufficient
for the convergence of the numerical results and the required
accuracy in this study. 

\begin{figure*}
\center
\includegraphics[scale=0.53]{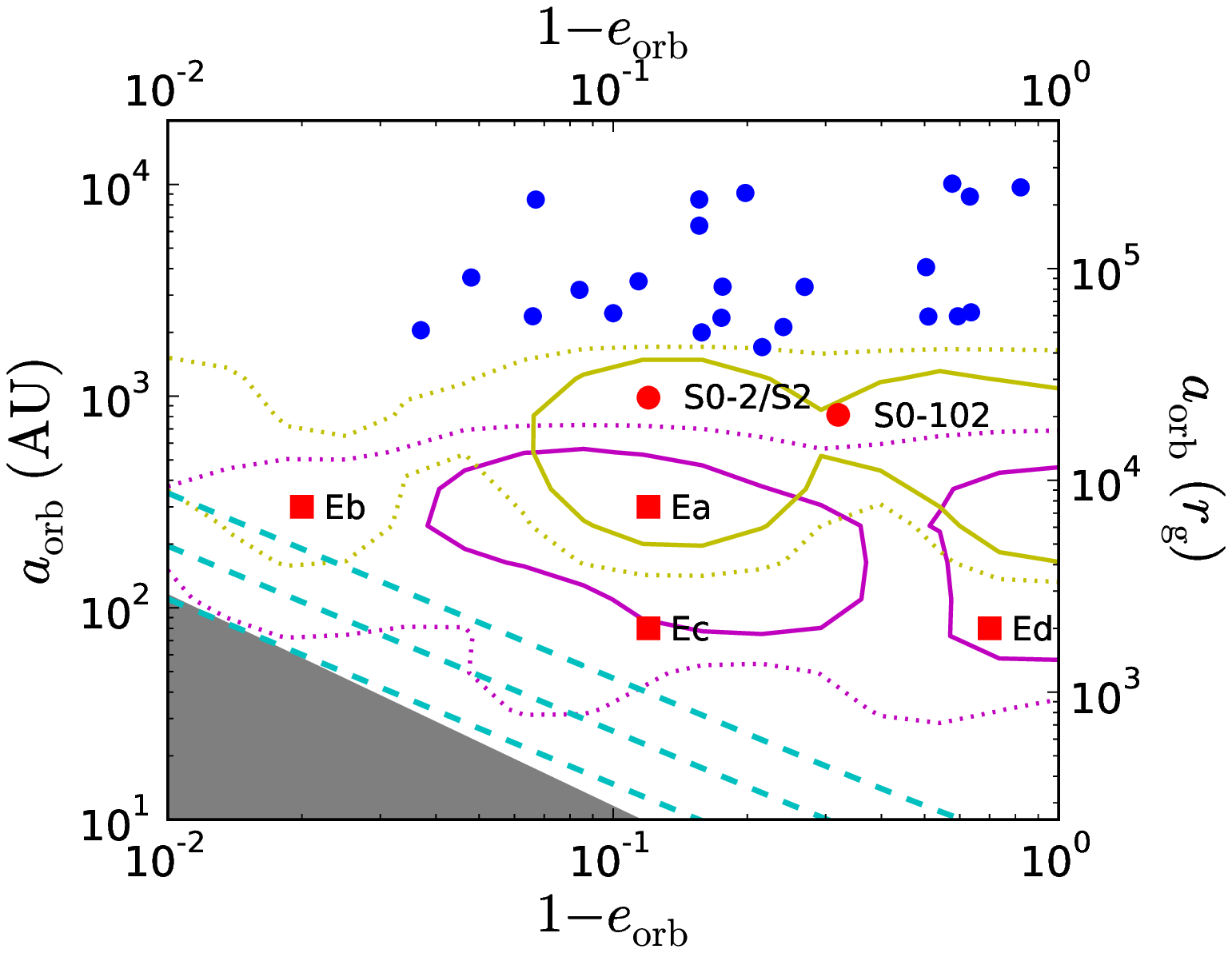}
\includegraphics[scale=0.53]{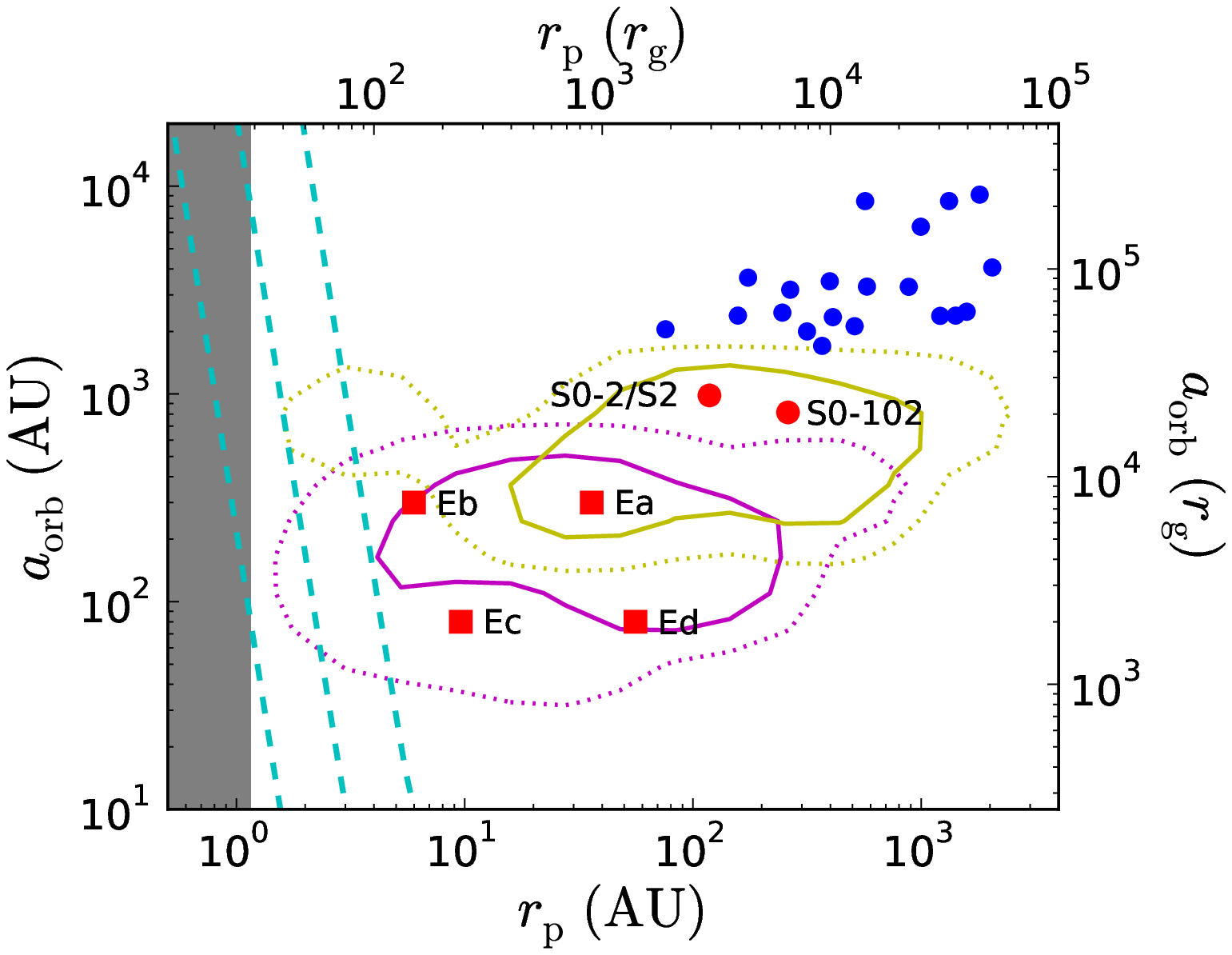}
\caption{Orbital distribution of some detected GC S-stars and the
possibly existing star that is closest to the GC MBH in the semimajor
axis versus eccentricity plane (left panel) and in the semimajor axis
versus the distance to the pericenter plane (right panel). The tick labels
at the right side of each panel and the top side of the right panel
are in units of gravitational radius $r_{\rm g} = G M_{\bullet}/c^2$,
while the tick labels at the left side of each panel and the bottom side
of the right panel are measured with the astronomical unit AU. The solid circles
represent those currently detected GC S-stars within a distance of
$4000$\,AU from the GC MBH. S0-2/S2 and S0-102 are the two with the
smallest semimajor axes among the GC S-stars \citep[see][]{Meyer12}.
The probabilities of a predicted star with mass in the range
$1-7\msun$ that is closest to the MBH within the yellow and the magenta
solid (dotted) curves are 85\% (98\%) and 66\% (98\%), resulting from
the model ``Disk-IM2'' and ``Disk-IM0'' in \citet{ZLY13},
respectively, by assuming that the GC S-stars are the captured
components of stellar binaries injected into the vicinity of the MBH.
The cyan dashed lines give the gravitational radiation timescale of a
star with mass $3\msun$ rotating around the MBH: $10^8$, $10^7$, and
$10^6$ years from top to bottom in the left panel (or from right to left
in the right panel).  The pericenter distances of the stars, if
existing, in the shaded region are smaller than the tidal disruption
radii for a star with mass $3\msun$, and thus stars almost cannot
exist in this region.  } 
\label{fig:stardistr} 
\end{figure*}

\subsection{Orbits of Example Stars }
\label{sec:GCstars}

Observations in the past two decades have revealed a number of GC
S-stars rotating around the MBH with semimajor axes in the range of
$\sim 4000$-$800$\,AU. Among them, the two with the smallest semimajor
axes are S0-2/S2 and S0-102 \citep{Ghezetal08, Gillessenetal09,
Meyer12}.  These stars have been very useful in determining the MBH
mass, and it is expected that S0-2/S2 (or a star within the orbit of
S0-2/S2) will be an important dynamical probe to the GR effects of the
GC MBH.  It is also anticipated that there are some stars fainter
than the GC S-stars existing within the orbits of S0-2/S2 and S0-102,
which could be even better probes for the GR effects. \citet{ZLY13}
predicted the probability density distribution in the semimajor axis
versus eccentricity plane of such stars (with mass in the range of
$1-7\msun$), by assuming that these stars and the GC S-stars are the
captured components of stellar binaries that are tidally broken up in
the vicinity of the GC MBH. 

Figure~\ref{fig:stardistr} shows both the orbital distribution of the
detected GC S-stars (solid circles) and the probability density
distribution of the expected innermost star\footnote{Here the
``innermost'' star means the star with the smallest semimajor axis.}
(color contours) in the semimajor axis versus eccentricity plane
(panel (a)) and the semimajor axis versus pericenter distance plane
(panel (b)). The magenta (yellow) solid and dotted contours represent a
probability for the expected innermost star with mass in the range of
$1-7\msun$ of $85\%$ and $98\%$ (or $66\%$ and $98\%$), resulted from
the ``Disk-IM0'' (or ``Disk-IM2'') model in \citet{ZLY13},
respectively. The semimajor axis and eccentricity of the expected
innermost star are in the range of $\sim 100-1000$\,AU and $0-0.99$,
respectively. The dotted lines in Figure~\ref{fig:stardistr} give the
gravitational radiation timescale of a star with mass of $3\msun$
rotating around the MBH with mass $4\times 10^6\msun$: $10^8$, $10^7$,
$10^6$ years from top to bottom (Equation~(5.6) in \citealt{Peters64} or
Equation (39) in \citealt{YT03}).  A star should be tidally disrupted
if it approaches the MBH within a distance of $r_{\rm tid} =\left(
\eta^2 M_{\bullet} /m_{\star} \right)^{1/3} r_{\star} \simeq \left(
\eta^2 M_{\bullet} /m_{\star} \right)^{1/3}(m_{\star}/\msun)^{0.47}
R_{\sun}$, where $m_{\star}$ is the mass of the star and $\eta$ is
set to be $2.21$ for a homogeneous, incompressible body
\citep[see][]{Magorrian99}. The upper (right) boundary of the shaded
region in the left (right) panel of Figure~\ref{fig:stardistr}
represents the tidal radius of stars ($r_{\rm tid}$) with mass
$3\msun$ rotating around the MBH. In principle, stars with masses $\geq
3\msun$ cannot exist in the shaded region. The probability of existing
stars in the region between  the top (right) dashed line and the upper
(or right) boundary of the shaded region in the left (or right) panel
is substantially suppressed because of the rapid orbital decay of a
star in this region due to gravitational wave radiation. 

According to the current observations and theoretical expectations, we
select six example stars as listed in Table~\ref{tab:t1}.  Two of the
example stars have similar orbital properties as those of S0-2/S2 and
S0-102, respectively, which have the smallest semimajor axes among the
detected GC S-stars \citep[see][]{Ghezetal08, Gillessenetal09,
Meyer12}. The other four example stars, i.e., Ea, Eb, Ec, and Ed (see
Table~\ref{tab:t1}), are set according to the probability distribution
of the expected innermost star shown in Figure~\ref{fig:stardistr}.
The semimajor axes, eccentricities, and pericenter distances of the
example stars Ea, Eb, Ec, and Ed are ($300$\,AU, $0.88$, $36$\,AU),
($300$\,AU, $0.98$, $6$\,AU), ($80$\,AU, $0.88$, $9.6$\,AU), and
($80$\,AU, $0.3$, $56$\,AU), respectively.  The probability for the
existence of a star like the example star Ea is probably high.  There are
some chances for the existence of a star like the example stars Eb, Ec,
or Ed in the vicinity of the MBH, though the chances are not
very high. Note that the semimajor axis and eccentricity listed there are
only taken as the Newtonian approximation to the orbit of each
example star, which are useful for setting the initial conditions.
\begin{table*}
\caption{Orbital parameters for example stars}
\centering
\begin{tabular}{lccccccccccc}\hline
\multirow{2}{1.0cm}{Name}    & \multicolumn{3}{c}{$a_{\rm orb}$} &
\multirow{2}{0.5cm}{$e_{\rm orb}$} & \multirow{2}{0.5cm}{I} &
\multirow{2}{0.5cm}{$\Omega'$} & \multirow{2}{0.5cm}{$\Upsilon'$} &
\multirow{2}{0.5cm}{$T_0^d$} & \multirow{2}{0.5cm}{$a$} &
\multirow{2}{0.5cm}{$i^e$} & \multirow{2}{0.5cm}{$\epsilon^e$} \\
\cline{2-4}
	&   AU\,$^a$ &  $r_{\rm g}\,^b$ & mas\,$^c$ &   & & & \\
\hline
S0-2/S2      & 984 & 24949 & 123 & 0.88 & 135$\arcdeg$ & 225$\arcdeg$
& 63$\arcdeg$ & 2.32 & 0.99 &  45$\arcdeg$ & 200$\arcdeg$ \\
S0-102  & 848 & 21500 & 106 & 0.68 & 151$\arcdeg$ & 175$\arcdeg$ &
185$\arcdeg$ & 9.5 & 0.99 &  45$\arcdeg$ & 180$\arcdeg$ \\
Ea       & 300 & 7606 & 37.5 &  0.88 & 45$\arcdeg$ & 0$\arcdeg$ &
0$\arcdeg$ & 0 &0.99 & 45$\arcdeg$ & 180$\arcdeg$  \\
Eb       & 300 &  7606 & 37.5 & 0.98 & 45$\arcdeg$ & 0$\arcdeg$ &
0$\arcdeg$ & 0& 0.99 & 45$\arcdeg$ & 180$\arcdeg$  \\
Ec       & 80  &  2028 & 10 & 0.88 & 45$\arcdeg$ & 0$\arcdeg$ &
0$\arcdeg$  & 0& 0.99 & 45$\arcdeg$ & 180$\arcdeg$ \\
Ed       & 80  &  2028 & 10 & 0.30 & 45$\arcdeg$ & 0$\arcdeg$ &
0$\arcdeg$ & 0& 0.99 & 45$\arcdeg$ & 180$\arcdeg$  \\ \hline 
\end{tabular}
\tablecomments{ \,$^a$ In units of AU.
\,$^b$ In units of the gravitational radius $r_{\rm
g}=GM_{\bullet}/c^2$ and the MBH mass $M_{\bullet}$ is assumed to be
$4\times 10^6\msun$.
\,$^c$ In units of mas by assuming the distance from the MBH to sun
$R_{\rm GC}= 8\kpc$.  
\, $^d$ Time of pericentric passage, with respect to the year $2000$ in
order to obtain mock observations for S0-2/S2, S0-102, and other
possibly existing stars by future facilities. 
\,$^e$ The MBH spin direction is set for the case for S0-2/S2 in order
to have an intermediate spin-induced effect; and it is set for the
cases of other stars arbitrary to the values listed in the table.  }
\label{tab:t1}
\end{table*}

\section{Photon Propagation: Ray-Tracing} \label{sec:cal_photon}

We consider a photon propagating from a star in the vicinity of an MBH
to a distant observer in this section.  The position of the star at
each moment $(t_{\star}, r_{\star}, \theta_{\star}, \phi_{\star})$ is
set by its orbital motion as obtained above in
Section~\ref{sec:cal_star} once the initial conditions are fixed. The
position of the distant observer is fixed at $(r_{\rm o}, \theta_{\rm
o}, \phi_{\rm o}) =  (R_{\rm GC}, i, 0\arcdeg)$, where $R_{\rm GC}$ is
the distance from the Sun to the GC and set to be $8$\,kpc if not
otherwise stated. The photon trajectory may be bended due to the
curved spacetime of the MBH, and the energy of the photon when it is
received by the distant observer is different from its original value
emitting from the star due to the gravitational redshift and the Doppler
shift.
To account for those effects,
we use the ray-tracing technique to trace back photons from the
distant observer to the star in the Boyer-Lindquist coordinate
system.  The ray-tracing technique is detailed in
Appendix~\ref{sec:Appendix1}, where both Jacobian elliptic functions
\citep{BF54, AS72} and the Gauss-Kronrod integration scheme
\citep[e.g., see][]{Rauch94} are adopted to integrate the motion
equations. 

The apparent position of the star in the distant observer's sky plane
at a given time $t_{\rm o}$ is described by two impact parameters,
($\alpha$, $\beta$), which map the position of the star $(t_{\star},
r_{\star}, \theta_{\star}, \phi_{\star})$ and are measured relative to
the center of the MBH. The impact parameter $\alpha$ is the apparent
displacement of the image perpendicular to the projected axis of
symmetry of the MBH, and $\beta$ is the apparent displacement parallel
to the projected axis of symmetry in the sense of the angular momentum
of the MBH. For photons, $\xi=0$, the values of the other two 
constants of motion, $\lambda$ and $q^2$ in Equations~(\ref{eq:Rr}) and
(\ref{eq:RT}), can be determined by  $\alpha$ and $\beta$ as 
\be
\lambda=-\alpha\sin i, 
\ee 
\be 
q^2=\beta^2+(\alpha^2-a^2) \cos^2 i.
\label{eq:ab} 
\ee 
For details, see \citet[][]{Cunningham73}, \citet{Chandra83}, and
\citet{Karas92}. The
star may have more than one image because of the gravitational lensing
effect. For the majority of the cases investigated in this study, however,
only one image is dominant because the star is normally far away from the
Einstein radius of the central MBH and the lensing effect is not
significant. We may trace a photon with given impact parameters
$(\alpha, \beta)$ from the distant observer to the vicinity of the MBH
by solving Equations~(\ref{eq:motionr})-(\ref{eq:motiont}), and check
whether the photon can reach the surface of the star at a given
moment.

According to the  motion Equations~(\ref{eq:motionr}) and
(\ref{eq:motiontheta}), we have  
\be
\int^{r}\frac{dr}{\sqrt{R}} = \int^{\theta}
\frac{d\theta}{\sqrt{\Theta}}.
\label{eq:reqtheta}
\ee
For a photon propagating from the star to the distant observer, the
$\theta$-integral at the right side of the above equation can be
obtained by integrating over $\theta$ from $\theta_{\rm o}$ to
$\theta_{\star}$ (for details, see Appendix~\ref{sec:thetaint}).  When
$\theta = \theta_{\star}$, the radial position of the photon is
denoted as $r_{\rm hit}$. Therefore, the $r$-integral at the left side
can be obtained by integrating over $r$ from $r_{\rm o}$ to $r_{\rm
hit}$ (see Appendix~\ref{sec:rint}). By equating the $r$-integral with
the $\theta$-integral, we can then get the solution of $r_{\rm hit}$
(see Appendix~\ref{sec:rhit}). Once $r_{\rm hit}$ is obtained, one can
subsequently obtain the azimuthal position of the photon, $\phi_{\rm
hit}$, when $\theta= \theta_{\star}$ and $r = r_{\rm hit}$, and the
time needed $\delta t_{\rm hit}=  t_{\rm hit} - t_{\rm o} $ for the
photon propagating from $(t_{\rm hit}, r_{\rm hit}, \theta_{\star},
\phi_{\rm hit})$ to the observer (see Appendix~\ref{sec:phitint}). 

The trajectory of a photon with arbitrary $(\alpha, \beta)$ from the
distant observer may not hit the surface of the star at any time
$t_{\star}$ when $\theta=\theta_{\star}$.  We adopt the following
procedure to judge whether a photon can reach the surface of the star.
In the pseudo-Cartesian coordinate system $xyz$, we have $\vec{R}_{\rm
hit}=(x_{\rm hit}, y_{\rm hit}, z_{\rm hit})$ and
\be
\left\{
\ba
x_{\rm hit} &=r_{{\rm hit}}\sin\theta_\star\cos\phi_{{\rm hit}},\\
y_{\rm hit} &=r_{{\rm hit}}\sin\theta_\star\sin\phi_{{\rm hit}},\\
z_{\rm hit} &=r_{{\rm hit}}\cos\theta_\star,
\ea
\right.
\label{eq:xyzhit}
\ee
according to Equation~(\ref{eq:xyz}). The position of the star at any
given time $t_{\star}$ can be described as $\vec{R}_{\star}= (x_\star,
y_{\star}, z_{\star})$, which are obtained similarly as $\vec{R}_{\rm
hit}$. The distance of the photon at $\theta=\theta_\star$ to the star
in the $xyz$ system is defined as
\begin{eqnarray}
d & = & \left | \vec{R}_{\rm hit} - \vec{R}_{\star} \right|  \nonumber
\\
   & = & \sqrt{ (x_{\rm hit} - x_{\star})^2 + (y_{\rm hit}-y_\star)^2+
(z_{\rm hit} -z_\star)^2}. 
\label{eq:dst}
\end{eqnarray}
We assume that a photon trajectory hits on the surface of the star if
$d < \xi r_{\star}$ and $\xi =10^{-6}-10^{-8}$.\footnote{For a star
with mass $\sim 3\msun$, its radius is $\sim 1.7\times 10^6$\,km$\sim
0.3 r_{\rm g}$ if $M_{\bullet}=4\times 10^6\msun$. The setting of $\xi
= 10^{-6}-10^{-8}$ enables sufficient accuracy for relating the
position of an example star to its image in the observer's sky plane,
as we have $d\la 0.05 r_{\rm g}$, substantially less than the star
radius, even for the example star with the largest semimajor axis.} In
this case, $(\alpha, \beta)$ represent the apparent position of the
image of the star at $(t_\star, r_\star, \theta_\star,\phi_\star)$ in
the observer's sky plane, which are re-denoted as $(\alpha_\star,
\beta_\star)$. The image $(\alpha_\star, \beta_\star)$ should be
detected at a time $t_{\rm o}=t_{\star} +\delta t_{\rm hit}$.
Therefore, the orbital motion of stars in the vicinity of the MBH
represent by the change of $\alpha_\star$ and $\beta_\star$ in a time
sequence of $t_{\rm o}$. 
%
%
In the meantime, the four-momentum of a
photon with motion parameters $\lambda$ and $q$ at any position, e.g.,
$\mathbf{p}_{\rm hit} = (p_{t,\rm hit},p_{r,\rm hit},p_{\theta,\rm
hit},p_{\phi,\rm hit})$ at $(r_{\rm hit}, \theta_\star, \phi_{\rm
hit})$ and $\mathbf{p}_{\rm o} = (p_{t,\rm o},p_{r,\rm
o},p_{\theta,\rm o},p_{\phi,\rm o})$ at $(R_{\rm GC}, i, 0\arcdeg)$,
can be obtained by
\be
\left\{ \ba
&p_t=-E, \\
&p_r=\pm E \sqrt{R}/\Delta,\\
&p_\theta=\pm E\sqrt{\Theta}, \\
& p_\phi=\lambda E,
\ea
\right.
\label{eq:photon_mom}
\ee 
in the Boyer-Lindquist coordinates.
%
%
We can then obtain the redshift of
the photon as 
\be 
Z=\frac{\mathbf{p}_{\rm hit}\cdot \mathbf{u}_{\star}}{\mathbf{p}_{\rm
o} \cdot \mathbf{u}_{\rm o}} -1 = -\frac{\mathbf{p}_{\rm
hit}\cdot\mathbf{u}_\star}{E_{\rm o}} -1.
\label{eq:redshift} 
\ee 
Here $\mathbf{u}_{\rm o}= (-1, 0, 0, 0)$ and $\mathbf{u}_\star$ are
the four-velocities of the distant observer and the star,
respectively, $E_{\rm o}$ is the energy of the photon received by the
observer. According to Equation~(\ref{eq:redshift}), the redshift
curve can be obtained for each star. This redshift can be measured by
the shift of the
center(s) of emission or absorption line(s).

In order to perform the ray-tracing calculations efficiently, the
integrations are started from the point at $(10^8\rg, i, 0)$ rather
than $(R_{\rm GC}, i, 0)$. This setting does not affect the
estimations of the impact parameters $\alpha$ and $\beta$ since the
bending of light is not important anymore at a distance larger
than $10^8\rg$ from the MBH. The final (mock) observations of the
apparent positions of a star on the observer's sky plane should be
given by $ \alpha  r_{\rm g}/ R_{\rm GC}$ and $\beta r_{\rm g} /
R_{\rm GC}$.

\section{Apparent Orbits and Redshift Curves of Example Stars}
\label{sec:orbred}

The apparent position of a star in the observer's sky and the redshift
of the star at any given time can be obtained
according to the procedure described in Sections~\ref{sec:cal_star}
and \ref{sec:cal_photon}. We illustrate the effects of the MBH spin
and its direction on the apparent orbits and redshift 
curves by using those example stars listed in
Table~\ref{tab:t1} in this Section.

\begin{figure*}
\centering
\includegraphics[scale=0.7]{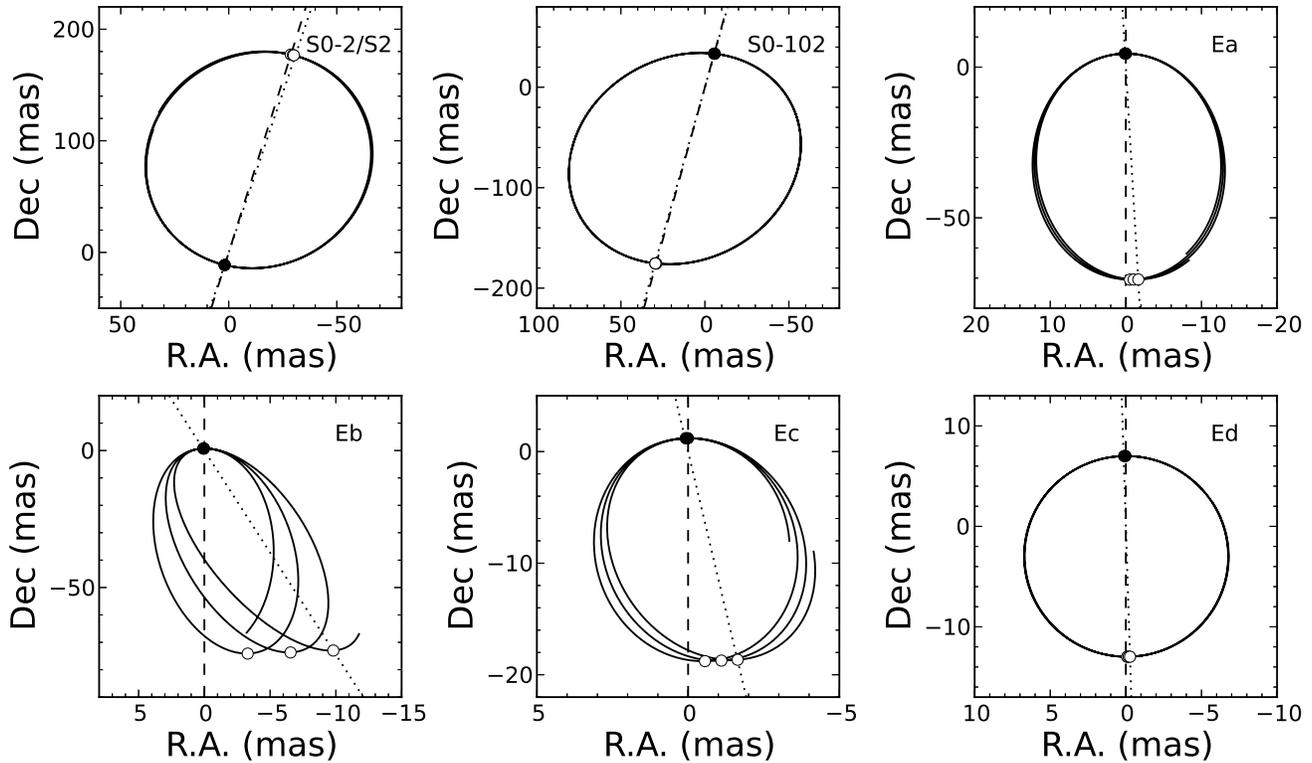}
\caption{Evolution of the apparent position of each example star on
the observer's sky plane. The apparent position of an example star is
described by the R.A. and Dec. Panels
from left to right, top to bottom, represent the example stars S0-2/S2,
S0-102, Ea, Eb, Ec, and Ed, respectively. The initial settings of the
orbital parameters for these stars are listed in Table~\ref{tab:t1}.
Three full orbits are shown for each example star. Open and solid
circles mark the locations of the apoapsides and periapsides,
respectively.  In each panel, the dashed and dotted lines represent
the eccentric vectors obtained by approximating the star orbit to a
Newtonian ellipse for the first and the third orbit, respectively. As
seen from the figure, the apparent orbital precession is most
significant ($\sim 3.1$\,mas per orbit) for the example star Eb, which
has the smallest pericenter distance.}
\label{fig:f3}
\end{figure*}

\subsection{Spin-induced Apparent Position Displacements}

\begin{figure*}
\centering
\includegraphics[scale=0.7]{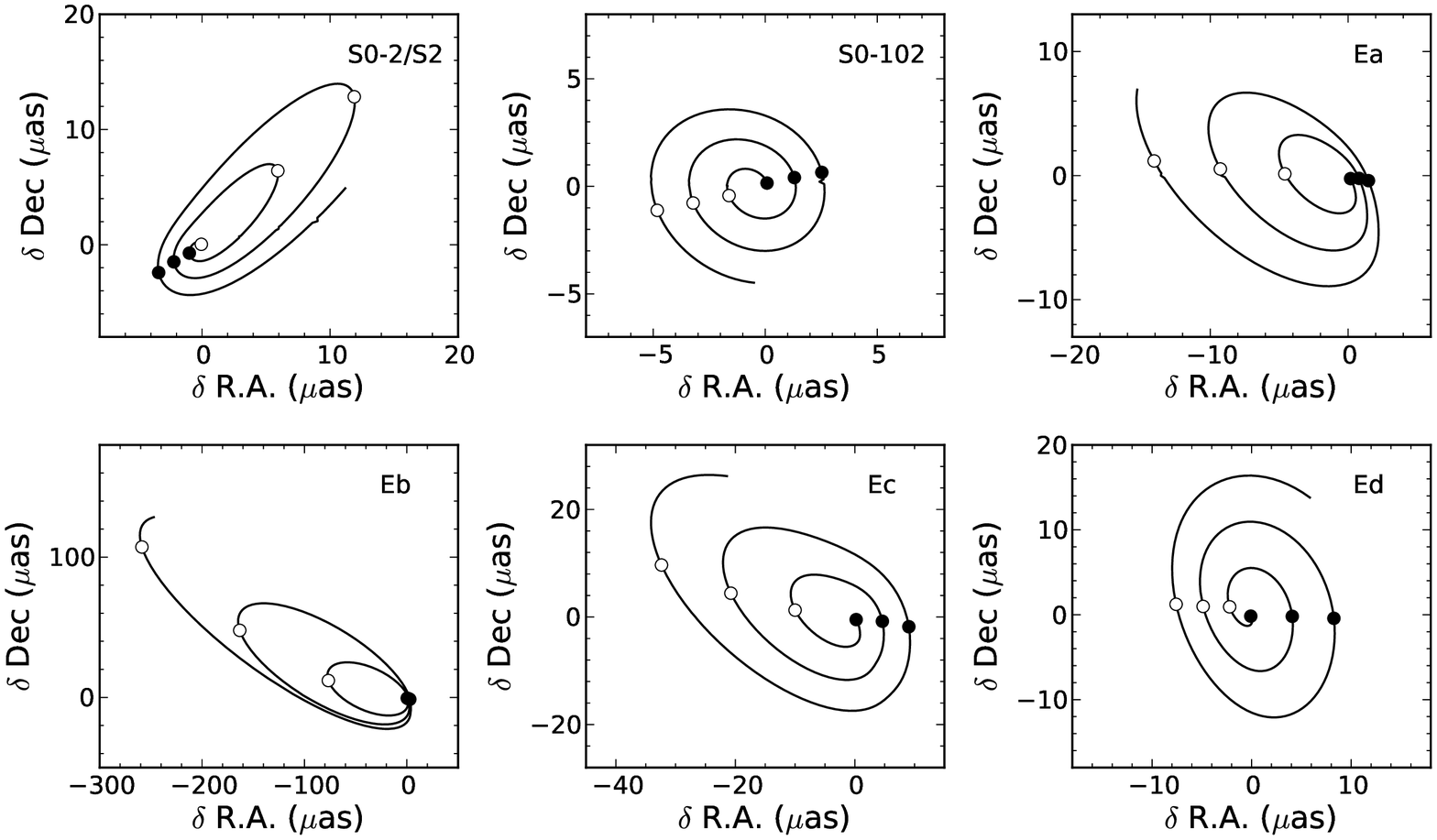}
\caption{Difference between the apparent position of each example star
on the observer's sky plane at each given moment for the case with a
rapidly spinning central MBH ($a=0.99$) and that with a non-spinning
MBH ($a=0$). The differences in R.A. and Dec. are denoted as $\delta$
R.A., and $\delta$Dec., respectively.  Panels from left to right, top
to bottom, represent the example stars S0-2/S2, S0-102, Ea, Eb, Ec, and
Ed, respectively.  Three total orbits for each star are shown here
and the first orbit of each star starts at ($\delta$R.A.,
$\delta$Dec.)$=(0, 0)$. Solid and open circles mark the periapsis and
apoapsis passage points, respectively.  This figure shows that the
difference in the apparent position of each example star increases
with increasing time because of the Lensing-Thirring precession and
the frame dragging and other higher order GR effects induced by the
MBH spin in the rapidly spinning MBH case.
 }
\label{fig:f4}
\end{figure*}

\begin{figure*}
\centering
\includegraphics[scale=0.7]{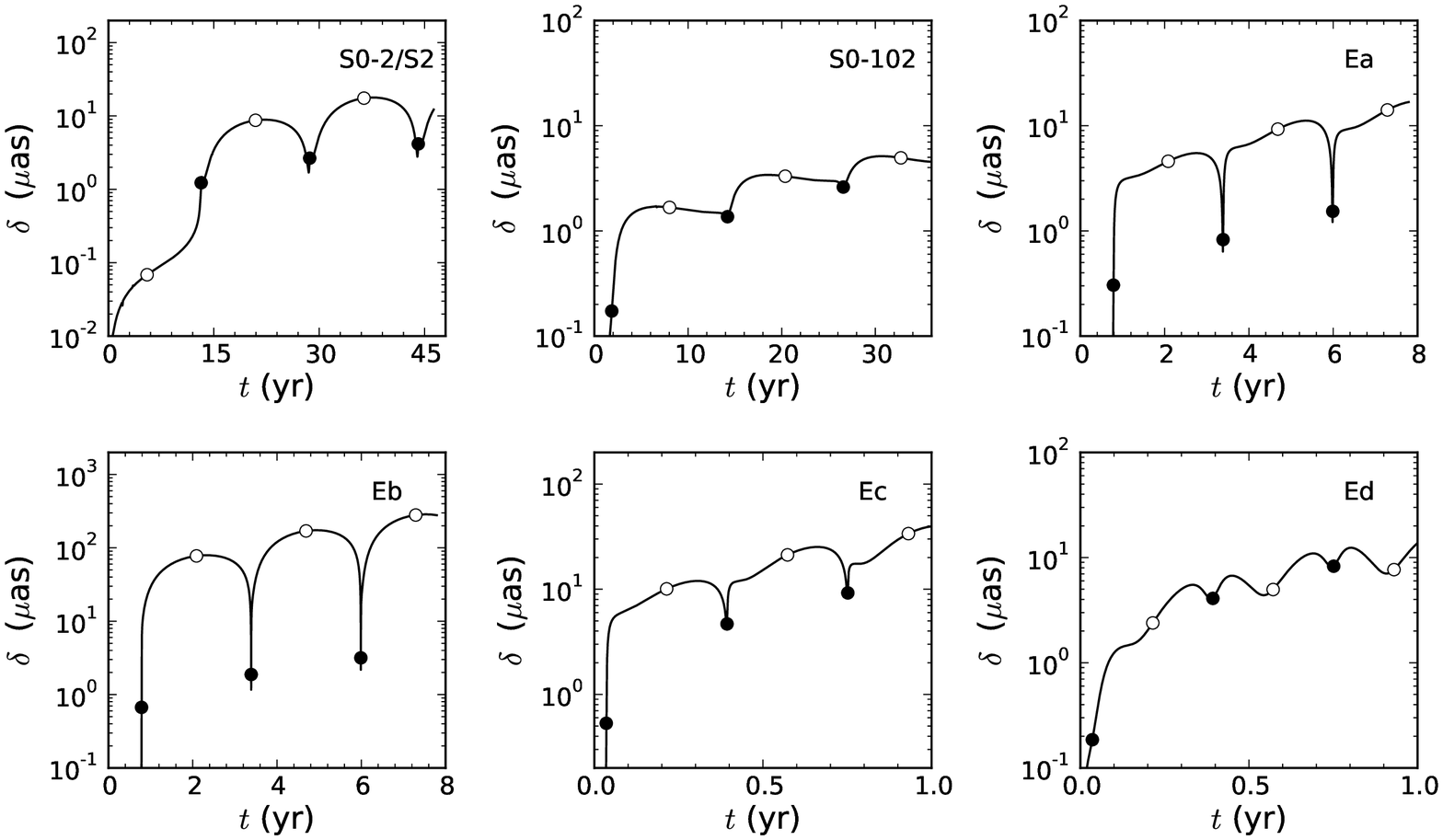}
\caption{Distance between the apparent position of each example star
on the observer's sky plane at each given moment for the case with a
rapidly spinning central MBH ($a=0.99$) and that with a non-spinning
MBH ($a=0$). The distance is defined as $\delta = \sqrt{ (\delta {\rm
R.A.})^2 + (\delta {\rm Dec})^2}$, and $\delta {\rm R.A.}$ and $\delta
{\rm Dec.}$ are shown in Figure~\ref{fig:f4}. Panels from left to
right, top to bottom, represent the example stars S0-2/S2, S0-102, Ea,
Eb, Ec, and Ed, respectively. Open and solid circles mark the
locations of the apoapsides and periapsides, respectively. 
Three total orbits for each example star are shown here and the first orbit
of each star starts at $t=0$.  }
\label{fig:f5} 
\end{figure*}

\begin{figure}
\centering
\includegraphics[scale=0.8]{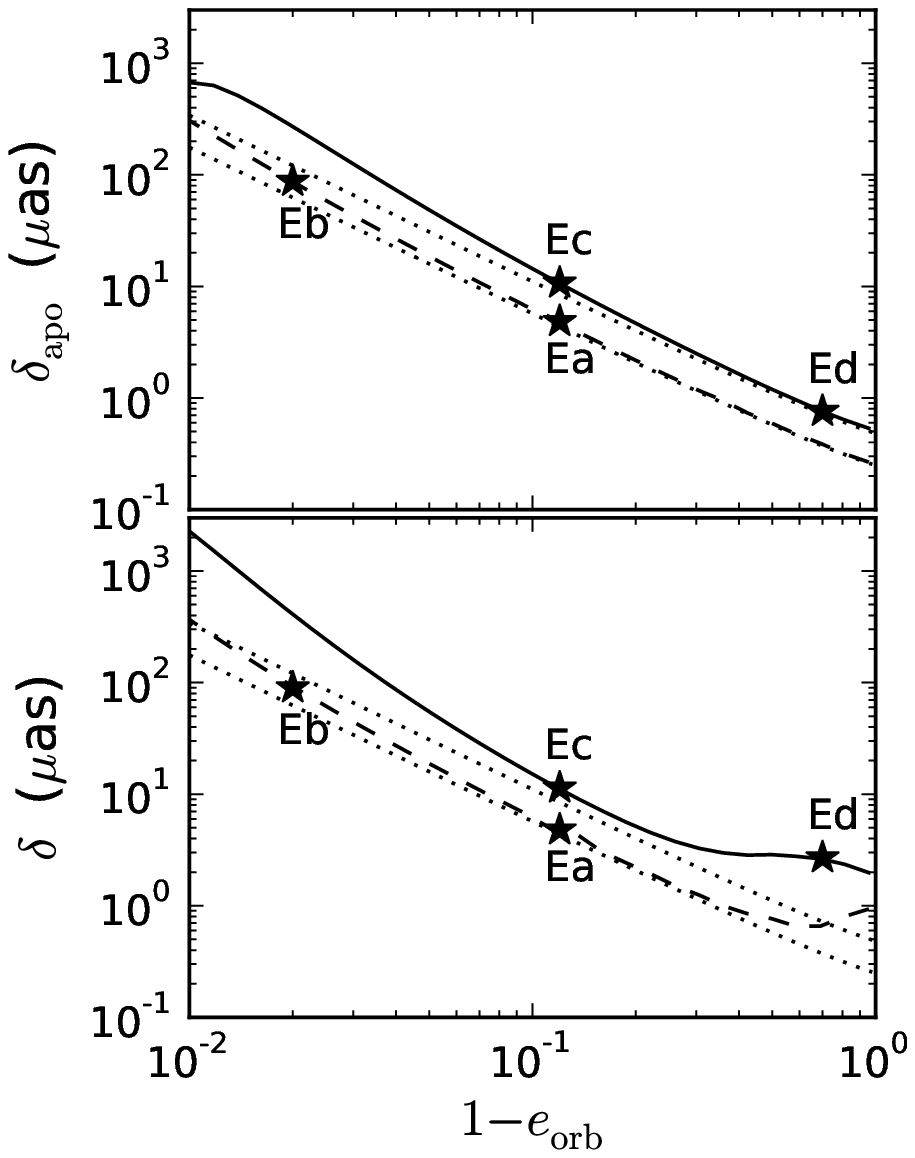}
\caption{Top panel: projected apoapsis displacements due to the MBH
spin after the motion of one full orbit. The solid (dashed) line
represents the results of apoapsis shift defined in the same way as
$\delta_{\rm apo}$ for those stars having the same initial orbital
elements as those of Ec (Ea), but with different eccentricities; the
spin and its direction of the MBH is fixed at $a=0.99$ and $(i,\,
\epsilon) = (45\arcdeg,\, 0\arcdeg)$. Dotted lines represent the
analytical estimates of $\delta_{\rm apo}$ for those stars by using
Equation~(\ref{eq:Delta}). Bottom panel: solid line (dashed line)
represents the difference of $\delta$ between two apoapsides of two
stars with the same initial orbital elements as those of Ec (Ea), with
one of the two rotating around a rapidly spinning MBH $(a=0.99)$ and
the other around a non-spinning MBH ($a=0$). Here $\delta$ is defined
in the same way as that shown in Figure~(\ref{fig:f5}) at apoapsis.
The dotted lines are the same as those in the top panel.  The star
symbols mark the location of the star Ea, Eb, Ec, and Ed,
respectively.  
}
\label{fig:f6}
\end{figure}

\begin{figure}
\centering
\includegraphics[scale=0.8]{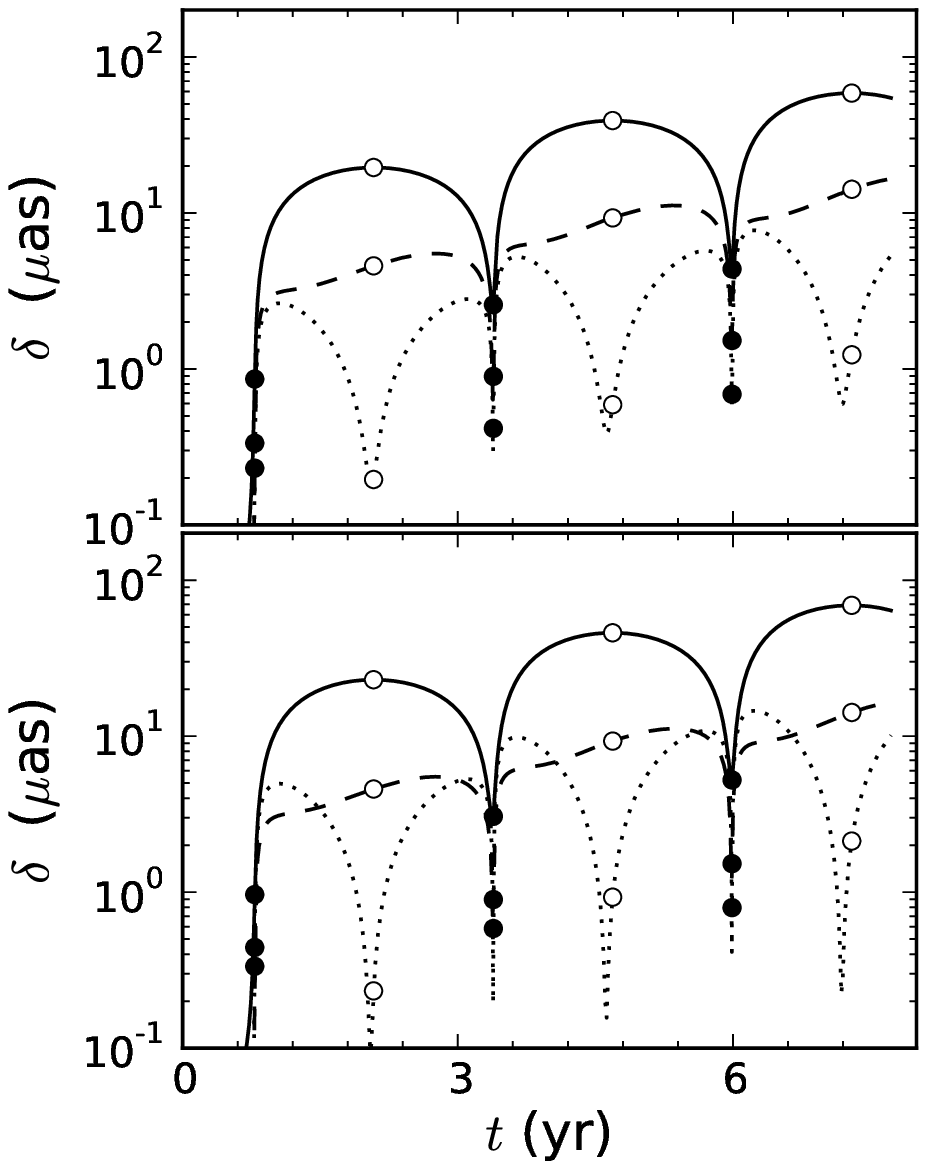}
\caption{Dependence of the spin-induced position displacements on the
spin direction and the star orbital orientation.  The top panel shows the
distance at any given moment between the apparent position of a star
rotating around an MBH with $a=0.99$ and that of the star around an
MBH with $a=0$,  and these two stars have the same initial orbital
elements as those of Ea.  The solid, dashed, and dotted lines represent the
cases for the MBH with a spin direction of $(i,\, \epsilon) =
(71\arcdeg,\, 90\arcdeg)$, $(45\arcdeg,\, 180\arcdeg)$, and
$(159\arcdeg,\, 63\arcdeg)$, respectively.  The bottom panel shows the
distance at any given moment between the apparent position of a star
rotating around an MBH with $a=0.99$ and that of a star rotating
around an MBH with $a=0$; and these two stars have the same semimajor
axis, eccentricity, and argument of periapsis as those of Ea, but with
different inclination ($I'$) and longitude of the ascending node
$(\Omega')$.  The solid, dashed, and dotted lines represent for the cases
with $(\Omega',\, I') = (22\arcdeg,\, 90\arcdeg)$,
$(45\arcdeg,\,0\arcdeg)$, and $(73\arcdeg,\, 62\arcdeg)$,
respectively.  The spin direction of the spinning MBH is fixed at
$(i,\, \epsilon) = (45\arcdeg,\, 0\arcdeg)$. Three total orbits are
shown here and the first orbit starts at $t=0$.
}
\label{fig:f7}
\end{figure}

\begin{figure*}
\centering
\includegraphics[scale=0.7]{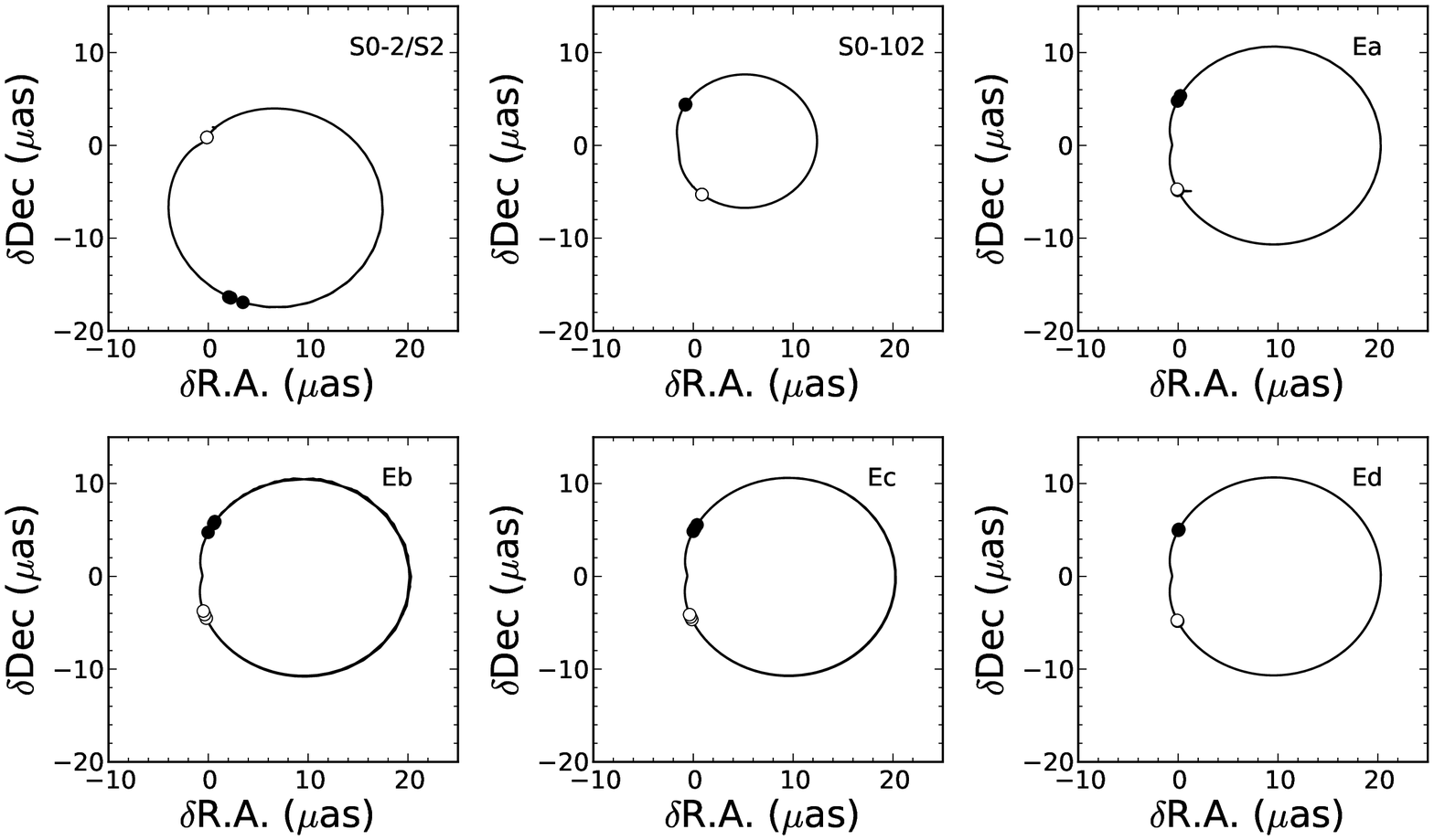}
\caption{Differences in the apparent position of each example star at
each given moment for the case with, and without, the consideration of the
bending of light (obtained from the ray tracing technique).  Panels
from left to right, top to bottom, represent the example stars S0-2/S2,
S0-102, Ea, Eb, Ec, and Ed, respectively. The open and solid circles
mark the locations of the apoapsides and periapsides, respectively.
Three total orbits for each example star are shown in this figure,
but they are difficult to be distinguished. 
}
\label{fig:f8}
\end{figure*}

Figure~\ref{fig:f3} shows the trajectory of each example star on the
observer's sky, which is described by the right ascension (R.A.) and
declination (Dec.) of the star with respect to the GC MBH (or Sgr A*).
The initial settings for the orbital parameters of each star are
listed in Table~\ref{tab:t1}.  As seen from Figure~\ref{fig:f3}, the
orbit of each star precesses mainly due to the advancement of
periapsis; Eb has the most significant advance precession, while
S0-102 has the least advance precession. The advance precession of
periapsis of a star per orbit is
\begin{eqnarray}
\delta \Upsilon'_{\rm S} & = & 3\pi \frac{2 r_{\rm g}}{a_{\rm orb}
(1-e_{\rm orb}^2)} \nonumber \\ & \simeq & 0.19\arcdeg
\left(\frac{4\times 10^6\msun}{\bh}\right) \left(\frac{10^3\AU}{a_{\rm
orb}}\right) \frac{1-0.88^2}{1-e_{\rm orb}^2},
\end{eqnarray}
where $a_{\rm orb}$ and $e_{\rm orb}$ are the semimajor axis and
eccentricity of the star in the Newtonian approximation
\citep[e.g.,][]{Misner73}. For those example stars listed in
Table~\ref{tab:t1}, $\delta \Upsilon'_{\rm S}$ are $0.19\arcdeg$,
$0.095\arcdeg$, $0.64\arcdeg$, $3.64\arcdeg$, $2.39\arcdeg$, and
$0.59\arcdeg$, respectively. Considering the projection of the orbital
plane of each star to the observer's sky plane, the apparent shifts of
the apoapsides of the example stars after three full orbits are $3
a_{\rm orb} (1+e_{\rm orb}) \delta \Upsilon'_{\rm S} (1-\cos^2\Upsilon'
\sin^2 I')^{1/2}\simeq 2.23$, $0.77$, $1.67$, $10.0$, $1.67$, and
$0.29$\,mas for S0-2/S2, S0-102, Ea, Eb, Ec, and Ed, respectively (see
Fig.~\ref{fig:f3}).  According to Figure~\ref{fig:f3}, the advancement
of periapsis can be accurately measured within a few orbits for all of
the example stars, including S0-2/S2 and S0-102, if the astrometric
precision can reach $100-10\mu$as scale.  The spin-induced effects,
including the Lense-Thirring precession and the frame dragging, cannot be
clearly seen in Figure~\ref{fig:f3} because they are more than one order of
magnitude smaller than the periapsis advancement, and these effects
are addressed in Figures~\ref{fig:f4}-\ref{fig:f8}.

If the MBH is non-spinning ($a=0$), the trajectories of those example
stars on the observer's sky plane at a given moment may slightly
deviate from those shown in Figure~\ref{fig:f3} for the case with
$a=0.99$. Figure~\ref{fig:f4} shows the evolution of this difference
for each star over three full orbits. The difference in the apparent
position is mainly due to the Lense-Thirring precession and the frame
dragging. Figure~\ref{fig:f5} shows the distance ($\delta$) between
the apparent position of each star on the observer's sky plane at each
given moment for the case with $a=0.99$ and that with $a=0$.  After
the motion of one full orbit of each star, the differences at apoapsis
are $\sim 8.7$, $1.6$, $4.7$, $92.5$, $11.1$, and $2.6\mu$as for
S0-2/S2, S0-102, Ea, Eb, Ec, and Ed, respectively (see
Fig.~\ref{fig:f5}).  These values are roughly consistent with simple
analytical estimates as illustrated below. This difference mounts up
with the increasing number of orbits as shown in Figures~\ref{fig:f4} and
\ref{fig:f5}.

The Lense-Thirring precession and the frame dragging cause the precession
of the orbital plane of a star, which leads to a difference between
the position of the star in the sky plane at apoapsis/periapsis for
the case with a rapidly spinning BH  and that with a non-spinning BH
after one full orbit. For perturbation theory, this difference is approximately
given by
\begin{eqnarray}
\delta_{\rm apo/peri} & \simeq & a_{\rm orb} (1 \pm  e_{\rm orb}) \left[
\delta^2 \Omega' (1-\sin^2 \Upsilon' \sin^2 I') + \right. \nonumber \\
& & \delta^2 \Upsilon' (1-\cos^2\Upsilon' \sin^2 I')
+\sin\Upsilon'^2\sin^2I' \delta^2 I' + \nonumber \\
&& 2\cos I'\delta \Omega'\delta\Upsilon' -2\sin\Upsilon'\cos\Upsilon'\sin
I'\cos I'\delta\Upsilon'\delta I'\nonumber\\
&& -\left.2\sin\Upsilon'\cos\Upsilon'\sin I'\delta\Omega'\delta I'
\right]^{1/2}.
\label{eq:Delta}
\end{eqnarray}
Here, the signs ``+" and ``-" are for the difference at apoapsis ($\delta_{\rm apo}$) and
periapsis ($\delta_{\rm peri}$), respectively, $\delta \Omega'$, $\delta \Upsilon'$, and
$\delta I'$ are the changes of $\Omega'$, $\Upsilon'$, and $I'$ due to the
Lense-Thirring precession over one full orbit, respectively. For the
detailed derivation of Equations~\ref{eq:Delta}, see
Appendix~\ref{sec:analytical}.  According to
Equation~(\ref{eq:Delta}), we obtain $\delta_{\rm apo}
= 8.6$, $1.5$, $4.4$, $62.5$, $8.4$, and $0.7\,\mu$as;
$\delta_{\rm peri} = 0.55$, $0.30$, $0.28$, $0.64$, $0.55$, and $0.40\,\mu$as
for the example stars S0-2/S2, S0-102, Ea, Eb, Ec, and Ed,
respectively, if $a=0.99$ for the rapidly spinning BH case. The
$\delta_{\rm apo}$ values are roughly consistent with numerical results
obtained above for most example stars, except that for Ed.  

We note here that the $\delta$ at periapsides shown in
Figure~\ref{fig:f5} is obtained in a way slightly different from the
$\delta_{\rm apo}$ estimated from Equation~(\ref{eq:Delta}). The values of
$\delta$ are calculated for the projected distances at any given time
between the star rotating around a rapidly spinning MBH $(a=0.99)$ and
a star with the same initial orbital elements but rotating around a
non-spinning MBH ($a=0$); while the value of $\delta_{\rm apo}$ is estimated
by considering the difference between the position differences of two
adjacent apoapsides of a star rotating around an MBH with $a=0.99$ and
those of a star with the same initial orbital elements rotating around
a non-spinning MBH.  As the orbital periods of these two stars are
slightly different, the shift of $\delta$ at two adjacent apoapsides
shown in Figure~\ref{fig:f5} should therefore be roughly the same as,
but different from, the analytical estimates $\delta_{\rm apo}$, and the
difference is most significant for those with small eccentricity.

For a consistency check between the numerical results obtained in this
study and the analytical ones, as an example, we show the apoapsis
shifts numerically estimated in two different ways for those stars
with the same semi-major axis of $300$\,AU (or $80$\,AU) but various
eccentricities after one full orbit motion in Figure~\ref{fig:f6}. The
other initial orbital elements of those stars are the same as those of
Ea (or Ec). The top panel of Figure~\ref{fig:f6} shows the numerical
results obtained in the same way as $\delta_{\rm apo}$, which are well
consistent with the analytical estimates at $e\la 0.8$, but higher
than the analytical ones at $e\ga 0.9$. There are three reasons for the difference
at high eccentricities: (1) the Newtonian
approximation to the orbit of a star for the initial conditions
becomes inaccurate if the star is close to the MBH (e.g., for those
orbits with extremely high eccentricities); (2) for an Ea-like star
with $e_{\rm orb}$ higher than $0.98$, its orbital precession can be
larger than $\sim 20^\arcdeg$ per orbit, therefore, the first order
approximation (for the projection) to obtain Equation~(\ref{eq:Delta})
is inaccurate; (3) the contributions from the higher order precessions
are ignored in the analytical estimates. As showed in
Figure~\ref{fig:f6}, the combinations of the above three factors can
almost lead to $20\%$ to a factor of two difference when $e_{\rm orb}
\ga 0.98$.  The bottom panel of Figure~\ref{fig:f6} shows the
numerical results of $\delta$ obtained in the same way as those shown
in Figure~\ref{fig:f5}, which apparently differ significantly from
the analytical estimates at both $e_{\rm orb}\la 0.7$ and $e_{\rm orb}
\ga 0.9$ because $\delta$ at apoapsis and $\delta_{\rm apo}$ are defined and
obtained in a slightly different way.

According to Figure~\ref{fig:f5}, we conclude that, if the accuracy in
determining the apparent positions of any of those example stars (except
S0-102) on the sky plane can reach $\la 10\,\mu$as, the spin of the GC
MBH can be then constrained by fitting the evolution of its positions
over several orbits as demonstrated by a Bayesian fitting method in
Section~\ref{sec:fit} (see also Tab.~\ref{tab:t2}). 

For a star with given initial orbital elements, the displacement of
apoapsis after one full orbit due to the spin effects is proportional
to the absolute value of the spin, and it also depends on the spin
direction (see eqs.~\ref{eq:domega}-\ref{eq:Delta_A}).  To illustrate
the dependence of the spin-induced position displacement on
the spin direction, the top panel of Figure~\ref{fig:f7} shows the
evolution of $\delta$ of Ea for three different
spin directions. The MBH spin is fixed at $a=0.99$.  As seen from this
figure, $\delta $ is the most significant when $(i,\, \epsilon) =
(72\arcdeg,\, 90\arcdeg)$, and the least significant when $(i,\,
\epsilon) = (159\arcdeg,\, 63\arcdeg)$. The difference in the spin
direction may lead to more than an order of magnitude difference in
the maximum position displacement over a full orbit,
which suggests that the MBH spin may be easier to be constrained by
the relativistic motion of a given star if the spin vector is close to
some special direction. We note here that $\delta_{\rm apo}$ of Ea is the
largest ($\delta_{\rm apo} \simeq 20\mu$as for $a=0.99$) if $(i, \epsilon) =
(72\arcdeg, 90\arcdeg)$ or $(108\arcdeg, 270\arcdeg)$, and it is
almost the smallest ($\delta_{\rm apo} \simeq 0.4\mu$as for $a=0.99$) if $(i,
\epsilon)=(159\arcdeg,  63\arcdeg)$ or $(21\arcdeg, 243\arcdeg)$.
(These values are obtained from Equation~\ref{eq:Delta}.)

The spin-induced (apoapsis) position displacement ($\delta$) also
depends on the orientation of the orbital plane of a star relative to
the sky plane, if the MBH spin and its direction are fixed (see
eqs.~\ref{eq:domega}-\ref{eq:Delta_A}). The bottom panel of
Figure~\ref{fig:f7} illustrates this dependence by plotting the
evolution of $\delta$ for Ea and two other stars with the same
semimajor axis and eccentricity as those of Ea but with different
orbital orientations. The MBH spin is fixed at $a=0.99$ and the spin
direction is $(i,\, \epsilon)= (45\arcdeg,\, 180\arcdeg)$.  As seen
from the figure, the amplitude of $\delta$ may be different by more
than an order of magnitude and the evolution pattern of $\delta$ can
also be different if the orientation of the star orbital plane is
different, simply because of the projection effect.

The apparent positions of a star, rotating around the MBH, on the
observer's sky plane, may be significantly affected by the bending of
light rays propagating from the star to the observer due to the
relativistic potential of the MBH. Figure~\ref{fig:f8} shows the
difference in the apparent position of each example star at each
moment for the case with, and without, the consideration of the bending of
light (obtained from the ray-tracing method detailed in
Appendix~\ref{sec:Appendix1}).  As seen from this figure, the position
difference can be as large as $10-20\mu$as, roughly on the order of
the Einstein radius of the MBH ($\sim 4GM_{\bullet} /c^2 \sim
20\mu$as), which must be accurately considered when constraining the
MBH spin, as the spin-induced position displacements are mostly on the
same order (see Fig~\ref{fig:f5}). The position displacements caused
by the bending of light are almost the same for the case with a
rapidly spinning MBH and the case with a non-spinning MBH, and thus
cannot be distinguished from each other (see the solid lines for cases
with $a=0.99$ in Figure~\ref{fig:f8}, the results obtained for cases
with $a=0$ coincide with those solid lines, and the differences
between these two cases are on the order of $0.1\mu$as).  These
differences, though small, are roughly on the same order as the
effects induced by the quadruple moment of the MBH for S0-2/S2.
Therefore, it is important to include them for future high-precision
measurements that may be used to constrain the quadruple moment and
test GR.
The cardioid-like shapes of the curves shown in Figure~\ref{fig:f8}
are due to the position change of the star with respect to the central
MBH, and consequently the change of the MBH potential and the
deflection angle. If the true position of the star is right in between
the MBH and the observer, the apparent position should be at the
coordinate origin in Figure~\ref{fig:f8}.  We note that the
trajectories of apparent positions of a distant star deflected by
Jupiter, which is moving around sun, have similar cardioid-like shapes
due to similar underlying physics \citep[see][]{Kopeikin07}. 

\begin{figure*}
\centering
\includegraphics[scale=0.7]{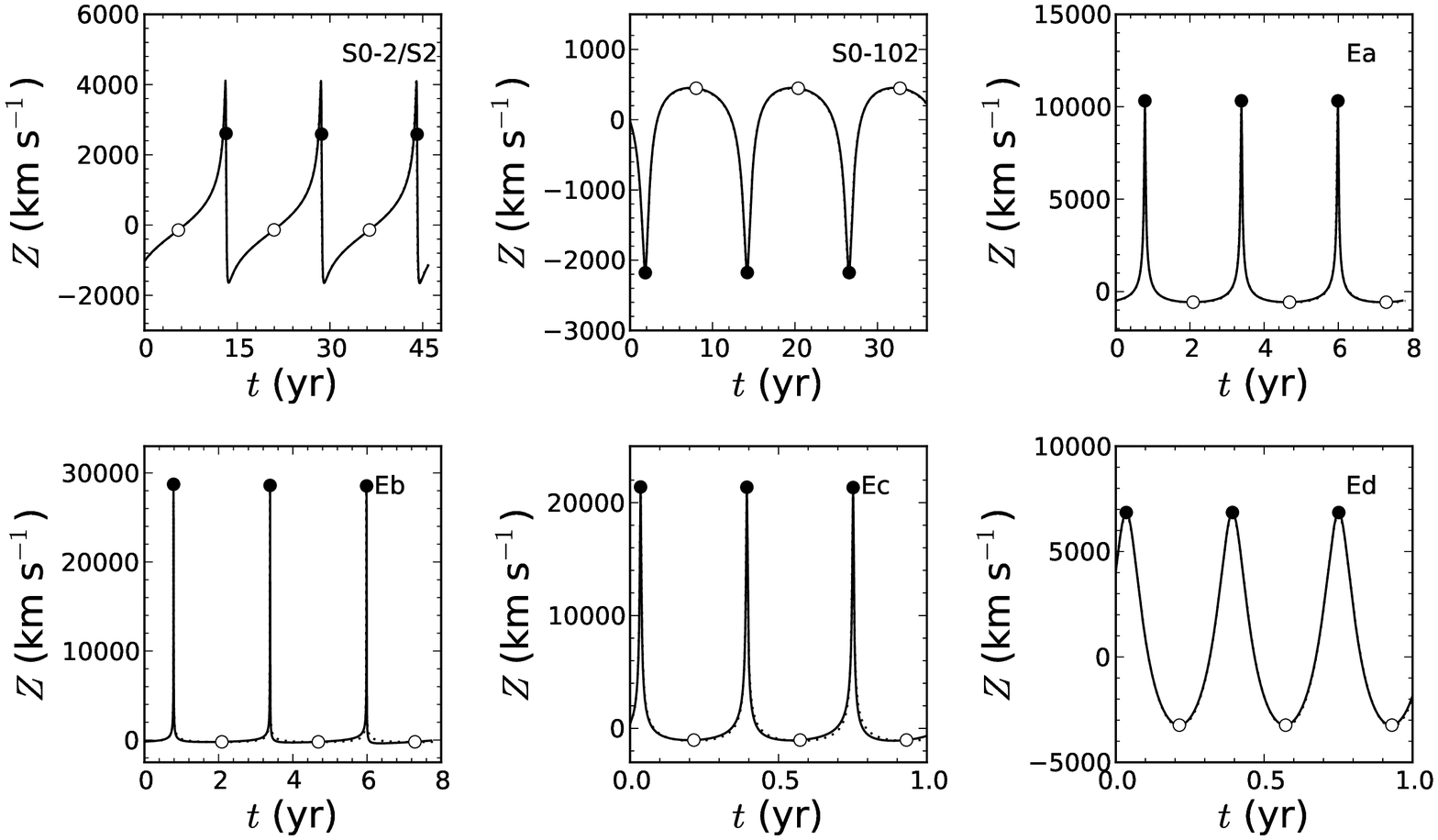}
\caption{Redshift evolution curves of example
stars. Panels from left to right, top to bottom, represent the example
stars S0-2/S2, S0-102, Ea, Eb, Ec, and Ed, respectively. In each
panel, the solid (open) circles mark the pericenter (apocenter)
passage of the star, and the solid and dashed lines represent the
numerical results and the approximation given by eq.~(\ref{eq:vzp}),
respectively. Totally three full orbits are shown here and the first
orbit starts at $t=0$.
}
\label{fig:f9}
\end{figure*}

\subsection{Spin-induced Redshift Differences}

Figure~\ref{fig:f9} shows the redshift evolution
of each example star monitored by a distant observer for three full
orbits. As shown in this figure, the maximum spin-induced redshift (or
blueshift) differences over a full orbit are about $0.4$, $0.05$,
$2.28$, $82$, $55$, and $3.7\,\kms$ for S0-2/S2, S0-102, Ea, Eb, Ec,
and Ed, respectively.

The redshift of a star is approximately given by
\be
Z \simeq - {\mathcal V}_{z'} \simeq  - {\mathcal{V}}_{{\rm K},z'} +
\frac{c}{2}({\mathcal V}_{\rm K}/c)^2 + c r_{\rm g}/r + \delta
{\mathcal V}(a),
\label{eq:vzp}
\ee
with the first, second, third, and fourth terms in the right-hand side
representing the Newtonian Doppler shift due to the Keplerian motion
of the star, the special relativity correction to the Doppler shift
(or the transverse Doppler shift), the gravitational redshift, and the
part of redshift due to the MBH spin, respectively. Here the redshift
$Z$ is defined to have the same unit ($\kms$) as that of velocity.  In
the Newtonian approximation,
\be
{\mathcal V}_{{\rm K},z'} = - \sqrt{\frac{G\bh}{a_{\rm orb}(1-e_{\rm
orb}^2)}} \sin I'[e_{\rm orb}\cos\Upsilon' + \cos( \Upsilon' +
\upsilon')],
\ee
%
%
\be
{\mathcal V}_{\rm K} \simeq \left(2G\bh/r - G\bh/a_{\rm orb}
\right)^{1/2},
\ee
\be
r \simeq  a_{\rm orb} (1-e_{\rm orb}^2)/(1+e_{\rm orb} \cos \upsilon'),
\ee
and $\delta {\mathcal V}(a)$ is negligible comparing with those other
terms. The second and third terms on the right side of
Equation~(\ref{eq:vzp}) are much less than the first term when $r\gg
r_{\rm g}$.  According to Equation~(\ref{eq:vzp}), ${\mathcal V}_{z'}$
approaches the maximum redshift and the maximum blueshift (or the
maximum blueshift and the maximum redshift) at $\upsilon' \sim
\Upsilon'$ and  $\pi -\Upsilon'$ if $\sin I' \cos \Upsilon'
> 0$ (or if $\sin I' \cos \Upsilon' <0$), respectively. At $\upsilon'
=\Upsilon'$ (or $\upsilon' = \pi -\Upsilon'$), the star crosses the sky
plane ($x'y'$). As seen from Figure~\ref{fig:f9}, the approximations
given by Equation~(\ref{eq:vzp}) are well consistent with the full GR
calculations. 

\begin{figure*}
\centering
\includegraphics[scale=0.70]{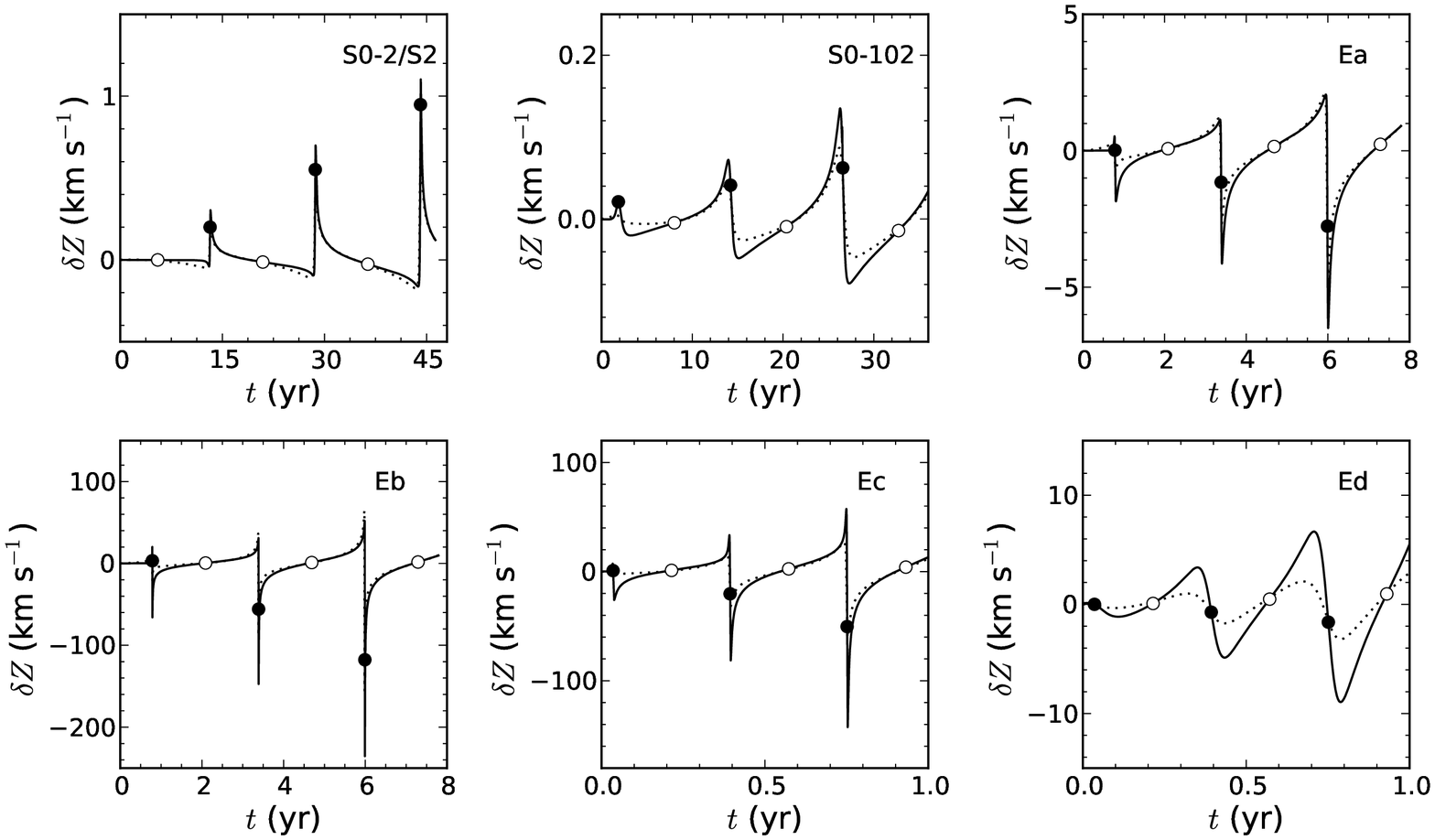}
\caption{Difference between the redshift of each example star at each
given moment for the case with a rapidly spinning MBH ($a=0.99$) and
that with a non-spinning MBH ($a=0$). Panels from left to right, top
to bottom, represent the example star S0-2/S2, S0-102, Ea, Eb, Ec, and
Ed, respectively. Open and solid circles mark the locations of the
apoapsides and periapsides, respectively. Solid lines and dashed lines
represent the numerical results and analytical approximations obtained
from Equation~\ref{eq:vz}, respectively.  Totally three orbits for
each example star are shown here and the first orbit of each star
starts at $t=0$. 
}
\label{fig:f10}
\end{figure*}

The gravitomagnetic field generated by the spinning MBH influences not
only the motion of a star close to it but also the propagation of
photons from the star to the distant observer
\citep[i.e.,][]{Thorne86}. Both of these effects are encoded in the
amount of shifts (or the redshift) of lines in the star spectrum
measured at different times.\footnote{The low-order effects, such as
the GR gravitational redshift, the relativistic Doppler shift, and the
R{\o}mer delay,  on the redshift have been investigated in the
literature separately through perturbative approximations
\citep[e.g.,][]{Zucker06, Angelil10b}.  In this study, we only focus
on the spin related line shifts (or redshift) and do not separately
study those low-order effects one by one as in previous studies. 
The low-order effects, as well as the high-order effects, are automatically involved in our full GR
calculations.} Figure~\ref{fig:f10} shows the redshift difference
($\delta Z$) of each example star at each given moment for the case
with a rapidly spinning MBH ($a=0.99$) and that with a non-spinning
MBH. For each star, as seen from this figure, the redshift difference
changes over the course of a single orbit, and it is most significant
near the periapsis and most insignificant near the apoapsis; the
amplitude of the differences also increases with increasing number of
the orbital periods passed. For S0-2/S2, the redshift difference is
about $\sim 0.3\kms$ at the periapsis of its first orbit, and it
mounts up to $\sim 1\kms$ after about three full orbits; for the Eb
and Ec, the redshift difference can be up to $\ga 100\kms$ at the
periapsis of the third orbit. The amplitude of the redshift
differences is the largest for Eb and the smallest for S0-102, which
is fully consistent with the dependence of the redshift difference due
to the Lense-Thirring precession and the frame dragging effect on the
distance ($r$) to the MBH revealed by \citet{KannanSaha09} and
\citet{Angelil10b}, i.e., $\delta Z \propto r^{-2}$ [and $\propto
a_{\rm orb}^{-2} (1-e_{\rm orb})^{-2}$ at the periapsis]. If the
accuracy of the redshift measurements can reach $\la 1\kms$ (or
$\la 50\kms$), the spin of the MBH may be well constrained by using an
observational redshift curve of S0-2/S2 (or the example star Eb or Ec)
over three or more orbits and correspondingly $\ga 45$ years (or $\ga
6$ years for Eb, or $\ga 1$ years for Ec).

The Lense-Thirring precession and the frame dragging can lead to a
change in the orbital orientation of a star, as shown above and in
Appendix~\ref{sec:analytical}. One significant component of the
redshift difference $\delta {\mathcal V}(a)$ must be that introduced
by the change of the orbital orientation, i.e.,
\begin{eqnarray}
\delta Z(v) & \simeq & - \sqrt{ \frac{G\bh}{a_{\rm orb}(1-e^2_{\rm
orb})}} \left\{ \cos I' [e_{\rm orb} \cos \Upsilon' +\cos (\Upsilon'
+\upsilon')]\right.  \nonumber \\ 
& & \left. \times \delta I'' + \sin I' [-e_{\rm orb} \sin \Upsilon'
-\sin (\Upsilon' + \upsilon')] \delta \Upsilon'' \right\},
\label{eq:vz}
\end{eqnarray} 
where $\delta I''$ and $\delta \Upsilon''$ are the changes of $I'$ and
$\Upsilon'$ induced by the Lense-Thirring effect since the start of
the orbital motion from $\upsilon'_0$  to $\upsilon'$ (or $E_0$ to $E$
from $t_0$ to $t$). Furthermore, 
\be 
\delta I'' \simeq [ (E-E_0) - e_{\rm orb}( \sin E- \sin E_0) ] \delta
I' /2\pi, 
\ee 
\be
\delta \Upsilon'' \simeq [(E-E_0)- e_{\rm orb}( \sin E- \sin E_0 )]
\delta \Upsilon'/2\pi, 
\ee
\be
E_0-e_{\rm orb}\sin E_0 =\sqrt{\frac{GM_{\bullet}}{a_{\rm
orb}^3}}(t_0-T_0),
\ee \be
E-e_{\rm orb}\sin E =\sqrt{\frac{GM_{\bullet}}{a_{\rm orb}^3}}(t-T_0).
\ee
Here $\delta I'$ and $\delta \Upsilon'$ are the spin-induced changes of
$I'$ and $\Upsilon'$ per orbit (see Appendix B), $T_0$ is the time of
pericentric passage with respect to the year of $2000$, and
$t_0=2020$ year is the starting time.  The dashed lines in
Figure~\ref{fig:f10} represent the analytical estimates obtained from
Equation~(\ref{eq:vz}), which seem to be roughly consistent with the
numerical results (dashed lines) obtained from the full GR
calculations, at least at the apoapsides and the periapsides.
According to Equation~(\ref{eq:vz}) and Appendix~\ref{sec:analytical},
$\delta Z $ is roughly proportional to $a^{-2}_{\rm orb} (1-e_{\rm
orb})^{-2}$ [or $a^{-2}_{\rm orb} (1- e_{\rm orb})^{-1}$] at periapsis
[or apoapsis], if $1-e_{\rm orb} \ll 1$.
According to Equation~(\ref{eq:vzp}), the maximum redshift (or
blueshift) difference induced by the MBH spin at periapsis are $0.2$
$(-0.06)$, $0.04$ $(-0.02)$, $0.9$ $(-2.0)$, $28$ $(-68)$, $13$
$(-28)$ and $1.0\kms$ ($-1.4\kms$) over a full orbit for S0-2/S2,
S0-102, Ea, Eb, Ec, and Ed, respectively, which are roughly consistent
with the full GR numerical results shown in Figure~\ref{fig:f10}. 
There are indeed some differences between the approximations given by
Equation~(\ref{eq:vz}) and those obtained from the full GR
calculations, which are due to (1) the adoption of the mean change
rates of $\Omega'$, $\Upsilon'$, and $I'$ (for the purpose of illustration
here, we do not intend to adopt the evolution forms of those change
rates as functions of $\upsilon'$); and (2) the neglect of the light
propagation effects and high-order precession in the approximations
and the slight differences in the estimates of  $\delta Z$ (as
discussed for the position displacements above; see
Fig.~\ref{fig:f6}).

\begin{figure}
\centering
\includegraphics[scale=0.8]{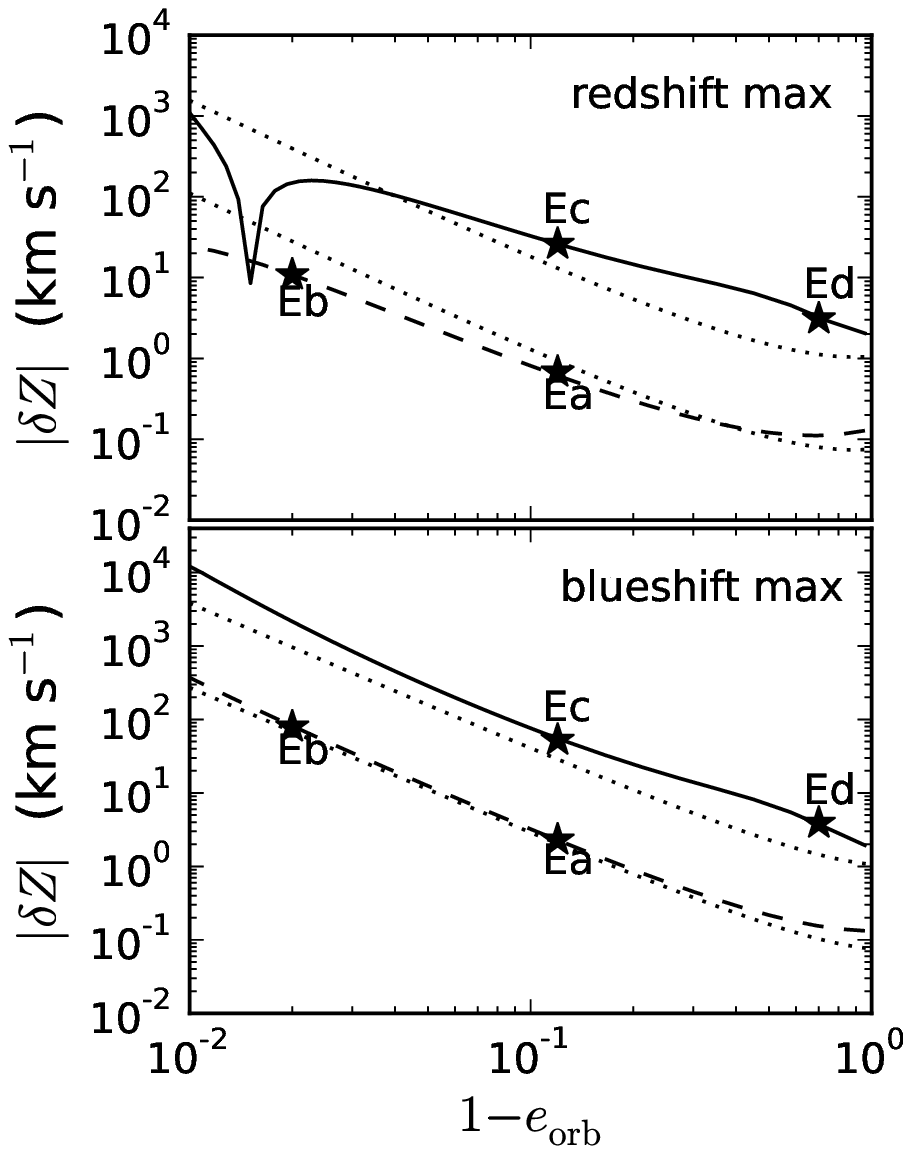}
\caption{Maximum redshift difference (top panel) and the maximum
blueshift difference (bottom panel) induced by the MBH spin. The MBH
spin and its direction are assumed to be $a=0.99$ and $(i,\, \epsilon)
= (45\arcdeg,\, 180\arcdeg)$.  Solid and dashed lines represent the
numerical results obtained in the same way as those shown in
Fig.~\ref{fig:f10} for stars with the same orbital elements as those
of Ec (solid lines) and Ea (dashed lines), respectively, except with
various eccentricities.  Dotted lines represent the estimates obtained
approximately by Equation~(\ref{eq:vz}). This figure indicates that
the spin-induced redshift is mainly contributed by the change of the
orbital orientation due to the Lense-Thirring precession and the frame
dragging. 
}
\label{fig:f11}
\end{figure}

Figure~\ref{fig:f11} shows the maximum redshift difference (top panel)
and the maximum blueshift difference (bottom panel) induced by the
MBH spin ($\delta Z$) obtained for stars with the same orbital
elements as those of Ec (or Ea) except with various initial
eccentricities. With an increasing number of orbits, the absolute value
of the maximum redshift (or blueshift) difference mounts up.  As seen
from this figure, the analytical approximations are quite consistent
with the numerical results, especially when the semimajor axis of the
star is larger, which suggests that the main contribution to the
redshift difference $\delta Z$ is the change of the star orbital
orientation and thus the change of the projection of the velocity to
the line of sight. For stars with small semimajor axes, the analytical
estimates deviate from the full GR numerical results, which
strengthens the necessity of using the full GR calculations for
accurately constraining the MBH spin.

Note that the redshift difference due to the spin effects on the
propagation of photons is about $10\sim 100$ times smaller than that
on the star orbit \citep[see also][]{Angelil10b}. 
For S0-2/S2, if the accuracy of redshift measurements is on the order
of $\ga 1\kms$, the spin effect on photon propagation may be
negligible (although the Schwarzschild effect on photon propagation is
significant, close to $10\kms$, and cannot be neglected; see
\citealt{Angelil10b}).  However, for stars with smaller semimajor axes
and higher eccentricities, such as the example stars Eb and Ec, it is
necessary to include the spin effect on photon propagation (on the
order of $\sim 3\kms$ for Eb and $\sim 1\kms$ for Ec, close to $10\%$
of the total spin-induced redshift differences) in
order to get an accurate constraint on the MBH spin.  Our full GR
calculations are fast and efficient, which enable us to automatically
have the high-order precession effects due the MBH spin and thus
provide a more accurate fitting scheme for future high-precision
measurements (see Section~\ref{sec:fit}) compared with those
approaches adopting the perturbative approximations
\citep[e.g.,][]{Angelil10a}.

\begin{figure}
\centering
\includegraphics[scale=0.8]{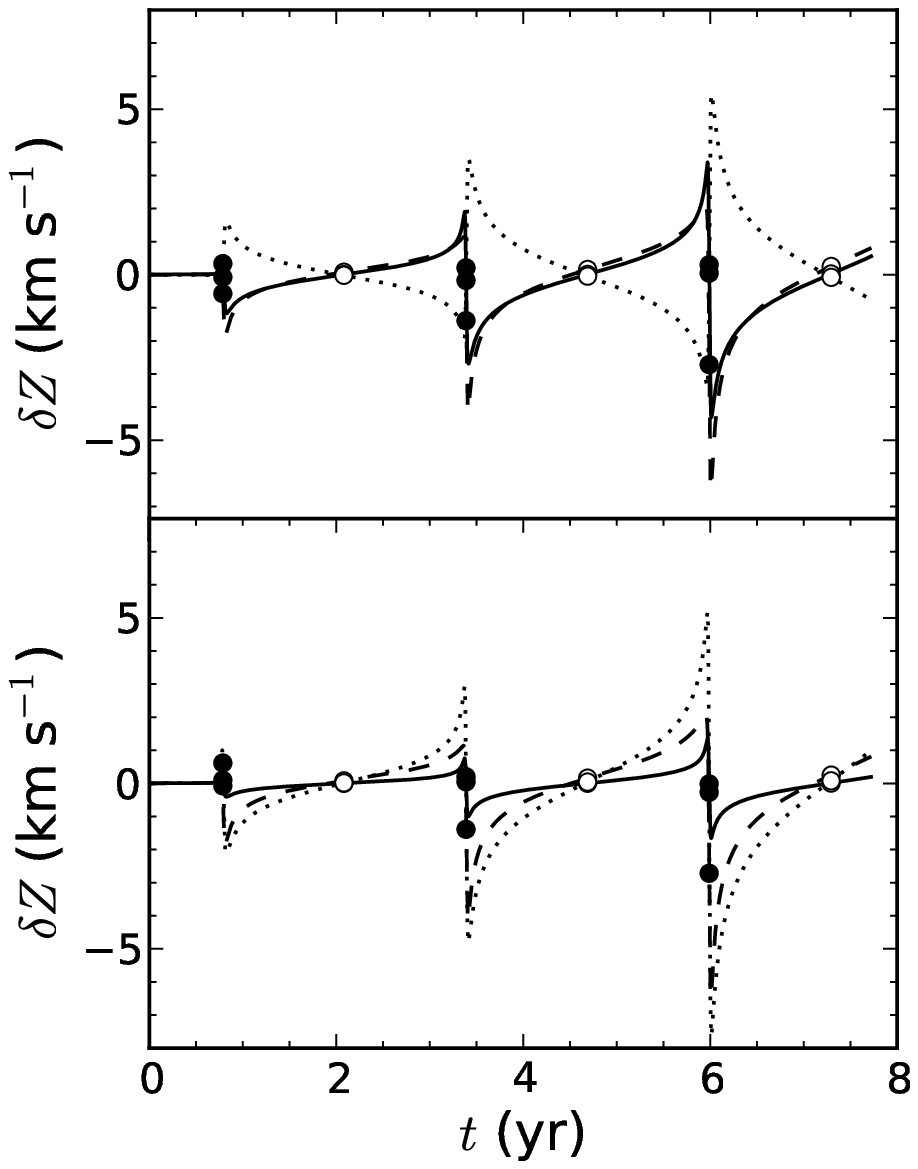}
\caption{Dependence of the spin-induced redshift difference on the
spin direction (top panel) and the star orbital orientation (bottom
panel).  Legend similar to Figure~\ref{fig:f7}.
}
\label{fig:f12}
\end{figure}

The spin-induced redshift difference $(\delta Z)$ also depends on the
spin direction and the orbital orientation of the star as shown in
Figure~\ref{fig:f12}, similar to that for the spin-induced position
displacement shown in Figure~\ref{fig:f8}. The top panel of
Figure~\ref{fig:f12} illustrates the dependence of $\delta Z$ on the
spin direction by showing the evolution of $\delta Z$ of Ea, for three
different spin directions.  The bottom panel of Figure~\ref{fig:f12}
illustrates the dependence of $\delta Z$ on the orbital orientation of
stars, by showing the evolution of $\delta Z$ of Ea and two other
stars with the same semimajor axis and eccentricity as those of Ea but
with different orbital orientations.  As seen from this figure, these
dependences help to simultaneously constrain both the absolute spin
value and the spin direction.

\section{A General Method to Constrain the MBH Spin Parameter }
\label{sec:fit}

In this section, we introduce a general method to constrain the spin
parameter and the metric of the GC MBH via the relativistic motion of
a star in the immediate vicinity of the MBH by using the MCMC
technique.

We first generate mock observations for each example star listed in
Table~\ref{tab:t1}, in order to investigate whether the spin parameter
of the GC MBH can be tightly constrained by monitoring the orbital
motion of a close-in star through the next generation facilities, such
as the GRAVITY on VLTI, TMT, and E-ELT. For this purpose, we assume
that the MBH mass is $\bh = 4\times 10^6\msun$ and the distance from
the Sun to the GC is $R_{\rm GC} = 8\kpc$. The MBH spin and its direction
are assumed as those listed in Table~\ref{tab:t1}.  With each given
set of initial parameters for the six orbital elements of a star, we
can obtain the trajectory of the star's apparent position on the sky
plane ($\alpha_{{\rm obs},j},\, \beta_{{\rm obs},j}$) and the redshift
curve ($Z_{{\rm obs},j}$) for any given time period through
the full general relativistic calculations described in
Sections~\ref{sec:cal_star} and \ref{sec:cal_photon}. Here the
subscript $j$ denotes a mock observation made at the observer's time
$t_{{\rm obs},j}$. The spin-induced position displacement is most
significant near the apoapsis and the spin-induced redshift difference
is most significant near the periapsis. The best strategy would be
monitoring the star near its apoapsis and periapsis more frequently
than at other locations in order to efficiently measure the
spin-induced effects and constrain the spin.  Nevertheless, here we
set the intervals between two consecutive observational times as
$\delta t_{\rm obs} \propto r^{-1.5}$ (in the observer's rest frame).
For each star, the total number of mock observations is $120$.
Further assuming a set of expected accuracies on the apparent position
($\sigma_{\rm p}$) and the redshift ($\sigma_Z$) measurements
of a star, e.g., $(\sigma_{\rm p},\, \sigma_Z) = (10\mu{\rm as},\,
1\kms)$, we obtain the mock observations on the apparent position and
the redshift of the star. The points in the top two panels of
Figures~\ref{fig:f13} and \ref{fig:f15}) show the mock observations of
the apparent positions (top left) and the redshifts (top right) for
S0-2/S2 and Eb, respectively.  With those mock data, we can use the
following procedures to constrain the MBH properties (mass and spin)
and the distance from the MBH to sun.

\subsection{MCMC Fitting Procedures}
\label{subsec:fitmethod}

\begin{enumerate}

\item For a given set of parameters for the  MBH spin parameters
($a,\, i,\, \epsilon$), the MBH mass ($\bh$), and the initial orbital
elements of a model star $(a_{{\rm orb},0},\, e_{{\rm orb},0},\,
\Omega'_0,\, I'_0,\, \Upsilon'_0,\, \upsilon'_0)$, the orbital motion
of the star is generated over a period time that is the same as that of
the mock one from
full GR calculations as described in Sections~\ref{sec:cal_star}. 
%
Note that the motion of the star is defined with respect to the
central MBH, which is approximated to be at rest in the Galactocentric
rest frame.
\footnote{
%
%
The acceleration of the MBH with respect to the observer are
negligible on the timescale considered in this paper ($\la 50$ years).
First, the acceleration of the Sun with respect to the GC is on the
order of $10^{-6}\kms$ per year, and the total change of the relative
velocity of the MBH due to this acceleration on the timescale $\la
50$ years is on the order of $ 10^{-4}\kms$. Second, the Brownian motion
of the MBH due to background stars is estimated to be on the order of
$0.1\kms$ and the change timescale may be larger than $ 10^3$ years
\citep[e.g.,][]{Merritt07}, so that the velocity change due to the
Brownian motion should be $\la 0.01\kms$ on a timescale of $50$ years.
Therefore, the initial position of the MBH and the relative motion of
the observer with respect to the MBH (to Sgr A*) can be approximately
described by the following six parameters: the distance from the MBH
(or Sgr A*) to the observer, the initial position of the MBH (or Sgr
A*) on the sky plane (two parameters, and the proper motion and the
radial motion of the observer with respect to the MBH (or Sgr A*;
three parameters). For an accurate reconstruction of the orbital
configuration of a star, it is necessary to also include all of those
parameters in the MCMC fitting procedures described below. For
the purpose of demonstration purpose in the present paper, however, we assume that
the MBH position is fixed (or the same as that of the Sgr A*) on the
sky plane and both the proper motion and the radial motion of the
distant observer with respect to the MBH (or Sgr A*) are zero in the
Galactocentric rest frame. Therefore, in the following MCMC fitting
procedures, the MBH position is described by only one parameter, i.e.,
the distance from the distant observer to the MBH (or Sgr A*).  With
future high-quality observations (e.g., on the position and the motion
of Sgr A*), it is possible to also constrain all of those six
parameters,
simultaneously, in practice.}

\item According to the motion of the model star, we can obtain the
apparent position $(\alpha_j,\ \beta_j)$ and the 
redshift $Z_{{\rm obs},j}$ of the star at each observation time
$t_{{\rm obs},j}$, for any given distance of $R_{\rm GC}$ from the MBH
to the Sun (see Section~\ref{sec:cal_photon}). 
In order to determine both the trajectory of the apparent positions of
a star on the observer's sky plane and the redshift curve, totally
$11$ total parameters, i.e., ${\bf \Theta} = (R_{\rm GC},\, \bh,\, a,\, i,\,
\epsilon,\, a_{{\rm orb},0},\, e_{{\rm orb},0},\, \Omega'_0,\, I'_0,\,
\Upsilon'_0,\, \upsilon'_0)$, are needed.  The exact values of the last
six parameters for the initial orbital elements of the star are not
interesting for the purpose of this study.

\item To check whether the model star can fit the mock observations,
we use the $\chi^2$-statistics for both the trajectory of the apparent
position $(\alpha_j,\, \beta_j)$ and the redshift curve ($Z_j$), i.e.,
\be
\chi^2_{\rm p}=\sum^{N}_{i=1} \left[\frac{(\alpha_{j}- \alpha_{{\rm
obs},j})^2+ (\beta_{j}-\beta_{{\rm obs},j})^2}{\sigma_{\rm
p}^2}\right], 
\ee
and
\be
\chi^2_{Z}=\sum^{N}_{j=1} \left[\frac{(Z_j- Z_{{\rm
obs},j})^2}{\sigma_{Z}^2}\right],
\ee
respectively, where $N$ is the total number of observations. If both
the trajectory of the apparent positions and the redshift curve of the
mock star are used for the fitting, we have $\chi^2= \chi^2_{\rm p} +
\chi^2_Z$; however, if only one of the two data sets is used, we
should have $\chi^2 = \chi^2_{\rm p}$ or $\chi^2_{Z}$.

\item For a given set of mock observations ${\bf D}$ of the apparent
positions, the redshifts, or both, of a star, the posterior
probability distributions of those parameters ${\bf \Theta}$ can be
constrained according to the Bayesian theorem, i.e., $P({\bf \Theta}
|{\bf D}) \propto P({\bf D} | {\bf \Theta}) P({\bf \Theta}) \propto
\exp(-\chi^2/2) P({\bf \Theta})$, where $P({\bf \Theta})$ is the prior
distribution of ${\bf \Theta}$ and assumed to be flat over an initial
guess range for each parameter. The MCMC method is adopted, with an
implement of the Metropolis-Hasting algorithm, in order to efficiently
get the best fit to the data ${\bf D}$ and obtain constraints on the
model parameters ${\bf \Theta}$ by searching the parameter space.

\end{enumerate}

\subsection{Fitting Results}
\label{subsec:fit_results}

\begin{table*}
\caption{Best-fit parameters for the MBH mass and spin and the
distance to the GC.}
\centering
\begin{tabular}{lccccc}\hline
\multirow{2}{2.0cm}{Example Star}    & \multirow{2}{0.5cm}{$a$} & $i$
& $\epsilon$ & $\delta R_{\rm GC}$\,$^a$ & $\delta \bh$\,$^b$ \\
& & $(\arcdeg)$ & $(\arcdeg)$ & (pc) & ($10^3\msun$)  \\ \hline 
S0-2/S2 & $0.812^{+0.187}_{-0.257}$ & $37^{+39}_{-27}$ &
$156^{+74}_{-89}$ & $-0.30^{+0.98}_{-0.97}$ & $-0.41^{+1.17}_{-1.15}$
\\[5pt]
S0-102  & $0.457^{+0.542}_{-0.457}$ & $92^{+78}_{-82}$ &
$191^{+169}_{-191}$ & $-0.31^{+1.69}_{-1.71}$ & $0.52^{+2.43}_{-2.47}$
\\[5pt]
Ea      & $0.922^{+0.077}_{-0.127}$ & $51^{+13}_{-13}$ &
$178^{+14}_{-15}$ & $-0.34^{+1.39}_{-1.52}$ & $-0.46^{+1.93}_{-2.04}$
\\[5pt]
Eb      & $0.989^{+0.010}_{-0.011}$ & $45^{+1}_{-1}$ & $181^{+1}_{-1}$
&  $-1.16^{+2.50}_{-2.34}$ & $-1.53^{+3.55}_{-3.29}$ \\[5pt]  
Ec      & $0.987^{+0.012}_{-0.015}$ & $46^{+2}_{-1}$ & $178^{+5}_{-6}$
& $-1.07^{+2.64}_{-2.61}$ & $-0.80^{+3.41}_{-3.21}$ \\[5pt]  
Ed      & $0.840^{+0.159}_{-0.233}$ & $61^{+25}_{-23}$ &
$180^{+35}_{-38}$ & $-1.54^{+2.69}_{-2.71}$ & $1.12^{+2.67}_{-2.79}$
\\[5pt]
\hline
\end{tabular}
\tablecomments{Notes. The error associated with each parameter represents
the $2$-$\sigma$ error of the best-fit value of the parameter. \,$^a$
The best fit of the distance to the GC is $ R_{\rm GC} = 8\kpc +
\delta R_{\rm GC}$.
\,$^b$ The best fit of the MBH mass is $\bh = 4\times 10^6\msun +
\delta \bh$.  
}
\label{tab:t2}
\end{table*}

\begin{figure}
\centering 
\includegraphics[scale=0.5]{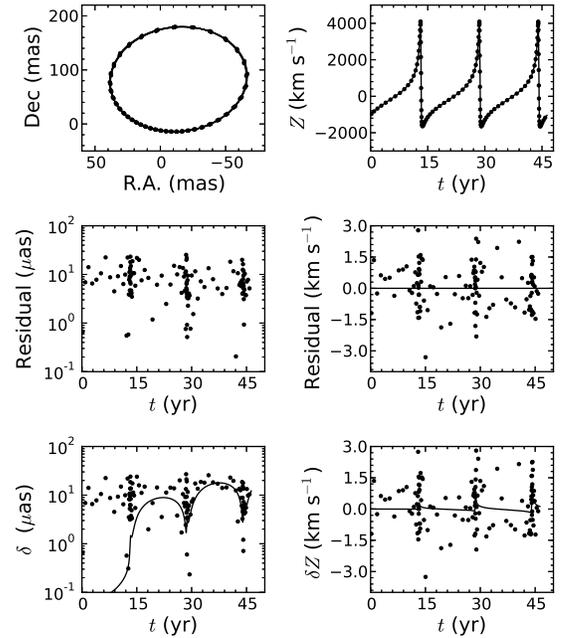}
\caption{Mock trajectory of the apparent positions and the redshift
curve of S0-2/S2 over three full orbital periods and their best fits.
Top two panels are for the mock apparent positions (solid circles in
the left panel) and the redshift curve (solid circles in the right
panel) and their best fits (solid lines); middle panels are for the
residuals of the best fits; and bottom panels represent the distances
(solid circles in the left panel) or the redshift difference (solid
circles in the right panel) between the mock observational positions
(or redshifts) of S0-2/S2 and the positions (or redshifts) of an
assumed star with the same initial orbital elements as the best fit of
S0-2/S2 but around a non-spinning MBH, and the solid lines therein
represent the differences between the predicted orbit of the best fit
to S0-2/S2 and that of the assumed star. 
}
\label{fig:f13}
\end{figure}

Figure~\ref{fig:f13} illustrates the best fits to the mock
observations of S0-2/S2, including both the apparent positions (solid
line in the top-left panel) and redshifts (solid line in the top-right panel)
The residuals of the fits are shown in the two middle panels. The
solid circles in the two bottom  panels show the difference between
the mock observations and those predicted from a star with the same
initial orbital elements around a non-spinning MBH, and the solid
lines represent the differences between the best fits and those
predicted from the star around a non-spinning MBH, similar to $\delta$
and $\delta Z$ shown in Figures~\ref{fig:f5} and \ref{fig:f10},
respectively. Obviously, the mock observations (points) are well fit by
the model. 

\begin{figure}
\centering
\includegraphics[scale=0.35]{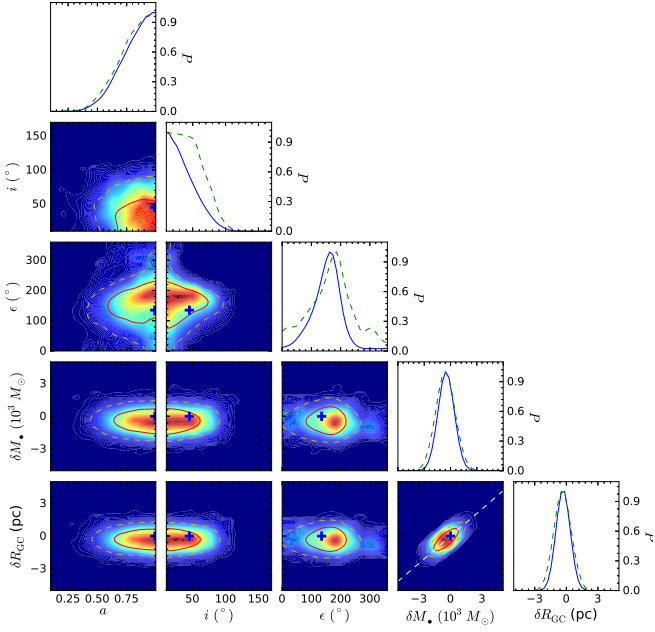}
\caption{Two-dimensional probability contours and one-dimensional
probability distributions of the best-fit parameters for S0-2/S2. For
the two-dimensional probability panels, the color contour maps represent
the mean likelihood of the MCMC sample, and the line contours represent
the marginalized distribution, with the red solid line and the yellow
dashed line as the $1$-$\sigma$ and $2$-$\sigma$ confidence level,
respectively.  The `+' symbol in each color map represents the
original set of the parameter value to produce the mock observations.
The white dashed line in the panel for $\delta R_{\rm GC}$ versus $\delta
M_{\bullet}$ represents $M_{\bullet} \propto R_{\rm GC}^{\Gamma}$ (and
$\delta M_{\bullet} / M_{\bullet} = \Gamma \delta R_{\rm GC}/ R_{\rm
GC}$) with $\Gamma \sim  2.4$. For those panels showing the
one-dimensional probability distributions, the blue solid line and the
green dashed line represent the one-dimension marginalized
distribution and the one-dimensional mean likelihood, respectively.
}
\label{fig:f14}
\end{figure}

Figure~\ref{fig:f14} shows the two-dimensional contours and
one-dimensional probability distributions of the parameters $\bh$,
$R_{\rm GC}$, $a$, $i$, and $\epsilon$, respectively. The parameters
of the best fit are listed in Table~\ref{tab:t2}. As seen from
Figure~\ref{fig:f14} and Table~\ref{tab:t2}, the absolute value of the
MBH spin can be reasonably constrained by using the orbital motion of
S0-2/S2 over two or three full orbits ($30-45$ years), if (1)
the astrometric and the redshift accuracies can reach $\sigma_{\rm p}
= 10\mu{\rm as}$ and $\sigma_Z = 1\kms$; (2) the MBH spin $a$ is close
to one; and (3) the spin is pointing toward some directions with
moderate or even maximum spin-induced effects on the trajectory of
apparent position and the redshift curve, e.g.,  $(i,\, \epsilon) =
(45\arcdeg,\, 200\arcdeg)$, $(49\arcdeg,\, 126\arcdeg)$ or
$(131\arcdeg,\, 306\arcdeg)$.
According to Figures~\ref{fig:f5} and \ref{fig:f10}, $\delta$ mounts
up with increasing number of orbits, therefore, the longer the time of
monitoring the motion of a star, the tighter the constraints on the
MBH spin obtained for a given set of $(\sigma_{\rm p},\, \sigma_Z)$.
If the astrometric and redshift accuracies can be a factor of a few
times, or more, higher, i.e., $\sigma_{\rm p} \la 3\mu{\rm as}$ and/or
$\sigma_Z \la 0.3\kms$, the MBH spin can be constrained by using the
motion of S0-2/S2 over only one full orbit (a period of $\sim
15$ year). However, if the astrometric and the redshift accuracies are
a factor of two times or more lower, i.e., $\sigma_{\rm p} > 20
\mu{\rm as}$ and/or $ \sigma_Z
> 2\kms$, then the MBH spin cannot be constrained by using the
orbital motion of S0-2/S2 within a few orbits. 

Note here that both the distance to the GC and the MBH mass can be
constrained simultaneously to an unprecedented high accuracy, i.e.,
$\delta \bh/\bh \sim $ a few times of $ 10^{-4}$ and $\delta R_{\rm
GC} /R_{\rm GC} \sim $ a few times of $ 10^{-4}$. Even if the
astrometric and redshift measurements are significantly less accurate,
e.g., $\sigma_{\rm p} = 50\mu{\rm as}$ and/or $\sigma_Z = 10\kms$, the
MBH mass and the distance to the GC can be still constrained to an
accuracy of $0.98\%$ and $0.36\%$, respectively, although the MBH spin
cannot be constrained, by monitoring the motion of S0-2/S2 over less
than three full orbits.
These results are consistent with those obtained by
\citet{Weinberg05}.

If the absolute value of the MBH spin is smaller than $0.99$ by a
factor of $C$, then it requires the astrometric and/or redshift
accuracies of $\sigma_{\rm p}\la 10 C^{-1}\mu{\rm as}$ and/or
$\sigma_Z \la C^{-1} \kms$ or a longer period for monitoring the star
($\ga (2-3)C$ full orbits), in order to distinguish the MBH from a
Schwarzschild MBH.

According to Figures~\ref{fig:f7} and \ref{fig:f11}, the amplitudes of
$\delta$ and $\delta Z$ (due to the spin effects) also depend on the
spin direction. However, the pattern of the dependence of $\delta$ on
$(i,\,\epsilon)$ is somewhat different from that of $\delta Z$. For
example, the amplitude of $\delta$ is the largest (or moderate) when
$(i,\, \epsilon) = (72\arcdeg,\, 90\arcdeg)$ [or $(45\arcdeg,\,
180\arcdeg)$]; while the amplitude of $\delta Z$ is moderate but not
the largest (or moderate but close to the largest) when $(i,\,
\epsilon) = (72\arcdeg,\, 90\arcdeg)$ [or $(159\arcdeg,\,
63\arcdeg)$].
If both measurements on redshifts and apparent positions with
compatible accuracies are available, it would improve the constraint
on the MBH spin and its direction.  If only measurements on apparent
positions with accuracy of $\sigma_{\rm p} =10\mu$as are available,
then the spin is easier (or harder) to constrain by using
the motion of S0-2/S2 within several orbits if $(i,\,\epsilon) =
(72\arcdeg,\, 90\arcdeg)$ [or $(159\arcdeg,\, 63\arcdeg)$] (see more
discussions in \citealt{Yuetal14}).

\begin{figure}
\centering
\includegraphics[scale=0.5]{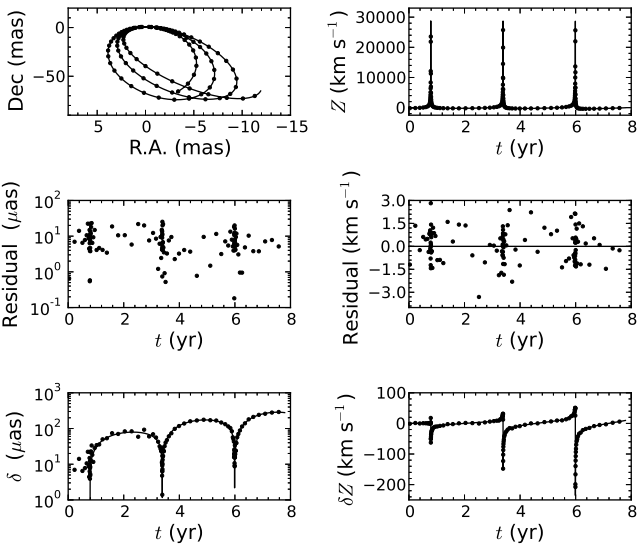}
\caption{Legend similar to that for Fig.~\ref{fig:f13} but for the
example star Eb. 
}
\label{fig:f15}
\end{figure}

\begin{figure}
\centering
\includegraphics[scale=0.35]{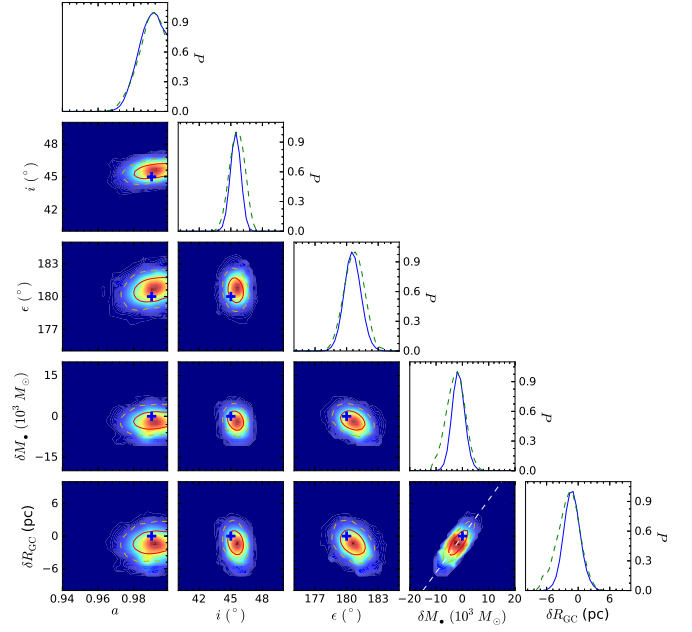}
\caption{Legends similar to those for Fig.~\ref{fig:f14} but for the
example star Eb.  The dashed line in the panel for $\delta R_{\rm GC}$
versus $\delta M_{\bullet}$ represents $M_{\bullet} \propto R_{\rm
GC}^{\Gamma}$  with $\Gamma \sim 3.0$.
}
\label{fig:f16}
\end{figure}

Figure~\ref{fig:f15} illustrates the best fits to the mock
observations of Eb, similar to that shown in Figure~\ref{fig:f13} for
S0-2/S2. The two-dimensional contours and one-dimensional probability
distribution of the parameters are shown in Figure~\ref{fig:f16}, and
the best-fit parameters are listed in Table~\ref{tab:t2}.  Since
Eb has a smaller semimajor axis ($300$\,AU) and a larger eccentricity
($0.98$) compared with S0-2/S2, the spin-induced effects on $\delta$
and $\delta Z$ for Eb are larger than those for S0-2/S2 by a factor of
several tens to a hundred, respectively (see Figs.~\ref{fig:f5} and
\ref{fig:f10}).  Therefore, the MBH spin and its direction can be more
accurately constrained by using the motion of Eb over three full
orbits ($\sim 10$ years) if $(\sigma_{\rm p},\, \sigma_Z) = (10\mu{\rm
as},\, 1\kms)$, and the accuracy can be improved by more than one
order of magnitude (see Tab.~\ref{tab:t2}).  For much lower 
astrometric and redshift precisions, e.g., $\sigma_{\rm p}
\sim 50\mu{\rm as}$ and $\sigma_Z \sim 10\kms$, which can be easily
achieved by the next generation extremely large telescopes, it is still
possible to constrain the MBH spin with considerable accuracy by using
the motion of a star like Eb over a period less than a few full orbits
($\la 10$ years).

To produce the mock observations, we have assumed that the MBH spin
$a=0.99$. If $a$ is set to be smaller than $0.99$  by a factor of $D$,
then the resulting amplitudes of $\delta$ and $\delta Z$ would be
smaller than those shown in Figures~\ref{fig:f5} and \ref{fig:f10} by
the same factor (see eqs.~\ref{eq:domega}-\ref{eq:Delta_A} and
\ref{eq:vz}). If $D$ is less than $10$ and $a \ga 0.1$, we may still
be able to constrain $a$ or distinguish it from a Schwarzschild BH  by
using the motion of a star like Eb within a few orbits provided
$(\sigma_{\rm p},\, \sigma_Z) = (10\mu{\rm as},\, 1\kms)$. If $a \la
0.5$ and $D \ga 2$, however, it is difficult to get an accurate
constraint on the spin value or distinguish it from a Schwarzschild BH
by using the orbital motion of S0-2/S2 over a few orbits,  even if the
spin vector is close to some specific direction [e.g., $(45\arcdeg,\,
200\arcdeg)$ or $(45\arcdeg,\,135\arcdeg)$] with which the
spin-induced effects on $\delta$ and/or $\delta Z$ are most
significant. In this case, a constraint on whether the GC MBH is an
extremely rotating MBH or not is possible. If the astrometric and
redshift measurements are much more accurate (e.g., $\sigma_{\rm p}
\la 10D^{-1}\mu{\rm as}$ and/or $\sigma_Z \la D^{-1}\kms$) as
discussed above, an accurate constraint on the MBH spin may still
be obtained and it can be distinguished from a Schwarzschild BH. 

\begin{figure*}
\centering
\includegraphics[scale=0.5]{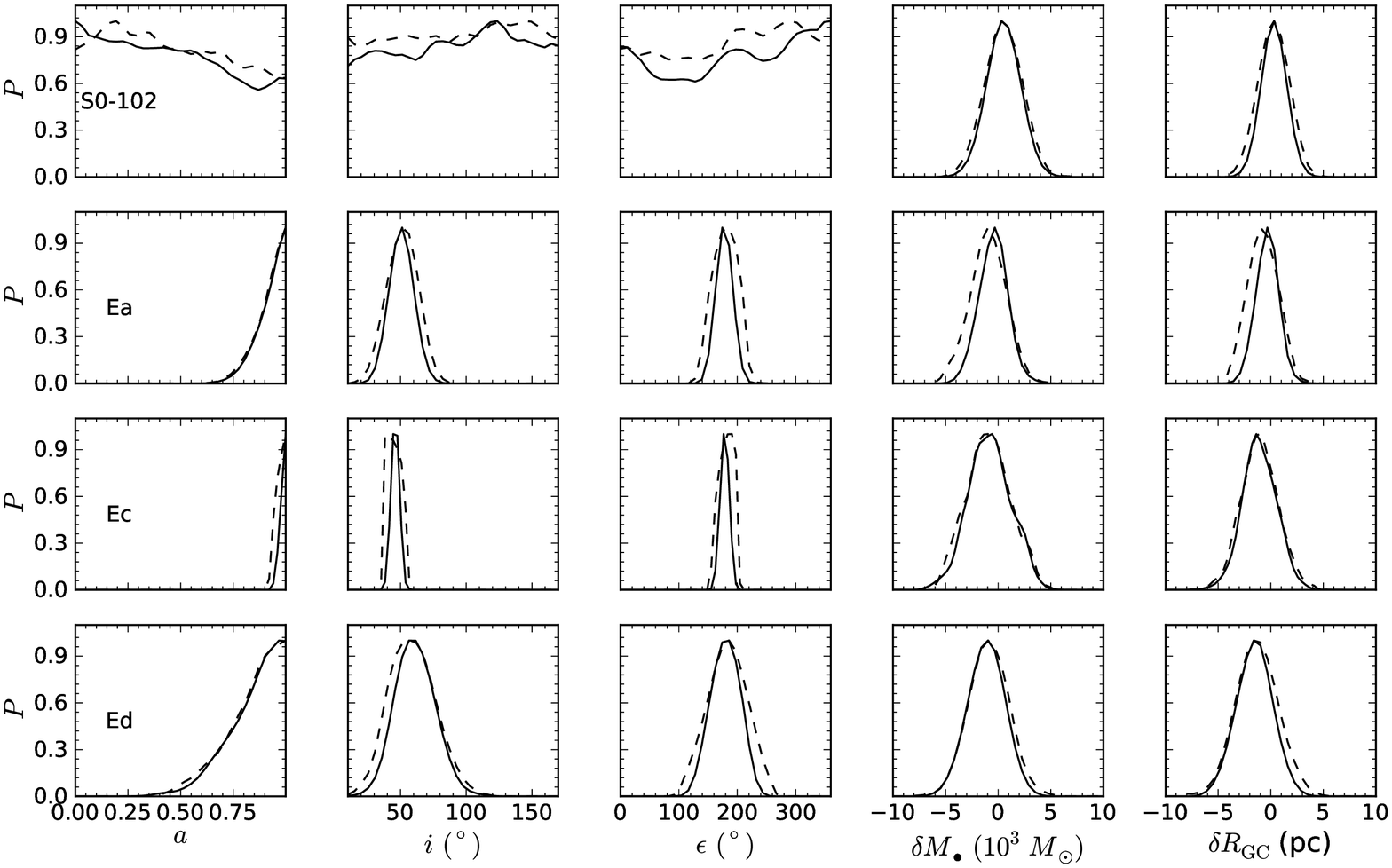}
\caption{One-dimensional probability distributions of the best-fit
parameters for other stars obtained in the same way as done in
Figs.~\ref{fig:f14} and \ref{fig:f16}. Top to bottom rows are for
S0-102, Ea, Ec, and Ed, respectively. The solid line and the dashed
line represent the one-dimension marginalized distribution and the
one-dimensional mean likelihood, respectively.
}
\label{fig:f17}
\end{figure*}

Figure~\ref{fig:f17} presents the one-dimensional probability
distributions of the best-fit parameters for other stars, i.e.,
S0-102, Ea, Ec, and Ed, respectively. The best-fit parameters are
listed in Table~\ref{tab:t2}. As seen from Figure~\ref{fig:f17} and
Table~\ref{tab:t2}, the MBH spin cannot be constrained by using the
relativistic orbital motion of S0-102 within a period of less than two to
three orbits ($\la 30-45$ years) for $\sigma_{\rm p} \ga 10\mu{\rm as}$
and $ \sigma_Z \ga  1\kms$, even if the spin direction is close to the
one with the largest effects on $\delta$ and $\delta Z$; it can be
well constrained by using other stars, i.e., Ea, Ec, or Ed. The reason
is that S0-102 has the largest pericenter distance and thus the least
changes of $\delta$ and $\delta Z$ per orbit caused by the spin
effects. The MBH spin is less well constrained by using Ed, compared
with those using Ea and Ec, because the pericenter distance of Ed is
larger, compared to Ea, Eb, and Ec, though its semimajor axis is
smaller than that of Ea and Eb.

\section{Discussions}
\label{sec:discussions}

\subsection{Perturbations from Other Stars or Stellar Remnants}

In our full general relativistic calculations, the possible
perturbation on the orbit and motion of a star due to other background
stars existing within the star orbit is ignored. By assuming several
different distributions of stars close to the GC MBH,
\citet{Merritt10} performed post-Newtonian simulations and showed that the
stellar perturbation may obscure the signals due to frame dragging and
quadruple moment for stars beyond $\sim 0.5-0.2$\,mpc ($\sim
100-400$\,AU); and \citet{Sadeghian11} also analytically investigated the
perturbing effects of background stars and found that it may
still be possible to detect relativistic precession of stars within a
few tenth milliparsecs.  Therefore, it may not be easy to use S0-2/S2 (or even
a star like Ea or Eb) to constrain the GC MBH spin even if the
accuracies of position and redshift measurements are sufficiently high
(e.g., $\sigma_{\rm p} \la 10\mu{\rm as}$, $\sigma_Z \la 1\kms$).
However, on the one hand, the exact number of stars or stellar
remnants existing in the vicinity of the central MBH is not clear. It
is not impossible that the stars or stellar remnants existing within
the S0-2/S2 (or Ea/Eb) orbit are very rare and their perturbations on
the orbit and motion of S0-2/S2 (or Ea/Eb) are small and do not affect
the spin signals much.  Furthermore, the orbit-averaged torques from
background stars are approximately constant in magnitude over year- to
decade-long timescales, it is possible to disentangle the stellar
perturbation effects from those due to the MBH spin as discussed in
\citet{Merritt10}.  On the other hand, it is an important step to
first check whether the spin-induced effects on the motion of a star,
as close to the GC MBH as those example stars, can be detected by
ignoring the perturbations from other possibly existing stars within
its orbit, with future facilities with sufficient astrometric and
redshift accuracies.

%
Even if the perturbations from background stars on a target star are not
negligible, the observational data collected around the pericenter,
where the perturbations are substantially less significant than those
near the apocenter, may still be useful to constrain the MBH
spin parameters \citep[see][]{Agnelil14}. By using the redshift data near
the pericenter, strong constraints on spin parameters may be obtained
since the spin-induced redshift differences are the most significant near
the pericenter. However, the constraints on spin parameters may be
much weaker if only the position data are used because the spin-induced
position displacements near the pericenter are substantially smaller
than those near the apocenter. As a check, we perform
additional MCMC fittings for S2/S0-2 and Eb by only using the mock
observations near the pericenter (i.e., $r<a_{\rm orb}$ or $r\ll
a_{\rm orb}$). Our results show that the MBH spin parameters are
poorly constrained by using the mock observations near the pericenter of
S2/S0-2 ($0 \la a \la 1$, $i= 61\arcdeg^{+66\arcdeg}_{-61\arcdeg}$, and
$\epsilon = 160\arcdeg^{+136\arcdeg}_{-113\arcdeg}$) because the
spin-induced maximum redshift difference can only be marginally
detected (see Figure~\ref{fig:f10}). For Eb, however, the spin
parameters can be well constrained ($a=0.99^{+0.009}_{-0.012}$,
$i=45\arcdeg^{+1\arcdeg}_{-1\arcdeg}$, and
$\epsilon=182\arcdeg^{+3\arcdeg}_{-3\arcdeg}$) because the
spin-induced redshift differences near the pericenter are
substantially higher than the detection limit and can be accurately
measured.
%

It is also possible that an intermediate mass black hole (IMBH) exists
in the vicinity of the GC MBH \citep[e.g.,][]{YT03, HM2003, Genzel10},
and this IMBH may perturb the orbit and motion of a star near the MBH.
\citet{Gualandris10} have suggested that future observations on the
S0-2/S2 orbit may be used to probe or constrain the existence of an
IMBH in the GC.

We also note here that other effects, such as those from gas dynamics,
wind mass loss, and tidal dissipation, may affect the orbit and motion
of a star close to the MBH. \citet{Psaltis12} and \citet{Psaltis13}
have shown that these effects may be negligible and do not preclude
the measurements of the MBH spin and quadruple moment.

\subsection{Accurately Constraining the MBH Mass and the Distance from
the Sun to the GC}

With the expected astrometric and redshift accuracies of the next
generation facilities [$(\sigma_{\rm p},\,\sigma_{Z})
= (10\mu{\rm as},\,1\kms)$ or $(50\mu{\rm as},\, 5\kms)$], the MBH
mass and the distance from the Sun to the GC can be constrained with
an unprecedented accuracy on the order of a few $0.01\%$ to $0.1\%$ or
a few $0.1\%$ to $1\%$ if using S0-2/S2 \citep[see early works
by][]{SalimGould99, Weinberg05}. Even with such high accuracies of
$\sigma_{\rm p}$ and $\sigma_{Z}$, the degeneracy between the MBH mass
and the distance from the Sun to the GC remains, i.e., $M_{\bullet}
\propto R_{\rm GC}^{\Gamma}$  ($\delta M_{\bullet} / M_{\bullet} =
\Gamma \delta R_{\rm GC} / R_{\rm GC}$) and $\Gamma \sim 2.4-3.0$
(e.g., see Figures~\ref{fig:f14} and \ref{fig:f16}), similar to that
found in \citet[][see Fig.~11 therein]{Ghezetal08} and \citet[][see
Fig. 15 therein]{Gillessenetal09}. 
Although the MBH spin can be better constrained by using the orbital
motion of a star with a small semimajor axis and a high eccentricity,
such as Eb, the constraints on $M_{\bullet}$ and $R_{\rm GC}$ are not
significantly improved. The tightness of the constraints on
$M_{\bullet}$ and $R_{\rm GC}$ depends on the ratio of the astrometric
accuracy to the semimajor axis and the ratio of the redshift accuracy
to the orbital velocity, but does not depend directly on the semimajor
axis and eccentricity of the test star.

\subsection{Usefulness of the Observations of Stars with Very High
Eccentricities during Their Pericentric Passages}

In the GC, some stars may move around the MBH on nearly parabolic
orbits ($e_{\rm orb} > 0.95$) as suggested by the existence of the
star S0-16/S14 (with $e_{\rm orb} = 0.963$ and pericenter distance of
$\sim 70$\,AU; \citealt{Gillessenetal09}). With such high
eccentricities, they may approach the MBH at a distance even
substantially smaller than those stars with smaller semimajor axes,
such as S0-2/S2, and thus more significantly affected by the spin
effects. As demonstrated in Section~\ref{sec:orbred}, the spin-induced
position displacements and redshift differences are most significant
near the apoapsis and periapsis, respectively. Since the orbital
periods of those stars on nearly parabolic orbits may be substantially
longer than those with smaller semimajor axes (e.g., S0-2/S2 and
S0-102), it is difficult to monitor them for a whole orbit. However,
it should be useful to monitor the motion of those stars during their
pericentric passage, when the spin-induced redshift differences (but
not the position displacements) are most significant. We further check
the validity of using the observations near the pericentric passage of
a star on nearly parabolic orbit and find that the MBH may be
accurately constrained if the pericenter distance of the star is
smaller than a few tens AU. If there are more than one such stars,
which would pass their periapsides, in the near future, monitoring
their motions would be helpful in constraining the MBH spin.

\subsection{Direction of the MBH Spin}

The spin-induced GR effects, e.g., the Lense-Thirring precession and
the frame dragging, on the orbit of a star depend on the direction of the
spin. For a star with a given orbital orientation, the spin-induced
effects on the changes of its apparent positions (or redshift) may
vary by a factor of several to more than an order of magnitude if
choosing significantly different spin directions, and the spin-induced
effects are negligible for some spin directions, even if the absolute
value of the spin is large and the semimajor axis of the star is
small. However, the patterns of the changes of the position
displacements and those of the redshift differences are different (see
Figures~\ref{fig:f7} and \ref{fig:f12}). Therefore, simultaneously
using both the trajectory of the apparent positions and the redshift
curve would help to obtain a better constraint on the spin, provided
that the maximum values of the spin-induced position displacements and
redshift differences over the observing period are substantially
larger than the astrometric and redshift accuracy, respectively
\citep[see discussions in][]{Yuetal14}. In this case, if only choosing
one of the two data sets, the constraint on the spin would become
losser because of the above orientation dependence. If the
relativistic orbital motions of more than one star with significant
different orbital orientations are available, the constraints on the
MBH spin would also be tightened.

\subsection{Test for Alternative Metrics}

In this study, the metric of the massive object (presumably an MBH) in
the GC is assumed to be described by a Kerr metric. In principle, this
object could be described by a metric,  different from the Kerr metric
\citep[see][]{Will93}, e.g., the Johannsen-Psaltis metric
\citep{Johannsen11}. Similarly, it is also possible to use the
relativistic orbits of stars in the immediate vicinity of the central
object to check  whether the central object is described by such
alternative metrics or not. To do this, a modification to the
numerical calculations of the orbits and the ray tracing in such a
metric is required, which is deferred to a future study.

\section{Conclusions} 
\label{sec:conclusions}

The innermost S-star discovered in the vicinity of the GC
MBH is believed to be able to provide strong constraints on the MBH
spin and metric, and thus provide a strong test to the no-hair
theorem of BHs and GR.  In this paper, we develop a
full general relativistic method to simultaneously constrain the MBH
mass, spin, and its direction, and the distance from the Sun to the GC,
by using the relativistic orbital motion of a star close to the MBH.

The spin information is encoded in the trajectory of the apparent
position of the star in the observer's sky plane and the redshift
curve measured by the observer, which include: (1)
the changes in the orbital motion induced by the MBH spin and (2) the
changes of the photon trajectory propagating from the star to a
distant observer due to the spin. We numerically solve the motion
equations in the Kerr metric for a star rotating around an MBH; and we
follow the propagation of photons from the star to the distant
observer by using a full general relativistic ray-tracing technique.
Armed with these numerical methods, we can obtain the trajectory of
the apparent position and the redshift curve of any star, with any
given initial conditions, rotating around an MBH, with any given mass
and spin (and the spin direction). We adopt six example stars with
different semi-major axes and eccentricities. Two of these stars have
orbital configurations similar to that of S0-2/S2 and S0-102,
respectively; and the other four are adopted according to the
probability distribution of the orbit of a star that is expected to be
the closest one to the GC MBH. We obtain the trajectories of the
apparent positions and redshift curves for those stars over a given
period, and further obtain the mock observations of both the
trajectories of their apparent position and their redshift curves by
assuming astrometric ($\sigma_{\rm p}$) and redshift accuracies
($\sigma_Z$) of future facilities.  According to our calculations, we
emphasize some points below.

First, the spin-induced effects on the orbital motion depend on both
the absolute spin value ($a$) and spin direction. The amplitude of the
spin-induced position displacements and redshift differences can
differ significantly (by an order of magnitude) for MBHs with
different spin directions but the same absolute spin value.

Second, for S0-2/S2, the spin-induced effects
would lead to a maximum change of $13\mu$as in the apparent position
of the star within a full orbit for a maximum spin with direction of
$(49\arcdeg, 125\arcdeg)$ [or $(131\arcdeg, 305\arcdeg)$].  These
values are consistent with previous estimates \citep[e.g.,][]{Will08,
Angelil10a}. The maximum change depends on the spin direction.  The
changes in the apparent positions are mainly due to the spin-induced
changes of the orbital motion, while the part due to the spin-induced
changes of the photon propagation path is on the order of $0.01\mu$as,
only $\sim 0.1\%$ of the total changes.

Third, for S0-2/S2, the spin-induced effects would also lead to a
maximum change of the redshifts ($\sim 0.3\kms$) over a full orbit for
a maximum spin with direction of $(129\arcdeg, 171\arcdeg)$ ]or
$(51\arcdeg, 351\arcdeg)$].  The dependence of the maximum changes in
the redshift curves on the spin direction is different from that of
the maximum changes in the apparent positions.  The spin-induced
maximum change of the redshift over a full orbit is also dominated by
the orbital change, and the contribution from the photon propagation
is on the order of $0.003\kms$, about the percentage level of the total
changes.

Fourth, for a star much closer to the MBH than S0-2/S2, such as Eb
with a semimajor axis of $\sim 300$~AU and an eccentricity of $\ga
0.98$, the spin-induced effects would lead to a maximum change of
$308\mu$as in its apparent position within a full orbit for a maximum
spin with a direction of $\sim (111\arcdeg, 275\arcdeg)$ [or $\sim
(69\arcdeg, 95\arcdeg)$]; and it would also lead to a maximum change
of $\sim 93\kms$ in the redshifts over a full orbit for a maximum spin
with direction of $\sim (150\arcdeg, 319\arcdeg)$ [or $\sim
(30\arcdeg, 139\arcdeg)$]. The spin signals of a star like Eb can be
more than an order of magnitude higher than those of S0-2/S2, and thus
monitoring the motion of such a star could provide much tighter
constraints on the MBH spin and metric. The maximum value of the
spin-induced redshift differences over a full orbit is still dominated
by the orbital change, and the contribution from the photon
propagation is on the level of 10\% of the total changes.  

Finally, in order to accurately constrain the spin and metric of the
GC MBH (on the percentage level), it is necessary to consider both the
spin effects on the stellar orbit and the photon propagation from the
star to the distant observer. Although the changes in the apparent
position of a star due to the spin effects on photon propagations are
only about $0.001\%-0.1\%$ of those caused by the change of the orbital
motion, the changes in the redshift of a star due to the spin effect
can be up to $\sim 1\%-10\%$ of those caused by the change of the
orbital motion if the semimajor axis is in the range from $1000$
to $100$\,AU.  Therefore, it is important to include the full GR
calculations of the photon propagations in order to accurately
constrain the MBH spin.

Adopting the MCMC method, we illustrate that the spin of the GC MBH
may be able to be constrained by using the orbital motion of S0-2/S2
over a period of $\sim 45$ years if (1) the spin value $a$ is close to
$1$, (2) $\sigma_{\rm p} \sim 10\mu$as and $\sigma_Z \sim 1\kms$,
which will be achieved by the new instrument GRAVITY on VLTI, and (3)
the spin is pointing toward some specific direction, with which the
spin-induced effects are likely moderate to most significant. 
If the astrometric and spectroscopic precisions can be further
improved by a factor of several, then the GC MBH spin may be well
constrained by using the relativistic motion of S0-2/S2 monitored over
a period of no more than $15$ years. In the mean time, the distance from
the Sun to the GC and the MBH mass can also be constrained to an
unprecedented accuracy ($0.01-0.1\%$).  If there exists a star with a
semimajor axis of less than several hundred AU and an eccentricity of
$\ga 0.9$ as expected \citep[see][]{ZLY13}, the MBH spin can be
constrained with high accuracy over a period of $\sim 15$ years for
$\sigma_{\rm p}\sim 10\mu$as and $\sigma_Z \sim 1\kms$, even if $a$ is
only moderately large.  The next generation facilities, such as the
GRAVITY on VLTI, TMT, and E-ELT, may reach an astrometric accuracy of
$10-50\mu$as.\footnote{http://www.tmt.org;
http://www.eso.org/sci/facilities/eelt/} Those extremely large
telescopes may reveal some stars, if existing, with semimajor axes
less than a few hundred AU and eccentricities $\ga 0.9$, within the
S0-2/S2 (or S0-102) orbit. The spin (and its direction) of the GC MBH
is likely to be tightly constrained by using the relativistic motions
of such stars.  Long time monitoring of the orbital motions of stars
in the vicinity of the GC MBH by the next generation facilities is
likely to provide an ultimate dynamical test for the first time to the
no-hair theorem and GR.

\acknowledgements

\noindent We thank Scott Tremaine for bringing our attention to the
work by A. Ang{\'e}lil and P. Saha.  This work was supported in part
by the National Natural Science Foundation of China under grant Nos.
11273004, 11373031, and 11390372, the Strategic Priority Research
Program ``The Emergence of Cosmological Structures'' of the Chinese
Academy of Sciences, grant No.  XDB09000000, the Bairen program from
the National Astronomical Observatories of China, Chinese Academy of
Sciences, and a grant from the John Templeton Foundation and National
Astronomical Observatories of Chinese Academy of Sciences. F.Z. is
supported by a postdoctoral fund 2014M550549.


\appendix

\section{Photon Orbits in a Kerr Spacetime}
\label{sec:Appendix1}

The motion of a photon in a Kerr spacetime is given by
Equations~(\ref{eq:motionr})-(\ref{eq:motiont}). We assume that an
observer, located at $(r_0, \theta_0, \phi_0)= (10^8\rg, i, 0)$ in the
BH rest frame, receives photons emitted from a star located at
$\vec{\bf R}_\star =(r_\star, \theta_\star, \phi_\star)$, which is
rotating around the BH.  Define $\mu=\cos\theta$, the motions
of those photons in the $r-\theta$ plane are given by
\citep{Bardeen72,Cunningham73,Chandra83}
\be
\int^{r_{\rm hit}}_{r_0}\frac{dr}{\sqrt{R}} =
\pm\int^{\mu_{\star}}_{\mu_0}\frac{d\mu} {\sqrt{\Theta_{\mu}}},
\label{eq:rtheta}
\ee
where $\mu_0=\cos i$, $\mu_{\star}=\cos \theta_\star$, and $r_{\rm
hit}\rightarrow r_\star$ is the redshift of a photon when
$\mu=\mu_{\star}$, i.e., the photon hits the surface of the star at
$(r_{\rm hit}, \theta_\star)$. In Equation~(\ref{eq:rtheta}), the left
side is integrated from a distant observer to the surface of the star
and it should be negative as $r_0 > r_{\rm hit}$. We should carefully
choose the sign in front of the integration at the right side, i.e.,
if $\mu_0<\mu_\star$, the sign should be ``-"; and if
$\mu_0>\mu_\star$, the sign should be ``+". Below, we denote the left
and right side of Equation~(\ref{eq:rtheta}) as $\mathscr{I}_{\mu}$
and $\mathscr{J}_{r}$, respectively.

\subsection{Integral over the $\theta$-direction}\label{sec:thetaint}

We denote the integral over $\tilde{\mu}$ on the left side of
Equation~(\ref{eq:rtheta}) as
\be
\mathscr{F}_{\mu}(\mu) =\int
\frac{d\tilde{\mu}}{\sqrt{\Theta(\tilde{\mu})}},
\label{eq:thetaint0}
\ee
where

\be
\Theta(\tilde{\mu}) = q^2+(a^2-q^2-\lambda^2) \tilde{\mu}^2-
a^2\tilde{\mu}^4 = a^2(\mu_-^2+\tilde{\mu}^2)(\mu_+^2-\tilde{\mu}^2),
\ee
\be
\mu_{\pm}^2=\frac{1}{2a^2} \left[\sqrt{(\lambda^2+q^2-a^2)^2+4a^2q^2}
\mp(\lambda^2+q^2-a^2)\right],
\ee
and $-\mu_+$ and $\mu_+$ are the two real roots of
$\Theta(\tilde{\mu})=0$. Because $\Theta(\tilde{\mu})$ must be $\geq 0$,
the physical allowed region for a photon is $-\mu_+\leq \tilde{\mu}
\leq \mu_+$. Once $\tilde{\mu}$ reaches the turning point (either
$\mu_+$ or $-\mu_+$) and bouncing back, the $\tilde{\mu}$-integral at
the right side of Equation~(\ref{eq:thetaint}) changes the sign. If
the integration range of the $\theta$-integral is from $\mu (\geq 0)$
to $\mu_+$, we may denote the function $\mathscr{F}_{\mu}(\mu)$
explicitly as
\be
\mathscr{F}_{\mu}(\mu) =\int^{\mu_+}_{\mu}
\frac{d\tilde{\mu}}{\sqrt{\Theta(\tilde{\mu})}}.
\label{eq:thetaint}
\ee
If $-\mu_+\leq \mu<0$, we have
\be
\int^0_{\mu} \frac{d\tilde{\mu}}{\sqrt{\Theta(\tilde{\mu})}}
=\int^{\mu_+}_0 \frac{d\tilde{\mu}}{\sqrt{\Theta(\tilde{\mu})}}
-\int^{\mu_+}_{-\mu} \frac{d\tilde{\mu}}{\sqrt{\Theta(\tilde{\mu})}}
=\mathscr{F}_{\mu}(0)-\mathscr{F}_{\mu}(-\mu).
\ee

According to \citet[][p48, eq. 213.00]{BF54}, $\mathscr{F}_{\mu}(\mu)$
can be expressed in terms of an inverse Jocabian elliptic function 
\be
\mathscr{F}_\mu(\mu)= g_{\mu} \text{cn}^{-1}(\cos\varphi|m_{\mu})=
g_{\mu}F\left(\varphi |m_{\mu}\right),
\ee
where
\be
g_{\mu}= \frac{1}{|a|\sqrt{\mu_+^2+\mu_-^2}},
\label{eq:gmu}
\ee
\be
m_\mu =\frac{\mu_+^2}{\mu_+^2+\mu_-^2},
\label{eq:mmu}
\ee
$\text{cn}^{-1}(\cos\varphi|m_{\mu})$ is the inverse Jacobian elliptic
function, and $F(\varphi|m_\mu)$ is the incomplete elliptic integral
of the first kind. Therefore, the integral $\mathscr{J_{\mu}(\mu)}$
obtained from the integration over any range of $\mu$ can be expressed
as a combination of inverse Jocabian elliptic functions.

For the purpose of tracing photons from a distant observer back to the
surface of a star located at $(r_\star,\theta_\star,\phi_\star)$, we
integrate the trajectory of each photon from $\mu_{0}$ to $\mu_\star$.
For those photons that can finally reach the star surface, some of
them may reach the star surface when they first approach $\mu=
\mu_\star$. For these photons, the $\mu$-integral
$\mathscr{I}_{\mu,1}$ can be obtained by
\be
\ba
\mathscr{I}_{\mu,1}=\left\{ \begin{array}{lc}
-[\mathscr{F}_{\mu}(\mu_0)-\mathscr{F}_{\mu}(\mu_\star)], & \text{if~}
\beta>0 \text{~and~} \mu_0<\mu_\star,\\
 -[\mathscr{F}_{\mu}(\mu_0)+\mathscr{F}_{\mu}(\mu_\star)], &
\text{if~} \beta>0 \text{~and~} \mu_0>\mu_\star,\\
 +[\mathscr{F}_{\mu}(\mu_0)-\mathscr{F}_{\mu}(\mu_\star)], &
\text{if~} \beta<0 \text{~and~} \mu_0>\mu_\star,\\
 +[-2\mathscr{F}_{\mu}(0)+\mathscr{F}_{\mu}(\mu_0)
+\mathscr{F}_{\mu}(\mu_\star)], & \text{if~} \beta<0 \text{~and~}
\mu_0<\mu_{\star}.
\end{array}
\right.
\ea
\ee

Some photons may not reach the star surface when they first reach $\mu
= \mu_\star$. However, they may reach the star surface after their
trajectory, turning back from $\mu_+$ or $-\mu_+$ to $\mu_\star$ again.
For this case, we have 
\be
\mathscr{I}_{\mu,2}= \mathscr{I}_{\mu,1} + \delta \mathscr{I}_{\mu,\rm
even},
\ee
where
\be
\ba
\delta \mathscr{I}_{\mu,\rm even}= \left\{
\begin{array}{lc}
-2[\mathscr{F}_{\mu}(0)-\mathscr{F}_{\mu}(\mu_\star)], & \text{if~}
\beta>0 \text{~and~} \mu_0<\mu_\star,\\
 -2[2\mathscr{F}_{\mu}(0)-\mathscr{F}_{\mu}(\mu_\star)], & \text{if~}
\beta>0 \text{~and~} \mu_0>\mu_\star,\\
 -2[2\mathscr{F}_{\mu}(0)-\mathscr{F}_{\mu}(\mu_\star)], & \text{if~}
\beta<0 \text{~and~} \mu_0>\mu_\star,\\
 -2[\mathscr{F}_{\mu}(0)-\mathscr{F}_{\mu}(\mu_\star)], & \text{if~}
\beta<0 \text{~and~} \mu_0<\mu_{\star}.
\end{array}
\right.
\ea
\ee
Consequently, some photons may reach the star surface after their
trajectories turn back to $\mu_\star$ for the third time. For this
case, we have
\be 
\mathscr{I}_{\mu,3}= \mathscr{I}_{\mu,2} +\delta \mathscr{I}_{\mu,\rm
odd},
\ee
where
\be
\ba
\delta \mathscr{I}_{\mu,\rm odd} = \left\{
\begin{array}{lc}
-2[2\mathscr{F}_{\mu}(0)-\mathscr{F}_{\mu}(\mu_\star)], & \text{if~}
\beta>0 \text{~and~} \mu_0<\mu_\star,\\
 -2[\mathscr{F}_{\mu}(0)-\mathscr{F}_{\mu}(\mu_\star)], & \text{if~}
\beta>0 \text{~and~} \mu_0>\mu_\star,\\
 -2[\mathscr{F}_{\mu}(0)-\mathscr{F}_{\mu}(\mu_\star)], & \text{if~}
\beta<0 \text{~and~} \mu_0>\mu_\star,\\
 -2[2\mathscr{F}_{\mu}(0)-\mathscr{F}_{\mu}(\mu_\star)], & \text{if~}
\beta<0 \text{~and~} \mu_0<\mu_{\star}.
\end{array}
\right.
\ea
\ee
Therefore, if some photons reach the star surface when their
trajectories reach $\mu_\star$ for the $2k-$th time [$k (>1)$ is a
integer], then we have 
\be
\mathscr{I}_{\mu,2k} = \mathscr{I}_{\mu,2k-1} + \delta
\mathscr{I}_{\mu,\rm even};
\label{eq:imu2k}
\ee
while we have 
\be
\mathscr{I}_{\mu,2k+1} = \mathscr{I}_{\mu,2k} + \delta
\mathscr{I}_{\mu,\rm odd},
\label{eq:imu2kp1}
\ee
if some photons reach the star surface when their trajectories reach
$\mu_\star$ for the $(2k+1)-$th time.  For most photons included in
this paper, $k\leq 2-3$.

\subsection{Integral Over the $r$-direction}\label{sec:rint}

We denote the integral over $r$ as
\bd
\ba
\mathscr{F}_{r}(r)=\int \frac{d\tilde r}{\sqrt{R(\tilde r)}},
\ea
\ed
which can also be expressed in the form of the inverse Jacobian
elliptic functions. To do so, we need to first solve equation $R(r)=0$
and find its four roots, denoted as $r_1$, $r_2$, $r_3$, and $r_4$,
respectively. Since $R(r\rightarrow \pm \infty)\rightarrow \infty$ and
$R(r=0)= - a^2 q^2$, the solution of $R(r)=0$ can be divided into two
cases: case A, $R(r)=0$ has two real roots and two complex roots; and
case B, $R(r)=0$ has four real roots \citep[e.g.,see][]{Chandra83,
Rauch94, Cadevz98, YL00, LY01, Li05, DA09}. Below, we consider these two cases
separately. 

Case A: we assume that $r_1$ and $r_2$ represent the two complex
roots, and $r_3$ and $r_4$ ($r_4<r_3$) represent the two real roots,
respectively. Therefore, we have $r_1= r^*_2$, where $r^*_2$
represents the complex conjugate of $r_2$. Since $r_1r_2r_3r_4 =
|r_1|^2r_3r_4=-a^2q^2\leq 0$, $r_4$ must be $\leq 0$ and $r_3\geq 0$,
but $r_3$ cannot be equal to $r_4$. Since $R(r)$ must be $\geq 0$, we
have either $r\geq r_3$ or $r\leq r_4$, and then the physically
allowed region for $r$ must be $r\geq r_3$. We assume that the real
part of the two complex roots $r_1$ and $r_2$ is $u$, while the
imaginary parts of $r_1$ and $r_2$ are $w$ and $-w$, respectively.
Therefore, the two complex roots can be expressed as $r_1=u + iw$ and
$r_2=u-i w$, respectively. The two real roots can be expressed as
$r_3=-u+v$ and $r_4=-u -v$, respectively, where $v$ is a real positive
number. Now we can denote the function $\mathscr{F}_r(r)$ explicitly
as
\bd
\ba
\mathscr{F}_{r2}(r)=\int^r_{r_3} \frac{d\tilde r}{\sqrt{R(\tilde r)}}
=\int^r_{r_3} \frac{d\tilde{r}}{\sqrt{(\tilde r-r_1)(\tilde r-r_2)
(\tilde r-r_3)(\tilde r-r_4)}}.
\ea
\ed
According to \citet[][p135, eq. 260.00]{BF54}, the above function can
be expressed in terms of an inverse Jacobian elliptic function, i.e.,
\be
\mathscr{F}_{r2} (r)=g_{2}\text{cn}^{-1} (\cos\varphi|m_2) =g_{2}
F(\varphi|m_2),
\label{eq:fr}
\ee
where
\be
g_{2} =\frac{1}{\sqrt{AB}}, 
\ee
\be
m_2=\frac{(A+B)^2-4v^2}{4AB},
\label{eq:Am2}
\ee
\be
A=\sqrt{(v-2u)^2+w^2}, 
\ee
\be
B=\sqrt{(v+2u)^2+w^2}
\ee
\be
\varphi
=\cos^{-1}\left[\frac{(A-B)r+(u+v)A-(u-v)B}{(A+B)r+(u+v)A+(u-v)B}\right].
\ee
It is easy to verify here that $0<m_2<1$ and $\cos^2\varphi\leq 1$.
Therefore, the elliptical integral $F(\varphi|m_2)$ in
Equation~(\ref{eq:fr}) is well defined for $r\geq r_3$. Therefore,
$\mathscr{J}_r$ may also be expressed as a combination of inverse
Jacobian elliptic functions. 

In this case, some photons may reach the star surface with $r$
monotonically decreasing from $r_0$ to $r_{\rm hit}$, and some other
photons may reach the star surface with $r$ first decreasing from
$r_0$ to $r_3$ and then turning back to $r_{\rm hit}$. Hence, we should
have
\be
\mathscr{J}_r = -[\mathscr{F}_{r2}(r_0)-\mathscr{F}_{r2}(r_{\rm
hit})],
\label{eq:Jr21}
\ee
and
\be
\mathscr{J}_r = -[\mathscr{F}_{r2}(r_0)+\mathscr{F}_{r2}(r_{\rm
hit})],
\label{eq:Jr22}
\ee
respectively.

Case B: assuming that $r_1\geq r_2 \geq r_3 \geq r_4$. Since
$r_1 r_2 r_3 r_4 = -a^2 q^2 \leq 0$ and $r_1 + r_2 + r_3 + r_4 =0$, we
have at least $r_4 \leq 0$ and $r_1 > 0$. Therefore, the allowed
physical regions for photons with $R(r) \geq 0$ are $r\geq r_1$
(region I) and $r_3\leq r \leq r_2$ (region II, if $r_3 >0$). For
those photons in region I, i.e., $r\geq r_1$, we denote the function
$\mathscr{F}_r(r)$ as
\be
\mathscr{F}_{r4,\rm I}(r)= \int^{r}_{r_1}
\frac{d\tilde{r}}{\sqrt{R(\tilde{r})}} =\int^r_{r_1} \frac{d\tilde{r}
}{(\tilde r-r_1)(\tilde r-r_2)(\tilde r -r_3)(\tilde r-r_4)}.
\label{eq:Aintr4_1}
\ee
According to \citet[][p128, eq. 258.00]{BF54},
Equation~(\ref{eq:Aintr4_1}) can be expressed in terms of an inverse
Jacobian elliptic function, i.e.,
\be
\mathscr{F}_{r4,\rm I}(r) = g_4 \text{sn}^{-1}(\sin \varphi|m_4)= g_4
F(\varphi|m_4),
\label{eq:Arintr1eqr2}
\ee
where
\be
g_4=\frac{2}{\sqrt{(r_1-r_3)(r_2-r_4)}},
\label{eq:Ag4}
\ee
\be
\varphi=\sin^{-1}\left(\sqrt{\frac{(r_2-r_4)(r-r_1)}{(r_1-r_4)(r-r_2)}}\right),
\ee
\be
m_4=\frac{(r_2-r_3)(r_1-r_4)}{(r_1-r_3)(r_2-r_4)}.
\label{eq:Am4}
\ee
For this region, if $r_1\neq r_2$, $0\leq m_4 \leq 1$ and thus
Equation~(\ref{eq:Arintr1eqr2}) is well defined. If $r_1 = r_2$,
$m_4=1$ and $\sin^2\varphi=1$, and thus
Equation~(\ref{eq:Arintr1eqr2}) $\propto \text{sn}^{-1}(1|1) =
\infty$, which is not well defined. If $r_1=r_2$, however, the integration
in Equation~(\ref{eq:Aintr4_1}) can be directly obtained as
\be
\int^{r}_{r_1}\frac{d\tilde r}{\sqrt{R(\tilde r)}} =
\mathscr{H}_{r4,\rm Ieq}(r) = \frac{1}{\sqrt{(r_1-r_3)(r_1-r_4)}} \ln
\left[ \frac{\sqrt{(r-r_3)(r-r_4)}}{r-r_1} +
\frac{r^2_1+r_3r_4+2r_1r}{(r-r_1) \sqrt{(r_1-r_3)(r_1-r_4)}}\right].
\label{eq:Arintr1eq2}
\ee

For photons in the region I, some photons may reach the star surface
with $r$ monotonically decreasing from $r_0$ to $r_{\rm hit}$, while
some other photons may reach the star surface with $r$ first
monotonically decreasing from $r_0$ to $r_1$ and then turning back to
the star surface. If $r_1\neq r_2$, hence, we should also have
\be
\mathscr{J}_r = -[\mathscr{F}_{r4,\rm I}(r_0)- \mathscr{F}_{r4,\rm
I}(r_{\rm hit})],
\label{eq:Ar41}
\ee
and
\be
\mathscr{J}_r = -[\mathscr{F}_{r4,\rm I}(r_0)+ \mathscr{F}_{r4,\rm
I}(r_{\rm hit})],
\label{eq:Ar42}
\ee
respectively; if $r_1=r_2$ otherwise, we should have
\be
\mathscr{J}_r = -[\mathscr{H}_{r4,\rm Ieq}(r_0)- \mathscr{H}_{r4,\rm
Ieq}(r_{\rm hit})],
\label{eq:Ar43}
\ee
and
\be
\mathscr{J}_r = -[\mathscr{H}_{r4,\rm Ieq}(r_0)+ \mathscr{H}_{r4,\rm
Ieq}(r_{\rm hit})],
\label{eq:Ar44}
\ee
respectively.

For those photons in the region II, i.e., $r_2\geq r \geq r_3$, we
denote $\mathscr{F}_r(r)$ as
\be
\mathscr{F}_{r4,\rm II}(r)= \int^{r_2}_{r}
\frac{d\tilde{r}}{\sqrt{R(\tilde{r})}} =\int_r^{r_2} \frac{d\tilde{r}
}{(\tilde r-r_1)(\tilde r-r_2) (\tilde r-r_3)(\tilde r-r_4)}.
\label{eq:Aintr4_2}
\ee
According to \citet[][p116, eq. 255.00]{BF54},
Equation~(\ref{eq:Aintr4_2}) can be expressed in terms of an inverse
Jacobian elliptic function, i.e.,
\be
\mathscr{F}_{r4,\rm II}(r) = g_4 \text{sn}^{-1}(\sin \varphi|m_4)= g_4
F(\varphi|m_4),
\label{eq:Arintr2vr3}
\ee
where
\be
\varphi=\sin^{-1}\left(\sqrt{\frac{(r_1-r_3)(r_2-r)}{(r_2-r_3)(r_1-r)}}\right),
\ee
$g_4$, and $m_4$ are the same as that given by
Equations~(\ref{eq:Ag4}) and (\ref{eq:Am4}), respectively. Similar to
Equation~(\ref{eq:Arintr1eqr2}), Equation~(\ref{eq:Arintr2vr3}) is
well defined if $r_1\neq r_2$ as $0\leq m_4\leq 1$ and
$\sin^2\varphi<1$. If $r_1 = r_2$, then Equation~(\ref{eq:Arintr2vr3})
can be directly replaced by
\be
\int^{r_1}_{r}\frac{d\tilde r}{\sqrt{R(\tilde r)}} =
\mathscr{H}_{r4,\rm IIeq}(r) = \frac{1}{\sqrt{(r_1-r_3)(r_1-r_4)}} \ln
\left[ \frac{\sqrt{(r-r_3)(r-r_4)}}{r_1-r} +
\frac{r^2_1+r_3r_4+2r_1r}{(r_1-r) \sqrt{(r_1-r_3)(r_1-r_4)}}\right].
\ee

For photons in the region II, some of them may reach the star surface
with $r$ monotonically decreasing from $r_0$ to $r_{\rm hit}$, while
some others may reach the star surface with $r$ first monotonically
decreasing from $r_0$ to $r_3$ and then turning back to the star
surface. For these two cases, if $r_0\leq r_2$ and $r_2\neq r_3$, we
should have
\be
\mathscr{J}_r = [\mathscr{F}_{r4,\rm II}(r_0)-\mathscr{F}_{r4,\rm
II}(r_{\rm hit})],
\ee
and
\be
\mathscr{J}_r =-2\mathscr{F}_{r4,\rm II}(r_3)+ [\mathscr{F}_{r4,\rm
II}(r_0)+\mathscr{F}_{r4,\rm II}(r_{\rm hit})],
\ee
respectively; if $r_0\leq r_1$ and $r_1=r_2$ otherwise, we should have
\be
\mathscr{J}_r = [\mathscr{H}_{r4,\rm IIeq}(r_0)-\mathscr{H}_{r4,\rm
IIeq}(r_{\rm hit})],
\ee
and
\be
\mathscr{J}_r = -2\mathscr{H}_{\rm r4,\rm II}(r_3)+
[\mathscr{H}_{r4,\rm IIeq}(r_0)+\mathscr{H}_{r4,\rm IIeq}(r_{\rm
hit})],
\ee
respectively.

For the problem considered in this paper, the photons that we are
interested in must be transported from the stars rotating around the
central MBH to a distant observer. Our calculations show that their
trajectories should not be bended too much to oscillate within the
region II (i.e., $r_3\leq r\leq r_2$) for many times. 


\subsection{$r_{\rm hit}$}\label{sec:rhit}

According to Equation~(\ref{eq:rtheta}), $\mathscr{I}_{\mu} =
\mathscr{J}_{r}$, so $r_{\rm hit}$ can be derived from this equation
once the two integrals are obtained as described above. If a photon
reaches $\mu=\mu_\star$ for $k$ times, then $\mathscr{I}_{\mu} =
\mathscr{I}_{\mu,k}$ (eqs.~\ref{eq:imu2k} and~\ref{eq:imu2kp1}). 

Case A of the $r-$integral: 

\begin{enumerate}

\item If a photon reaches the star surface before it encounters the
turning point $r_3$, then 
\be
\mathscr{I}_{\mu,k} = -[\mathscr{F}_{r2}(r_0)-\mathscr{F}_{r2}(r_{{\rm
hit},k})].
\ee
Therefore, we have the solution
\be
r_{{\rm hit},k} = \mathscr{F}^{-1}_{r2}(\mathscr{I}_{\mu,k} +
\mathscr{F}_{r2}(r_0)).
\ee
We define $s_{21}=\mathscr{I}_{\mu,k} + \mathscr{F}_{r2}(r_0)$, then
we may express $r_{{\rm hit},k}$ as
\be
r_{{\rm hit},k}=\mathscr{F}_{r2}^{-1}(s)=\frac{b_1-b_2\cos\varphi}{
a_2\cos\varphi-a_1},
\label{eq:rinverse}
\ee
where
\be
a_1=A-B,
\label{eq:CaseAa1}
\ee
\be
a_2=A+B,
\ee
\be
b_1=-(u+v)A+(u-v)B,
\ee
\be
b_2=-(u+v)A-(u-v)B,
\label{eq:CaseAb2}
\ee
and
\be
\cos\varphi =\text{cn}\left(\left.\frac{s_{21}}{g_{2}}\right|
m_2\right),
\label{eq:CaseAphi}
\ee
and $m_2$ is given by Equation~(\ref{eq:Am2}).

\item If a photon reaches the star surface after it encounters the
turning point $r_3$, according to Equation~(\ref{eq:Jr22}), then we
have
\be
r_{{\rm
hit},k}=\mathscr{F}_{r2}^{-1}(s_{22})=\frac{b_1-b_2\cos\varphi}{
a_2\cos\varphi-a_1},
\ee
where
\be
s_{22}=- \mathscr{I}_{\mu,k} - \mathscr{F}_{r2}(r_0),
\ee
and $a_1$, $a_2$, $b_1$, $b_2$, and $\cos\varphi$ are given by
Equations~(\ref{eq:CaseAa1})-(\ref{eq:CaseAphi}).  If $s_{22}
>\mathscr{F}_{r2}(\infty)$, then there is no solution for $r_{{\rm
hit},k}$, and therefore it is not necessary to go to larger $k$ to
search for the solution of $r_{{\rm hit},k}$.

\end{enumerate}

Case B of the $r$-integral:

\begin{enumerate}

\item For photons in the allowed region $r\geq r_1$ (but $r_1\neq
r_2$), if they reach the star surface before they encounter the
turning point $r_1$, according to Equation~(\ref{eq:Ar41}) we should
have
\be
r_{{\rm hit},k}=\mathscr{F}^{-1}_{r4,\rm I}(s_{41})= \frac{r_2 C -
r_1}{ C -1},
\ee
where 
\be
s_{41}=\mathscr{I}_{\mu,k}+\mathscr{F}_{r4,\rm I}(r_0),
\ee
\be
C=\frac{r_1-r_4}{r_2 - r_4} \sin^2\varphi,
\ee
\be
\sin\varphi=\text{sn} \left(\left.\frac{s_{41}}{g_4}\right|
m_4\right),
\label{eq:Asinphi41}
\ee
$g_4$ and $m_4$ are the same as that given by Equations~(\ref{eq:Ag4})
and (\ref{eq:Am4}), respectively. 

\item For photons in the allowed region $r\geq r_1$ (but $r_1\neq
r_2$), if they reach the star surface after they encounter the turning
point $r_1$, according to Equation~(\ref{eq:Ar42}) we should have
\be
r_{{\rm hit},k}=\mathscr{F}^{-1}_{r4,\rm I}(s_{42}) = \frac{r_2 C -
r_1}{ C -1},
\ee
where 
\be 
s_{42}= -\mathscr{I}_{\mu,k} - \mathscr{F}_{r4,\rm I}(r_0),
\ee
\be
\sin\varphi=\text{sn} \left(\left.\frac{s_{42}}{g_4}\right|
m_4\right),
\ee
$C$, $g_4$, and $m_4$ are the same as that given by
Equations~(\ref{eq:Asinphi41}), (\ref{eq:Ag4}) and (\ref{eq:Am4}),
respectively. If $s_{42} > -\mathscr{I}_{\mu,k} -\mathscr{F}_{r4,\rm
I}(\infty)$, then there is no solution for $r_{{\rm hit},k}$, and it
is not necessary to go to larger $k$ to search for the solution of
$r_{{\rm hik},k}$.

\item For photons in the allowed region $r\geq r_1$ and $r_1 = r_2$,
if the reach the star surface before they encounter the turning point
$r_1$, according to Equation~(\ref{eq:Ar43}) we should have
\be
r_{{\rm hit},k} = \mathscr{H}^{-1}_{r4,\rm Ieq}(s_{43}),
\ee
where 
\be
s_{43}= \mathscr{I}_{\mu,k} + \mathscr{H}_{r4,\rm Ieq}(r_0).
\ee

\item For photons in the allowed region $r\geq r_1$ and $r_1 = r_2$,
if they reach the star surface after they encounter the turning point
$r_1$, according to Equation~(\ref{eq:Ar43}) we should have
\be
r_{{\rm hit},k} = \mathscr{H}^{-1}_{r4,\rm Ieq}(s_{44}),
\ee
where
\be
s_{44}= -\mathscr{I}_{\mu,k} - \mathscr{H}_{r4,\rm Ieq}(r_0).
\ee
If $s_{44} > -\mathscr{I}_{\mu,k} -\mathscr{H}_{r4,\rm II}(\infty)$,
then there is no solution for $r_{{\rm hit},k}$, and consequently it
is not necessary to go to larger $k$ to search for the solution of
$r_{{\rm hit},k}$.

\item For photons in the allowed region $r_2\geq r \geq r_3$, they
cannot transport to the distant observer if $r_0\gg r_2$, and
therefore are not interested in this study.

\end{enumerate}

\subsection{Longitude and Time}~\label{sec:phitint}

The motion of a photon in the $\phi$- and $t$-directions can be also
obtained as \citep[e.g., see][]{Rauch94}
\be
\phi = r_{\rm sign}\int^{r} \frac{\lambda
r^2+2r(a-\lambda)}{r^2-2r+a^2}\frac{dr}{\sqrt{R(r)}} +\theta_{\rm
sign}\int^{\mu}\frac{\lambda\mu^2}
{1-\mu^2}\frac{d\mu}{\sqrt{\Theta_{\mu}}}
\label{eq:phiint}
\ee
and
\be
t = r_{\rm sign}\int^{r}\frac{r^4+a^2r^2
+2ar(a-\lambda)}{r^2-2r+a^2}\frac{dr}{\sqrt{R(r)}} +\theta_{\rm
sign}\int^{\mu} a^2\mu^2\frac{d\mu} {\sqrt{\Theta_{\mu}}},
\label{eq:tint}
\ee
provided that the motion of the photon in the $r$- and
$\theta$-directions are given (see Sections~\ref{sec:thetaint} and
\ref{sec:rint}). In Equations~(\ref{eq:phiint}) and (\ref{eq:tint}),
$r_{\rm sign}$ and $\theta_{\rm sign}$ are the signs of the $r$- and
$\theta$-integrals, which can be obtain by solving
Equation~(\ref{eq:rtheta}) (see details in Sections~\ref{sec:rint} and
\ref{sec:thetaint}). The integrations on the right sides of
Equations~(\ref{eq:phiint}) and (\ref{eq:tint}) over either $r$ or
$\theta$ can be expanded similar to those done in
Sections~\ref{sec:rint} and \ref{sec:thetaint} according to $r_{\rm
sign}$ and $\theta_{\rm sign}$ and how many times that they change
sign; we do not show it here in detail. The $r$-terms in
Equations~(\ref{eq:phiint}) and (\ref{eq:tint}) can be accurately
integrated by adopting the Gauss-Kronrod integration scheme (see also
\citealt{Rauch94}). The $\theta$-terms can be obtained analytically by
the elliptical functions as
\begin{eqnarray}
\mathscr{K}_{\mu}(\mu) &\equiv& \int^{\mu_+}_{\mu}\frac{\tilde\mu^2}{1
-\tilde\mu^2}\frac{d\tilde \mu} {\sqrt{\Theta_{\mu}(\tilde\mu)}} =
\int^{\mu_+}_{\mu} \frac{1}{1-\tilde{\mu}^2}
\frac{d\tilde{\mu}}{\sqrt{\Theta_{\mu}(\tilde{\mu})}}
-\int^{\mu_+}_{\mu}
\frac{d\tilde{\mu}}{\sqrt{\Theta_{\mu}(\tilde{\mu})}} \nonumber \\
 & = & \frac{1}{1-\mu_+^2} g_\mu
{\Pi}\left(\left.\cos^{-1}\frac{\mu}{\mu_+}\right|n_\mu, m_\mu\right)
-g_\mu
\mathscr{F}\left(\left.\cos^{-1}\frac{\mu}{\mu_+}\right|m_\mu\right),
~\label{eq: phi_theta}
\end{eqnarray}
according to \citet[][p48, eq. 213.02]{BF54}, and
\begin{eqnarray}
\mathscr{L}_{\mu}(\mu) & \equiv &\int^{\mu_+}_{\mu}\tilde\mu^2
\frac{d\tilde\mu}{\sqrt{\Theta_{\mu}(\tilde\mu)}} =  \frac{1}{|a|}
\int^{\mu_+}_{\mu}
\sqrt{\frac{\mu^2_-+\tilde{\mu}^2}{\mu^2_+-\tilde{\mu}^2}}
d\tilde{\mu} - {\mu}^2_- \int^{\mu_+}_{\mu}
\frac{d\tilde{\mu}}{\sqrt{\Theta_{\mu}(\tilde{\mu})}} \nonumber \\
 & = & (\mu_+^2+\mu_-^2) g_\mu
\mathscr{E}\left(\left.\cos^{-1}\frac{\mu}{\mu_+}\right|m_\mu\right)
-\mu_-^2 g_\mu
\mathscr{F}\left(\left.\cos^{-1}\frac{\mu}{\mu_+}\right|m_\mu\right),
\end{eqnarray}
according to \citet[][p48, eq. 213.01]{BF54}, where $\mathscr{F}$,
$\mathscr{E}$, and ${\Pi}$ are the elliptical integral of the first,
second, and third kind, respectively, $g_\mu$ and $m_\mu$ are given by
Equations~(\ref{eq:gmu}) and (\ref{eq:mmu}), and $n_\mu$ is given by
\be
n_\mu =\frac{\mu_+^2}{1-\mu_+^2}.
\ee

\subsection{Apparent Position of a Star Close to the GC MBH in the
Observer's Sky Plane}
It is necessary to develop a fast and accurate method to find those
rays propagating from the observer to the star. The most simple and
direct method is to perform ray tracing calculations for a large
number of rays with various ($\alpha$, $\beta$) and search for the
minimum of the distance function $d(\alpha, \beta) = d(r_{\rm hit},
\theta_\star,\phi_{\rm hit})$ (defined by Eq.~\ref{eq:dst}). However,
this method may be not efficient in finding the ray that actually hits
on the star at any given position because (1) not every ray with
arbitrary $(\alpha, \beta)$ can end up at the surface of the cone with
$\theta=\theta_\star$; and (2) the contours for $d(\alpha,\beta)$
appear to be extremely narrow elliptically in some cases; thus, the
convergence for numerically finding the local minimum of $d$ is slow. 

In this study, we implement a new method to find the local minimum of
$d(\alpha,\beta)$. We assume a flat space with coordinate
$(r',\theta',\phi')$ as an approximation to the Kerr space. In this
flat space, those rays from the distant observer that can hit the
position $(r', \theta_\star, \phi')$ have 
\be
\alpha=r'\sin\theta_\star \sin\phi', \,\,\,\,\beta= r'\cos\theta_\star
\sin\theta_{\rm o}- r'\sin\theta_\star \cos\phi' \cos\theta_{\rm o}.
\label{eq:alphabeta0}
\ee
Here and hereafter the superscript $'$ indicates that a quantity is
defined in the flat space. Since the example stars are still far away
from the event horizon, the trajectories of photons emitted from those
stars do not deviate away from the corresponding cases in a
flat space too much.  Therefore, the local minimum of the distance function $d$
can be found by the following procedures.

\begin{enumerate}
\item We first assume $(r'_0,\phi'_0)=(r_{\star},\phi_{\star})$ to
obtain $(\alpha_0,\beta_0)$, which is taken as the guess position of
the star [$(r_\star,\theta_\star,\phi_\star)$ in the Kerr space] on
the observer's sky plane. Adopting $(\alpha,\beta) =
(\alpha_0,\beta_0)$, we perform ray tracing calculations in the Kerr
metric to find the position of the ray hitting the surface
$\theta=\theta_\star$, i.e., $(r_{0}, \theta_\star, \phi_{0})$, which
may deviate from the real position of the star
$(r_\star,\theta_\star,\phi_\star)$ because of the curved spacetime.
We calculate the distance between $(r_\star,\theta_\star,\phi_\star)$
and $(r_{0},\theta_\star,\phi_{0})$ as $d_0=d(\alpha_0,\beta_0)$. 

\item We select a number of points (e.g., eight points) on a circle with
radius of $d'_0 = d_0$ around $(r'_0,\theta_\star,\phi'_0)$.  For the
$j$-th point ($j=1, ..., 8$) selected on the circle, i.e.,
$(r'_{1,j},\theta_*,\phi'_{1,j}$), we can obtain
$(\alpha_{1,j},\beta_{1,j})$ according to
Equation~(\ref{eq:alphabeta0}).  Using $(\alpha_{1,j},\beta_{1,j})$,
we perform ray tracing calculations again and find the position
$(r_{1,j},\theta_\star,\phi_{1,j})$ that the ray hits  on the surface
$\theta= \theta_\star$, and consequently, we obtain the distance of
that position from the star as $d_{1,j}=d(\alpha_{1,j},\beta_{1,j})$.
Among those (eight) points on the circle, we find the one [denoted by
$(r_{1,\rm m},\theta_\star,\phi_{1,\rm m})$] with the smallest
distance $d$ (denoted by $d_{1,\rm m}$), and the corresponding point
in the flat space is denoted by $(r'_{1,\rm
m},\theta_\star,\phi'_{1,\rm m})$.

\item If $d_{1,\rm m} \leq \xi r_\star$ ($\xi = 10^{-6}\sim 10^{-8}$),
then we find the apparent position of the star and the calculation is
terminated; if $d_{1,\rm m} > \xi r_\star$ and $d_{1,\rm m} > d_0$,
then we adopt the same point $(r'_0,\theta_*,\phi'_0)$ in the flat
space as the center of a new circle with a radius of $d'_1=d'_0/5$;
otherwise, we choose the point $(r'_{1,\rm m},\theta_*,\phi'_{1,\rm
m})$ in the flat space as the center of the new circle and its radius
is set to be $d'_1=d'_0$.

\item We select a number of points
$(r'_{2,j},\theta_\star,\phi'_{2,j})$ on the new circle and obtain
$(\alpha_{2,j},\beta_{2,j})$. We further perform ray tracing
calculations and find the position $(r_{2,j},\theta_\star,\phi_{2,j})$
that the ray hits on the surface $\theta= \theta_\star$. We calculate
the distance of that position from the star
($r_\star,\theta_\star,\phi_\star$), i.e.,
$d_{2,j}=d(\alpha_{2,j},\beta_{2,j})$ in the Kerr space. Among those
selected points, we find the one with the smallest $ d$, i.e.,
$d_{2,\rm m}$, and the corresponding point in the flat space is
denoted as $(r'_{2,\rm m},\theta_*,\phi'_{2,\rm m})$.

\item Similar to the above procedures 3 and 4, we do iterations until
we find the apparent position of the star at
($r_\star,\theta_\star,\phi_\star$) in the observer's sky plane.
\end{enumerate}

The above method is fast in searching for the apparent position
$(\alpha,\beta)$ of a star at $(r_\star, \theta_\star,\phi_\star)$ in
the Kerr spacetime on the distant observer's sky plane, which enables
an efficient MCMC fitting to the mock data
obtained for those example stars. 

\section{Analytical Estimation of the Shift of Star Apparent Positions
due to the Lense-Thirring Precession and the Frame Dragging}
\label{sec:analytical}

The Lense-Thirring effect causes the precession of the orbital plane
via the change of the longitude of the ascending node at a mean rate
of
\be
<\dot{\Omega}> =\frac{ 2 a M_{\bullet} }{ a^3_{\rm orb} (1- e^2_{\rm
orb})^{3/2}},
\ee
and the change of the argument of periapsis at a mean rate
\be
<\dot{\Upsilon}> = - 3 \dot{\Omega} \cos \zeta,
\ee
where $\Omega$ and $\Upsilon$ are the longitude of the ascending node
and the argument of periapsis defined for the stellar orbital plane
with the equatorial plane of the BH as the reference plane, and
$\zeta$ is the angle between the orbital angular momentum of the star
and the spin direction and given by $\zeta = \arccos (\cos I' \cos i +
\sin I' \sin i \sin (\Omega' -\epsilon))$.  The accumulated changes of
$\Omega$ and $\Upsilon$ per orbit are
\be
\delta \Omega = 2\pi \frac{2 a r^{3/2}_{\rm g}}{a_{\rm orb}^{3/2}
(1-e_{\rm orb}^2)^{3/2}},
\label{eq:domegap}
\ee
and
\be \delta \Upsilon = - 2\pi \frac{6 a r^{3/2}_{\rm g}}{a_{\rm
orb}^{3/2} (1-e_{\rm orb}^2)^{3/2}}\cos \zeta.
\label{eq:dUpsilonp}
\ee
Note that most of the changes in $\Omega$ and $\Upsilon$ take place
near periapsis.  The values of the changes are $\delta \Omega =
0.0017\arcdeg$, $0.00059\arcdeg$, $0.010\arcdeg$, $0.14\arcdeg$,
$0.074\arcdeg$, $0.0092\arcdeg$, and $\delta \Upsilon =
0.0044\arcdeg$, $0.0010\arcdeg$, $-0.015\arcdeg$, $-0.21\arcdeg$,
$-0.11\arcdeg$, and $-0.014\arcdeg$ for S0-2/S2, S0-102, Ea, Eb, Ec,
and Ed, respectively. Considering the projection effect, the changes
of $\Omega'$, $\Upsilon'$, and $I'$ can be obtained by
\be
\delta \Omega' =\left[ \cos i - \frac{\cos I' \sin i \sin
(\Omega'-\epsilon)}{\sin I'} \right]\delta \Omega,
\label{eq:domega}
\ee
\be
\delta \Upsilon' = - \frac{\sin i \sin(\Omega' - \epsilon)}{\sin I'}
\delta \Omega + \delta \Upsilon,
\label{eq:dUpsilon}
\ee
and
\be
\delta I' = \sin i\cos(\Omega'-\epsilon)\delta \Omega.
\label{eq:dI}
\ee
If $\zeta \rightarrow 0\arcdeg$, i.e., the star orbit is close to
being
on the MBH equatorial plane, $I' \rightarrow i$, $\Omega'
- \epsilon \rightarrow \pi/2$, then $\delta \Omega' \rightarrow 0$,
  $\delta \Upsilon' \rightarrow -2 \delta \Omega$, and $\delta I'
\rightarrow 0$.

After the motion of one full orbit, the difference between the
position of an example star in the sky plane at apoapsis/periapsis for
the case with a rapidly spinning BH ($a$) and that with a non-spinning
BH ($a=0$) is
\begin{eqnarray}
\delta_{\rm apo/peri} & \simeq & a_{\rm orb} (1 \pm  e_{\rm orb})
\left[ \delta^2 \Omega' (1-\sin^2 \Upsilon' \sin^2 I') + 
\delta^2 \Upsilon' (1-\cos^2\Upsilon' \sin^2 I')
+\sin\Upsilon'^2\sin^2I' \delta^2 I' \right. \nonumber \\
& & + 2\cos I'\delta \Omega'\delta\Upsilon'
-2\sin\Upsilon'\cos\Upsilon'\sin I'\cos I'\delta\Upsilon'\delta I'
 -\left.2\sin\Upsilon'\cos\Upsilon'\sin I'\delta\Omega'\delta I'
\right]^{1/2},
\label{eq:Delta_A}
\end{eqnarray}
where the signs `+' and `-' are for the difference at apoapsis
($\delta_{\rm apo}$) and periapsis ($\delta_{\rm peri}$),
respectively.

We may also include the high-order precession due to the quadruple
moment of the MBH by replacing $\delta \Omega$ and $\delta \Upsilon$
in Equations~(\ref{eq:domegap}) and (\ref{eq:dUpsilonp})  by
$ \delta \Omega + \delta \Omega_{\rm Q}$ and $\delta \Upsilon +
\delta \Upsilon_{\rm Q}$, respectively, where 
\be
\delta \Omega_{\rm Q} = - \cos \zeta  \frac{3\pi a^2 r^2_{\rm g}}
{a^2_{\rm orb} (1-e^2_{\rm orb})^2},
\ee 
and
\be
\delta \Upsilon_{\rm Q} = \frac{1- 5 \cos^2 \zeta}{2}  \frac{3\pi a^2
r^2_{\rm g}} {a^2_{\rm orb} (1-e^2_{\rm orb})^2}.
\ee See detailed derivations from \citet[][]{Barker75} and
\citet{Wex99}.


\begin{thebibliography}

\bibitem[Abramowitz \& Stegun(1972)]{AS72} Abramowitz, M., \& Stegun,
I.\ A.\ 1972, Handbook of Mathematical Functions (New Yorker: Dover)

\bibitem[Ang{\'e}lil \& Saha(2010)]{Angelil10b} Ang{\'e}lil, R., \&
Saha, P.\ 2010, ApJ, 711, 157

\bibitem[Ang{\'e}lil et al.(2010)]{Angelil10a} Ang{\'e}lil, R., Saha,
P., \& Merritt, D.\ 2010, ApJ, 720, 1303 

\bibitem[Ang{\'e}lil \& Saha(2011)]{Angelil11} Ang{\'e}lil, R., \&
Saha, P.\ 2011, ApJL, 734, 19

\bibitem[Ang{\'e}lil 
\& Saha(2014)]{Agnelil14} Ang{\'e}lil, R., \& Saha, P.\ 2014, \mnras, 444, 3780

\bibitem[Bardeen et al.(1972)]{Bardeen72} Bardeen, J.\ M., Press, W.\
H., \& Teukolsky, S.\ A.\ 1972, ApJ, 178, 347 

\bibitem[Bardeen(1973)]{Bardeen73}Bardeen, J.\ M.\ 1973, in Black
holes, ed. C. DeWitt \& B.\ S., DeWitt (New York: Gordon \& Breach),
215

\bibitem[Barker \& O'Connell(1975)]{Barker75} Barker, B.~M., \&
O'Connell, R.~F.\ 1975, \prd, 12, 329 

\bibitem[Boyer \& Lindquist(1967)]{Boyer67} Boyer, R.\ H., \&
Lindquist, R.\ W.\ 1967, Journal of Mathematical Physics, 8, 265 

\bibitem[Byrd \& Friedman(1954)]{BF54}Byrd, P.\ F., \& Friedman, M.\
D.\ 1954, Handbook of Elliptic Integrals for Engineers and Physicists
(Berlin: Spinger)

\bibitem[{\v C}ade{\v z} et al.(1998)]{Cadevz98} {\v C}ade{\v z}, A.,
Fanton, C., \& Calvani, M.\ 1998, New Astronomy, 3, 647 

\bibitem[Chandrasekhar(1983)]{Chandra83}Chandrasekhar, S.\ 1983, The
Mathematical Theory of Black holes (Oxford: Oxford Univ. Press)

\bibitem[Cunningham \& Bardeen(1973)]{Cunningham73} Cunningham, C.\
T., \& Bardeen, J.\ M.\ 1973, ApJ, 183, 237 

\bibitem[Dexter \& Agol(2009)]{DA09} Dexter, J., \& Agol, E.\ 2009,
\apj, 696, 1616 

\bibitem[Doeleman et al.(2009)]{Doeleman09} Doeleman, S. S., Fish V.
L., Broderick, A. E., Loeb, A., \& Rogers, A. E. E. 2009, ApJ, 695, 59

\bibitem[Dormand \& Prince(1980)]{DP80}Dormand, J.\ R., \& Prince, P.\
J.\ 1980, J.\ Comp.\ Appl.\ Math., Vol.6, p.19

\bibitem[Fragile \& Mathews(2000)]{Fragile00} Fragile, P.~C., \&
Mathews, G.~J.\ 2000, \apj, 542, 328 

\bibitem[Genzel et al.(2010)]{Genzel10} Genzel, R., Eisenhauer, F., \&
Gillessen, S.\ 2010, Reviews of Modern Physics, 82, 3121 

\bibitem[Ghez et al.(2008)]{Ghezetal08}Ghez, A.,  Salim, S., Weinberg,
N.\ N., Lu, J.\ R., Do, T., Dunn, J.\ K., Matthews, K., Morris, M.\
R., Yelda, S., Becklin, E.\ E., et al.\ 2008, ApJ, 689, 1044

\bibitem[Gillessen et al.(2009)]{Gillessenetal09}Gillessen, S.,
Eisenhauer, F., Trippe, S., Alexander, T., Genzel, R., Martins, F., \&
Ott, T.\ 2009, ApJ, 692, 1075

\bibitem[Gualandris et al.(2010)]{Gualandris10} Gualandris, A.,
Gillessen, S., \& Merritt, D.\ 2010, MNRAS, 409, 1146

\bibitem[Hairer et al.(1993)]{Hairer93}Hairer, E., Norsett, S.\ P., \&
Wanner, G.\ 1987, Solving Ordinary Differential Equations I.\ Nonstiff
Problems, Springer Series in Comput. Mathematics, Vol. 8
(Springer-Verlag)

\bibitem[Hansen \& Milosavljevi{\'c}(2003)]{HM2003} Hansen, B.~M.~S.,
\& Milosavljevi{\'c}, M.\ 2003, \apjl, 593, L77 

\bibitem[Iorio(2011a)]{Iorio11a}Iorio, L.\ 2011a, \prd, 84, 124001 

\bibitem[Iorio(2011b)]{Iorio11b}Iorio, L.\ 2011b, MNRAS, 411, 453

\bibitem[Jaroszynski(1998)]{Jaroszynski98} Jaroszynski, M.\ 1998, Acta
Astron., 48, 653

\bibitem[Johannsen \& Psaltis(2011)]{Johannsen11} Johannsen, T., \&
Psaltis, D.\ 2011, \prd, 83, 124015 

\bibitem[Kannan \& Saha(2009)]{KannanSaha09} Kannan, R., \& Saha, P.\
2009, \apj, 690, 1553 

\bibitem[Karas et al.(1992)]{Karas92}Karas, V., Vokrouhlicky, D., \&
Polnarev, A.\ G.\ 1992, MNRAS, 259, 569

\bibitem[Kerr(1963)]{Kerr63} Kerr, R.\ P.\ 1963, Physical Review
Letters, 11, 237

\bibitem[Kopeikin \& Makarov(2007)]{Kopeikin07} Kopeikin, S.~M., \&
Makarov, V.~V.\ 2007, \prd, 75, 062002 

\bibitem[Kormendy \& Ho(2013)]{KH13}Kormendy, J., \& Ho, L.\ C.\ 2013,
ARA\&A, 51, 511

\bibitem[Li et al.(2005)]{Li05}Li, L.-X., Zimmerman, E.\ R., Narayan,
R., \& McClintock, J.\ E.\ 2005, ApJS, 157, 335

\bibitem[Lu \& Yu(2001)]{LY01}Lu, Y., \& Yu, Q.\ 2001, ApJ, 561, 660

\bibitem[Magorrian et al.(1998)]{Magorrian1998} Magorrian, J.,
Tremaine, S., Richstone, D., et al.\ 1998, AJ, 115, 2285 

\bibitem[Magorrian \& Tremaine(1999)]{Magorrian99}Magorrian, J., \&
Tremaine, S.\ 1999, MNRAS, 309, 447

\bibitem[McClintock et al.(2011)]{McClintock11}McClintock, J.\ E.,
Narayan, R., Davis, S.~W., et al.\ 2011, CQGra, 28, 114009

\bibitem[Merritt et al.(2007)]{Merritt07} Merritt, D., Berczik, 
 P., \& Laun, F.\ 2007, \aj, 133, 553

\bibitem[Merritt et al.(2010)]{Merritt10} Merritt, D., Alexander, T.,
Mikkola, S., \& Will, C.~M.\ 2010, \prd, 81, 062002

\bibitem[Meyer et al.(2012)]{Meyer12} Meyer, L., Ghez, A.\ M.,
Sch{\"o}del, R., et al.\ 2012, Science, 338, 84 

\bibitem[Misner et al.(1973)]{Misner73}Misner, C.\ W., Thorne, K.\ S.,
\& Wheeler, J.\ A.\ 1973, Gravitation (San Francisco: W.\ H.\ Freeman
and Co.)

\bibitem[Narayan \& McClintock(2013)]{NM13}Narayan, R., \& McClintock,
J.\ E.\ 2013, arXiv:1312.6698

\bibitem[Peters(1964)]{Peters64}Peters, P.\ C.\ 1964, Phy. Rev., 136,
1224

\bibitem[Preto \& Saha(2009)]{PS09}Preto, M., \& Saha, P.\ 2009, \apj,
703, 1743

\bibitem[Psaltis(2012)]{Psaltis12} Psaltis, D.\ 2012, \apj, 759, 130 

\bibitem[Psaltis et al.(2013)]{Psaltis13} Psaltis, D., Li, G., \&
Loeb, A.\ 2013, \apj, 777, 57 

\bibitem[Rauch \& Blandford(1994)]{Rauch94} Rauch K. P., \& Blandford
R. D.\ 1994, ApJ, 421, 46

\bibitem[Reynolds(2013)]{Reynolds13}Reynolds, C.\ S., 2013,
CQGra, 30, 244004

\bibitem[Rubilar \& Eckart(2001)]{RE01}Rubilar, G.\ F., \& Eckart, A.\
2001, A\&A, 374, 95

\bibitem[Sadeghian \& Will(2011)]{Sadeghian11} Sadeghian, L., \& Will,
C.~M.\ 2011, Classical and Quantum Gravity, 28, 225029 

\bibitem[Salim \& Gould(1999)]{SalimGould99} Salim, S., \& Gould, A.\
1999, \apj, 523, 633 

\bibitem[Sch{\"o}del et al.(2012)]{Schodel12} Sch{\"o}del, R., Ott,
T., Genzel, R., et al.\ 2002, \nat, 419, 694 

\bibitem[Thorne(1986)]{Thorne86}Thorne, K.\ S.\ 1986, "Black Holes:
The Membrane Viewpoint," in Highlights of Modern Astrophysics:
Concepts and Controversies, eds. Stuart L. Shapiro and Saul A.
Teukolsky (John Wiley \& Sons, New York, 1986), pp. 103-161

\bibitem[Weinberg et al.(2005)]{Weinberg05}Weinberg, N.\ N.,
Milosavljevi\'{c}, M., \& Ghez, A.\ M.\ 2005, ApJ, 622, 878

\bibitem[Wex \& Kopeikin(1999)]{Wex99} Wex, N., \& Kopeikin, S.\ M.\
1999, ApJ, 514, 388

\bibitem[Will(1993)]{Will93}Will, C.\ M.\ 1993, Theory and Experiment
in Gravitational Physics (Cambridge: Cambridge Univ. Press)

\bibitem[Will(2008)]{Will08}Will, C.\ M.\ 2008, ApJ, 674, L25

\bibitem[Yu \& Lu(2000)]{YL00}Yu, Q., \& Lu, Y.\ 2000, MNRAS, 311, 161

\bibitem[Yu \& Tremaine(2003)]{YT03}Yu, Q., \& Tremaine, S.\ 2003,
ApJ, 599, 1129

\bibitem[Yu et al.(2015)]{Yuetal14} Yu, Q., Zhang, F., \& Lu, Y.\ 2015, 
submitted

\bibitem[Zhang et al.(2013)]{ZLY13}Zhang, F., Lu, Y., \& Yu, Q.\ 2013,
ApJ, 768, 153

\bibitem[Zhang et al.(2014)]{ZLY14}Zhang, F., Lu, Y., \& Yu, Q.\ 2014,
ApJ, 784, 106

\bibitem[Zucker et al.(2006)]{Zucker06} Zucker, S., Alexander, T.,
Gillessen, S., Eisenhauer, F., \& Genzel, R.\ 2006, \apjl, 639, L21 

\end{thebibliography}
\end{document}